\documentclass[a4paper,11pt]{report}

\usepackage{eucal}
\usepackage[all]{xy}
\usepackage[centertags]{amsmath}
\usepackage{latexsym}
\usepackage{hyperref}

\usepackage{amsfonts}
\usepackage{amssymb}
\usepackage{amsthm}
\usepackage{fancyhdr}
\usepackage [dvips]{epsfig}
\usepackage{amssymb,amsfonts,amstext,amsmath,graphicx,epic,amsthm,color,times}

\usepackage{graphicx}
\usepackage{t1enc}

\usepackage[latin1]{inputenc}

\usepackage[french]{minitoc}
\usepackage[english]{babel}
\makeatletter
\newcommand{\thechapterwords}
{ \ifcase \thechapter\or One\or Two\or Three\or Four\or
Five\or Six\or Seven\fi}
\def\thickhrulefill{\leavevmode \leaders \hrule height 1ex \hfill \kern \z@}
\def\@makechapterhead#1{%
  \vspace*{15\p@}%
  {\parindent \z@ \centering \reset@font
        \thickhrulefill\quad
        \scshape \@chapapp{} \thechapterwords
        \quad \thickhrulefill
        \par\nobreak
        \vspace*{15\p@}%
        \interlinepenalty\@M
        \hrule
        \vspace*{15\p@}%
        \Huge \bfseries #1\par\nobreak
        \par
        \vspace*{15\p@}%
        \hrule
    \vskip 60\p@
  }}
\def\@makeschapterhead#1{%
  \vspace*{15\p@}%
  {\parindent \z@ \centering \reset@font
        \thickhrulefill
        \par\nobreak
        \vspace*{15\p@}%
        \interlinepenalty\@M
        \hrule
        \vspace*{15\p@}%
        \Huge \bfseries #1\par\nobreak
        \par
        \vspace*{15\p@}%
        \hrule
    \vskip 60\p@
  }}
  \def\@makechapterhead#1{%

  \vspace*{15\p@}%
  {\parindent \z@ \centering \reset@font
        \thickhrulefill\quad
        \scshape \@chapapp{} \thechapterwords
        \quad \thickhrulefill
        \par\nobreak
        \vspace*{15\p@}%
        \interlinepenalty\@M
        \hrule
        \vspace*{15\p@}%
        \Huge \bfseries #1\par\nobreak
        \par
        \vspace*{15\p@}%
        \hrule
    \vskip 60\p@
    }}
\usepackage{fancyhdr}
 \usepackage [dvips]{epsfig}
\usepackage{amsmath,amssymb}

\fancypagestyle{plain}{ 
\fancyhead{} 

}

\usepackage{newlfont}

\newlength{\defbaselineskip}
\newcommand{\setlinespacing}[1]%
           {\setlength{\baselineskip}{#1 \defbaselineskip}}

\newenvironment{prof}[1][Proof]{\textbf{#1.} }{\ \rule{0.5em}{0.5em}}
\theoremstyle{plain}
\newtheorem{thm}{Theorem}[chapter]
\newtheorem{cor}[thm]{Corollary}
\newtheorem{lem}[thm]{Lemma}
\newtheorem{pro}[thm]{Proposition}
\newtheorem{dfn}[thm]{Definition}
\newtheorem{rap}{Rappel}[chapter]
\newtheorem{rmq}[thm]{Remark}
\newtheorem{nota}[thm]{Notations}
\newtheorem{expl}[thm]{Examples}

\newcommand{\p}{\partial}

\newcommand{\ben}{\begin{enumerate}}
\newcommand{\een}{\end{enumerate}}
\newcommand{\brap}{\begin{rap}}
\newcommand{\erap}{\end{rap}}
\newcommand{\bnota}{\begin{nota}}
\newcommand{\enota}{\end{nota}}
\newcommand{\beqs}{\begin{eqnarray}}
\newcommand{\eeqs}{\end{eqnarray}}

\newcommand{\beq}{\begin{eqnarray}}
\newcommand{\eeq}{\end{eqnarray}}
\newcommand{\bpro}{\begin{pro}}
\newcommand{\epro}{\end{pro}}
\newcommand{\blem}{\begin{lem}}
\newcommand{\elem}{\end{lem}}
\newcommand{\bdfn}{\begin{dfn}}
\newcommand{\edfn}{\end{dfn}}
\newcommand{\bcor}{\begin{cor}}
\newcommand{\ecor}{\end{cor}}
\newcommand{\bthm}{\begin{thm}}
\newcommand{\ethm}{\end{thm}}
\newcommand{\bex}{\begin{expl}}
\newcommand{\eex}{\end{expl}}
\newcommand{\brmq}{\begin{rmq}}
\newcommand{\ermq}{\end{rmq}}
\newcommand{\beproof}{\begin{proof}}
\newcommand{\eproof}{\end{proof}}
\newcommand{\bitem}{\begin{itemize}}
\newcommand{\eitem}{\end{itemize}}
\begin{document}
\begin{large}
\dominitoc
\begin{titlepage}
 \begin{center}

\begin{center}

\vspace{.5cm} {\bf Universit\'e d'Abomey-Calavi (UAC), B\'enin} \\[.4cm]
\textbf{Institut de Math\'ematiques et de Sciences Physiques (IMSP), Porto Novo}\\[.8cm]
  A dissertation submitted in partial fulfillment of the requirement for the degree of DOCTOR es-Sciences \\
       to the \\
 Universit\'{e} d'Abomey Calavi\\
by \\
 \bf Ancille NGENDAKUMANA \\[.7cm]
\end{center}
\begin{minipage}{.8\textwidth}
\hrulefill
 \begin{center}
\bf{ Group Theoretical Construction of Planar Noncommutative Systems }
 \end{center}
\hrulefill
\end{minipage}
\vspace{0.7 cm}

\vspace{.4cm}

Jury

\vspace{.2cm}
\begin{tabular}{lclclc}
President &: & Prof. Norbert HOUNKONNOU, CIPMA/UAC, B\'enin \\
Rapporters & : &Prof. Peter HORVATHY, Universit\'e de Tours, France \\[.2cm]
 & : & Prof. Partha GUHA, S. N. Bose National Center for Basic Sciences, Inde\\[.2cm]
& : &  Prof. Joachim NZOTUNGICIMPAYE, University of Rwanda, Rwanda\\[.2cm]
& : & Prof. Leonard TODJIHOUNDE, IMSP/UAC, B\'enin\\[.2cm]
Examiner& : & Prof. Joel TOSSA, IMSP/UAC, B\'enin\\[.2cm]

Directors & : & Prof. Joachim NZOTUNGICIMPAYE, University of Rwanda, Rwanda\\[.2cm]
& : & Prof. Leonard TODJIHOUNDE, IMSP/UAC, B\'enin \\
  &  &  \\
   &  &
\end{tabular}
 December 2013
 \end{center}
 \end{titlepage}

\chapter*{DEDICATION}
 \vspace{5cm}

\begin{center}
To
 \vspace{1cm}

\begin{center}
Cyprien HORUGAVYE,\\
 \vspace{1cm}

Don Dorel MUHINTAHE,\\
 \vspace{1cm}

  Cyriel HORUGAVYE,\\
  \vspace{1cm}

D\'elissa NIYERA,\\
 \vspace{1cm}

 Josh-Micky IHEZAGIRE.\\
\end{center}

\end{center}
\chapter*{ACKNOWLEDGEMENTS}
I first and foremost express my heartfelt thanks to my advisors
 Professor Joachim NZOTUNGICIMPAYE and Professor Leonard TODJIHOUNDE whose guidance, unlimited patience and constant encouragement
made this thesis possible.\;\;Their insight, passion for
 Mathematics and their quest for perfection inspired me all along our collaboration.\;\;
Furthermore, their critical feedback and input all contributed very fundamentally to this accomplishment.\\

I am thankful to my thesis reporters: Professor Peter HORVATHY, Professor Partha GUHA, Professor Joachim NZOTUNGICIMPAYE
and Professor Leonard TODJIHOU\-NDE
for their careful work in reading this thesis.\;\;Their remarks and suggestions have considerably contributed to the
improvement of this work.\;\;
To my thesis committee members Professor Norbert HOUNKONNOU and Professor Joel TOSSA, I wish to express my gratefulness.\\

By being at the \textquotedblleft Institut de Math\'ematiques et de Sciences Physiques\textquotedblright (IMSP), I met a huge
number of researchers, Physicists and Mathematicians.\;\;All interesting interactions and exchanges
that we had had helped me to become a Mathematician.\;\;
I thank all the IMSP group of professors for contributing to the very positive work environment
I found.\;\;Thank you so very much
for your constant encouragement.\;\;I would also like to thank Mrs Gis\`ele FOHOUNHEDO BANKOLE for always being helpful
with administrative
issues.\;\;I also wish to express my gratefulness to my current officemates: Joseph Salomon MBATAKOU, Jonas DOUMATE, Oumar SOW and
Van Borhen NKOU with whom I had interesting conversations about Mathematics and much more.  To
Dr Frank DJIDEME, Dr Stephane TCHUIAGA, Joel KPLE and Toussaint OKE for your help in LaTeX, I thank
you all very much.\;\;I would also like to thank all research students I met at IMSP
for making my stay in Benin enjoyable.\;\;God bless you all.\\

I would also like to thank people who have participated most directly in my formation and initiated me into the fascinating world
of Mathematics.\;\;I think about Professor Joachim NZOTUNGICIMPAYE, Professor Jean NDIMUBANDI, Professor Jean-Bosco KAYOYA, Professor
 Gaspard BANGEREZAKO and Professor Isidore MAHARA.\\

I am also grateful to the Government of BURUNDI for its constant financial support.\;\;I also wish to express my gratefulness to
Dr Gaspard NTAHONKIRIYE and his family for supporting my regular stays in Kigali (Rwanda) during my formation.\\

I am thankful also to Burundian students in B\'enin for their constant encouragements.\;\;In Particular I would like to thank Dr Donatien
GAPARAYI, Charles GATURAGI and Jean-Berchmans BIZIMANA for making my stay in B\'enin so enjoyable.\\

Last but not least, I thank my family who has endured my absence for years.\;\;They have always shown a lot of
patience and devotion.\;\;Thank you so much dear mother, husband, children, sisters, sisters and brothers-in-law.\\

\chapter*{ABSTRACT}
In this thesis, we construct and classify planar noncommutative phase spaces by the coadjoint orbit method on
the anisotropic and absolute time kinematical groups.\;\;We show that noncommutative symplectic
structures can be generated in the framework of centrally
extended anisotropic kinematical groups as well as in the framework of noncentrally abelian extended absolute time kinematical groups.\\

However, noncommutative phase spaces realized with noncentral abelian extensions of the kinematical groups are algebraically
more general than those constructed on their central extensions.\;\;As the coadjoint orbit construction has not been
carried through some of these planar kinematical groups before, physical interpretations of new generators of
those extended structures are given.\;\;Furthermore, in all the cases discussed here, the noncommutativity
is measured by naturally introduced fields, each corresponding to a minimal coupling. \\

This approach allows to not only construct directly a dynamical system when
of course the symmetry group is known but also permits to eliminate the non minimal couplings in that system.\;\;Hence, we show also
that the planar noncommutative phase spaces arise naturally by introducing minimal coupling.\;\;We introduce here new kinds of
couplings.\;\;A coupling of position with a dual
potential and a mixing model (that is minimal coupling
of the momentum with a magnetic potential and of position with a dual potential). \\

Finally we show that this group theoretical discussion can be recovered by a linear deformation of the Poisson bracket.\;\;The reason
why linear deformation of
Poisson bracket is required here is that the noncommutative parameters (which are fields) are constant (they are coming from
central and noncentral abelian extensions of kinematical groups).

\dominitoc
\tableofcontents
\newpage
 \rhead{\textbf{\thepage}}
\lhead{\textsl{\leftmark}}
\chapter*{INTRODUCTION}\addcontentsline{toc}{chapter}{INTRODUCTION}
\section{Introduction}
\pagenumbering{arabic}
  This thesis takes place within the framework of the titled domain {\it noncommutative geometry}.\;\;Noncommutativity appeared in
 nonrelativistic mechanics first in the work of Peierls \cite{peierls} on the diamagnetism of conduction
electrons.\;\;In relativistic quantum mechanics, the suggestions to use noncommutative coordinates goes back to Heisenberg
and was firstly formalized in $1947$  by Snyder \cite{snyder} at small length scales.\;\;Some time later, Von Neumann
 \cite{vonnuwman} introduced the term {\it noncommutative geometry} to refer in general to a geometry
 in which
an algebra of functions is replaced by a more general associative algebra called {\it noncommutative algebra}.\;\;For him, operator
algebra theory was a noncommutative outgrowth of measure theory.\;\;As in the quantization of classical
phase space, coordinates are replaced by generators of the algebra. \\

The correspondence between \textquotedblleft spaces\textquotedblright and \textquotedblleft commutative
algebras\textquotedblright is familiar
in Mathematics and in theoretical Physics.\;\;This correspondence allows an algebraic translation of various geometrical concepts on
spaces in
appropriate algebras of functions on these spaces.\;\;Replacing these commutative algebras
by noncommutative algebras, i.e forgetting commutativity, leads then to noncommutative
generalizations of geometries where notions of \textquotedblleft spaces of points \textquotedblright are not involved. \\

Interest in Snyder's idea was revived much later when Mathematicians, notably Connes \cite{connes,connes10} and Woronowicz
\cite{woro}, succeeded in
generalizing the notion of differential structure to noncommutative geometry.\;\;Such a noncommutative generalization
was a need in Physics for the formulation of quantum theory and the understanding of its relations
with classical Physics.\;\;Indeed, as the role of symplectic geometry and hence symplectic structures has increased its
importance in both Mathematics and Physics
to constitute nowadays an essential technique of describing and modeling natural phenomena, then noncommutative
symplectic structures offer a novel and promising framework for the construction of
physical theories.\;\;Particularly, noncommutative phase spaces provide mathematical backgrounds for the study of magnetic fields in
Physics.\\

As applied to Physics, noncommutative geometry is understood mainly in two approaches.\;\;The first one is the spectral triple of
A. Connes \cite{connes1} with the Dirac operator playing a central role in unifying, through the universal action principle, gravitation
with
standard model of fundamental interactions.\;\;The second one is the quantum field theory on noncommutative spaces \cite{nekrasov}
with Moyal
product as main ingredient.\;\;Besides these, a proposition by several authors \cite{horvathy6, horvathy02} corresponds to space
coordinates that no longer commute.\;\;This was implemented by an extension of the Poisson structure on the cotangent
space such that the
brackets satisfy $\{x^k,x^l\} \neq 0$.\;\;Upon quantization, the corresponding operators should then also be noncommutative.\\

One motivation for this work is to demonstrate that this extension of the Poisson structure
is achieved when we consider a Lie group $G$.\;\;Indeed, models
associated with a given symmetry group can be conveniently constructed using the coadjoint orbit method also called
Souriau's method.  His theorem says
in fact that when a symmetry group $G$ acts transitively on a phase space, then the latter is a
coadjoint orbit of $G$ equipped with its canonical symplectic form \cite{souriau, kostant, kirillov}.\;\;In other words, the classical
phase spaces
of elementary systems correspond to coadjoint orbits of their symmetry groups.\;\;Thus, by considering a Lie group $G$, the problem is
to find
a symplectic manifold $X$ whose symmetry group is $G$.\;\;Under some assumptions, this problem has a regular solution according to
the Souriau's approach. \\

The first applications that Souriau presented in his book \cite{souriau} concern both the Poincar\'e and the Galilei groups for
which coadjoint orbits represent elementary particles characterized by the invariants $m$ (mass) and $s$ (spin).\;\;Souriau
himself goes one step further as he considers massless particles with spin, $m=0$, $s\neq0$ identified as relativistic and
nonrelativistic spin respectively.\;\;Souriau's ideas were later extended to larger groups.\;\;Taking $G=Poincare \times H_0$ where $H_0$
is an internal symmetry group \\(e. g $SU(2), SU(3),...$) yields relativistic particles
with internal structure (for more details see \cite{duval2, duval3}).\\

The nonrelativistic kinematical groups admit nontrivial central extensions by one-dimensional algebra in dimension
$d\geq 3$.\;\;But in
the plane, they admit an exotic \cite{horvathy7} two-parameter central extension.\;\;The one-parameter
central extension of the spatial Galilei group has been considered by Souriau in his book \cite{souriau}, the
two-parameter central extension of the planar Galilei group was studied in \cite{ horvathy6, horvathy7}.\;\;Furthermore, the
coadjoint orbit method has recently also been applied to smaller spacetime symmetry groups.\;\;For
example, in \cite{duval1} a classical \textquotedblleft photon \textquotedblright model was constructed, based entirely
on the Euclidean group $E(3)$, a subgroup of both the Poincar\'e and the Galilei groups.\;\;An other application
of the Souriau's method is found in \cite{gonera} where the most general dynamical systems on which the
nonrelativistic conformal groups act transitively as symmetries are constructed. \\

A related motivation comes from a curiosity about a general solution to the above problem, that is, to find more general
symplectic manifolds whose symmetry groups are the planar kinematical groups.\;\;This takes place in
a well known fact that if a free particle is coupled with an external field, this reduces
the symmetry group to a subgroup (of the Galilei or Poincar\'e group) and conversely, this reduced symmetry is consistent
with the symmetry subgroup \cite{souriau}. \\

This thesis is devoted to realize classical dynamical systems associated with anisotropic kinematical groups
(kinematical groups without rotation parameters \cite{3derome}) by use of the reciprocal schema.\;\;Precisely, we
start with a model with anisotropic kinematical symmetry and by applying
Souriau's method, we obtain models with additional terms (in the
symplectic 2-form) interpreted as fields.\;\;The latter are linked to the free particle so as to preserve the anisotropic
kinematical symmetries.\;\;More specifically, we study the
maximal coadjoint orbit (all invariants are nonvanishing) of the all planar anisotropic kinematical groups
(oscillating and expanding  Newton-Hooke Lie groups,
Galilei, Para-Galilei, Carroll and Static Lie groups) according to the classification in \cite{5bacry}. \\

Furthermore, equivalently to Souriau's theorem, the dual ${\cal{G}}^*$ of the Lie algebra $\cal{G}$ of $G$ has a natural
Poisson structure whose symplectic leaves are the coadjoint orbits.\;\;Depending on the Lie group,
these orbits may provide noncommutative phase spaces.\;\;Note that on its general form, a noncommutative phase space allows
for nonzero commutator among the coordinates and among the momenta.\;\;Thus, by applying the Souriau's method, we construct and classify
noncommutative phase spaces (which are effectively generalized or modified symplectic structures) associated to the extended
kinematical groups.\;\;As already argued we consider the case where
the symmetry groups are the kinematical groups according to the classification in \cite{5bacry} and realize noncommutative phase spaces
on their maximal coadjoints orbits by using
central extensions of all anisotropic kinematical algebras (by relaxing the isotropy condition or dropping
the rotation generators \cite{3derome}), the latter
being the Lie algebras for the kinematical groups.\\

Note that for the one-parameter centrally extended kinematical algebras, the
nontrivial Lie bracket which contains the only central extension parameter $m$ (i.e the mass of the system) is
\begin{eqnarray*}
 [K_i,P_j]=M\delta_{ij}
\end{eqnarray*}
which means that the generators of space translations as well as pure kinematical group transformations
commute.\;\;One can not then
associate noncommutative phase spaces to both planar kinematical groups and their one-fold
centrally extended kinematical groups.\;\;So, it is necessary to
work with the two-fold central extensions of the kinematical algebras and of their
corresponding Lie groups and it is then the absence of the symmetry rotations (i.e anisotropy of the space)
which guaranties the existence of noncommutative phase space for planar anisotropic kinematical groups.\\

However, it is possible to associate a noncommutative phase space to absolute
time groups by considering their noncentral abelian extensions.\;\;Thus, we enlarge this theory by considering also
the noncentral abelian extensions of the absolute time kinematical algebras associated to the Lie groups classified in
\cite{mcrae}.\;\;Explicitly, we show that noncommutative symplectic structures can be generated in
the framework of centrally extended anisotropic kinematical algebras as well as in the framework of
noncentrally abelian extended isotropic kinematical algebras (rotations included). \\

Meanwhile, physical theories with noncommuting coordinates have become
the focus of recent research (see, e.g.,\cite{horvathy6, horvathy02, horvathy7, horvathy01,
grigore},\dots ), the notion of noncommutativity having different physical interpretations.\;\;
For example, it is well known that in the presence of an electromagnetic field, momenta do not commute.\;\; This
has been realized in this thesis for the first time as the maximal coadjoint orbits of the
centrally and noncentrally abelian extended planar Para-Galilei
groups \cite{ancille1, ancilla1}.\\ Also, it has been proved
that in the presence of the dual electromagnetic field the positions do not commute  and
the maximal coadjoint orbits of the extended planar Galilei groups have been shown to be models of
noncommutative phase spaces for this case \cite{ancilla1}.\\

Also, we realize a phase space with noncommutative positions and momenta
by Souriau's method on the anisotropic Newton-Hooke groups \cite{ancilla}.\;\;In \cite{zhang-horvathy}, the authors have found a similar
symmetry in the so-called Hill problem (the latter is studied in celestial mechanics), which is effectively an
anisotropic harmonic oscillator in a magnetic field.\;\;This system has
no rotational symmetry while translations and generalized boosts still act as symmetries.\;\;The noncommutative
version of the Hill problem has been discussed in \cite{zhang-horvathy1}.\;\;In this work,
we study also the classical dynamical systems associated with Aristotle group \cite{ancille}, a subgroup of the Galilei group
to highlight the fact that it is always possible to construct noncommutative phase spaces on noncentrally abelian extended Lie structures.\\

  Furthermore, we classify the obtained nonrelativistic models which are noncommutative phase spaces \cite{ancilla1}.\;\;Through
our constructions the coordinates of the phase spaces do not commute due to the presence of naturally introduced fields
giving rise to minimal couplings.\;\;Thus, this group theoretical method allows not only a direct construction of
a dynamical system when
of course the symmetry group is known but also permits to eliminate the non minimal couplings in that system.\;\;Hence, to be
complete in this work, we show also that planar noncommutative phase spaces arise naturally
by introducing minimal couplings, each type of noncommutative phase space realized here corresponding to a specific minimal coupling.\\

As the coadjoint orbit construction has not been carried through some of these planar kinematical groups before, physical
interpretations of new generators of those extended structures are given by symplectic realization methods.\;\;Note also
that noncommutative phase spaces realized with
noncentral abelian extensions of the kinematical groups are algebraically more general than those constructed on their
central extensions.\;\;It is also shown in this work that the noncommutativity of momenta implies
some modification of the second Newton law \cite{ancille1, romero, wei}.\;\;We prove finally that the group
theoretical discussion above can be recovered by a linear deformation of Poisson brackets.\;\;One simple reason
why linear deformation is required here is that the fields which are responsible to the noncommutativity
are all constant (because coming from central or noncentral abelian extensions).\\

In all these approaches, for simplicity in notation and for a clear physical interpretation, we work in two-dimensional space
(except for the anisotropic Newton-Hooke groups case where we consider also the three-dimensional space to highlight the way
of finding the invariant) although the extension to higher dimensions is straightforward.\\

This thesis is organized in the following way. \\

 Chapter $1$ is devoted to the construction of central extensions of planar anisotropic kinematical groups and
noncentral abelian extensions of planar absolute time groups according to the classification in \cite{5bacry} and \cite {mcrae}
respectively.\;\;We consider the (maximal) central extensions of kinematical algebras which exponentiate to the corresponding
kinematical groups and then appear to have a clear physical interpretation.\;\;In the application
to the nonrelativistic particles, the central extensions of planar anisotropic kinematical groups considered here
all have a common property.\;\;They have two central extension generators, one of them
is related to the particle's mass ($m$) and an other can be related to the particle's spin ($s$)
 \cite{jackiw, bose, levyleblond2, Duval-horvathy}.\;\;Note that the noncentral abelian extensions
of absolute time planar kinematical groups have a different number of extension parameters.\\

In Chapter $2$, we construct noncommutative phase spaces by introducing minimal couplings.\;\;We start by the usual coupling of momentum
with a magnetic potential.\\Then, we introduce a new kind of coupling.\;\;That of position with a dual potential and
we finish by a mixing model.\;\;In other words, we study the planar mechanics in the following three situations: when a
charged massive particle is in an electromagnetic field, when a massless
spring is in a dual magnetic field and finally when a pendulum is in an electromagnetic field and in its
dual counterpart.\;\;It is shown
in this Chapter that under the presence of these fields, the charged massive particle acquires an oscillation state of
motion with a certain frequency, the massless spring acquires a mass and that the pendulum looks like two synchronized
oscillators.\;\;The
second and third results above are quite new and expressed the formula of minimal couplings in symplectic terms as done by
Souriau for the first time \cite{souriau}.\;\;Furthermore, as it will be shown in the next Chapter, each kind
of minimal coupling is realized by coadjoint orbit method on a specific kinematical group.\\

The aim of Chapter $3$ is to construct and classify noncommutative phase spaces by coadjoint orbit method
in the framework of centrally and noncentrally abelian extended Lie groups we have encountered so far.\;\;We study, first
of all, the maximal coadjoint orbits of the Aristotle group.\;\;We obtain in this case, phase spaces equipped
with modified symplectic structures by using a noncentrally abelian extended Aristotle group.\;\;This example
proves that one can always obtain a noncommutative phase space on the noncentrally abelian extended
Lie group by the coadjoint orbit method.\;\;We do a similar construction on
the nonrelativistic anisotropic kinematical groups (i.e Newton-Hooke groups, Galilei, Para-Galilei groups,
Static and Carroll groups).\;\;The obtained orbits
are physically interpreted as the phase spaces of accelerated particles moving in the
respective kinematical spacetimes.\;\;Finally, we construct maximal coadjoint orbits on noncentrally abelian extended
absolute time kinematical
groups as classified in \cite{mcrae}.\;\;The noncommutative phase spaces obtained in this case are algebraically general than
those obtained in the anisotropic kinematical case.\\

Through all these constructions, the coordinates
of the phase spaces do not commute due to the presence of naturally introduced fields
giving rise to minimal couplings defined in the previous Chapter.\;\; The main result here is that the extended Galilei groups give
rise to phase spaces with noncommuting position coordinates, the extended Para-Galilei groups
and noncentrally abelian extended Aristotle group yield noncommuting momenta, and the remaining groups
all determine completely noncommutative phase spaces.\;\;
By symplectic realization methods, physical
interpretations of generators coming from the obtained extended structures are given.\\
\\

 In Chapter $4$, we show that the group theoretical construction above leads to similar results as in the
case of a linear deformation of Poisson brackets.\;\;Under some assumptions, it is shown that
the linear deformation of the Poisson bracket gives rise to the same classification of planar
noncommutative phase spaces as by the Souriau method.\;\;Linear deformation is required here because
the noncommutative parameters (fields) introduced in the previous Chapter
are all constant (they come from centrally or noncentrally abelian extended structures).\;\;Thus, to establish a relationship between the group
theoretical construction used previously and the deformation of the Poisson bracket developed in
this Chapter, the latter may be a linear one. \\

Finally, we conclude and briefly list directions for future research.

\chapter{PLANAR KINEMATICAL GROUPS, THEIR CENTRAL AND NONCENTRAL EXTENSIONS}
Lie groups are frequently introduced in Physics as groups of transformations acting on manifolds.\;\;In this Chapter we
revisit all possible kinematical groups and their corresponding Lie algebras according to
the classification in \cite{5bacry}.\;\;The former are assumed to be simply connected to single out
a Lie group for the given Lie algebra.\;\;We then extract those for which rotation generators can be
dropped producing anisotropic kinematical algebras (and groups) \cite{3derome}.
Recall that kinematical algebras are Lie algebras that are generated by spatial displacements, time translations, rotations and inertial
transformations.\\

Since the structure of $(2+1)$-dimensional kinematical groups is significantly different from that of the ($3+1$)-dimensional
ones \cite{bose}, then it is interesting to study the problem of finding central extensions
of the planar anisotropic kinematical groups and noncentral abelian extensions of planar absolute time kinematical groups according to
the classifications in \cite{5bacry} and in \cite{mcrae} respectively. \\

The aim of this Chapter is to solve the above problem: we compute central (and noncentral) extensions of
the planar anisotropic kinematical algebras (and the absolute time kinematical algebras) and their corresponding Lie
groups with the assumption that an abelian extension (central or noncentral) of a Lie algebra should integrates to
an abelian extension (central or noncentral) of its corresponding Lie group \cite{hekmati}.\\

In other words, we restrict our attention to the central and noncentral abelian extensions of Lie algebras which exponentiate to the
corresponding Lie groups, i.e through this thesis, we disregard extensions which correspond to
introducing a central generator that measures the noncommutativity
between $J$ and $H$ (we will consider $[J,H]=0$ i.e rotation invariance).\;\;In the nonrelativistic anisotropic case, the central
charges that have denoted $M$ and $S$, related respectively to the particle's mass ($m$)
and spin ($s$) \cite{jackiw, bose, levyleblond2, Duval-horvathy},
survive the exponentiation to the corresponding groups.\;\;Physically, $S$ is interpreted as the intrinsic angular momentum operator
representing rotation in the rest-frame.\\

This Chapter is organized as follows.\\

In the next section, we review the Bacry and Levy-Leblond approach of classifying kinematical algebras and groups
 \cite{5bacry} and extract the anisotropic kinematical ones (without rotations).\;\;In sections two and three, we determine
central extensions of anisotropic kinematical algebras and noncentral abelian extensions of
absolute time kinematical algebras (rotations included) respectively using dimensional analysis.\;\;Some of these extensions have not
appeared previously in the litterature.

\section{Possible kinematical groups}
Consider a manifold $M$ on which a transformation group $G$ acts transitively (this action is usually the left one) meaning that
$M$ is a $G$-homogeneous space.\;\;When $M$ describes spacetime, $G$ is the kinematical group of $M$.\;\;All $d$-dimensional
spacetimes
with a constant curvature have a kinematical group of dimension $\frac{1}{2}d(d+1)$ \cite{4defkingroup}. \\

Here, spacetime denotes a
mathematical model that, of a physical dynamic system, unifies space and time into a manifold of four-dimensions.\;\;Note
that traditionally in physics, spacetimes are described by (pseudo-) Riemann spaces, i.e: by smooth manifolds and with a tensor metric
$g_{ij}(x)$ and are of the form $T\times \Sigma$ where the one-dimensional space $T$ is the universally defined time line (admitting
the preferred coordinate $t$ and its one-form $dt$) while the manifold $\Sigma$ is with possibly punctures and a bounded (this last
assumption is identically satisfied for the spacetimes we examine).

Furthermore, in general relativity, it is assumed that spacetime is
curved by the presence of matter (energy), this curvature being represented by the Riemann
tensor and in nonrelativistic classical mechanics, the use of Euclidean space instead of spacetime is appropriate as the time is
treated as universal and constant, being independent of the state of motion of the observer.\\

It is well known that there are different kinds of kinematics on homogeneous ($3+1$)-dimensional spacetimes \cite{5bacry} and all
of them can be contracted from Minkowski, de Sitter and Anti-de Sitter spacetimes under group
contraction \cite{inonu} respectively.\;\;The complete kinematical
group, whatever it may be, will always have a subgroup accounting for the isotropy of space (rotation group)
and the equivalence of inertial transformations (boosts of different kinds for each case).\;\;The remaining transformations are
translations which may be either commutative or not and are responsible of homogeneity of space and time.
Roughly speaking, the point-set of the corresponding spacetime is, in each case, the point-set
of these translations.\;\;This holds for
usual special relativistic kinematics but also for Galilean and other conceivable
nonrelativistic kinematics \cite{5bacry}, which differ from each other precisely by being grounded on different kinematical groups
(depending on the different laws of composition of the transformations groups).\;\;The best known
examples are the Poincar\'e group $P$, a group naturally associated to Minkowski spacetime as its group of motion
and the Galilei group $G$ which represents the classical mechanics.\\

Therefore, kinematical spacetime is defined as the quotient space of the whole kinematical group by the subgroup including
rotations and boosts.\;\;This means that local spacetime is always a homogeneous manifold.\;\;We shall thus be
concerned with such very special kinds of spacetimes: the homogeneous spacetimes of groups which
can be called kinematical. \\

More specifically, let $G$ denotes the kinematical group and $\cal{G}$ its Lie algebra.\;\;We choose a basis for $\cal{G}$ in which the
infinitesimal generators of rotation, those of boosts and spatial translations along the spatial directions and that of time translation
are denoted respectively as $J_a, K_i, P_i, H, ~a=1,2,\dots ,\frac{(d-2)(d-1)}{2}$, $~i=1,2,\dots,d-1 $.\\

Before we proceed, we point the reader's attention to the choice of the above quantities and at the same time recall their physical
significances.\;\;First of all, note that if a general element of a one-parameter Lie group generated
by $X$ takes the form $\exp(xX)$ then this means that $xX$
must be dimensionless and then that the physical dimension of the parameter $x$ must be the inverse
of that of $X$.\;\;For that reason, the parameters $x^i, v^i$ and $t$ (that will be used later in this work) associated
respectively to $P_i$ (i.e the generators of spatial translations in the $i^{th}$ spatial direction), $K_i$ (i.e generators
of boosts in the $i^{th}$ spatial direction) and $H$ (i.e generator of time translation) have
dimension of a length, a velocity and a duration.\;\;Their duals (in the dual Lie algebra) are respectively linear momenta $p_i$, static
 momenta $k_i$ and energy $E$.\\

Thus, if $L$ and $T$ denote respectively the dimension of a length and of a duration, then the physical dimensions of
$P_i$, $H$ and $K_i$ are $L^{-1}$, $T^{-1}$ and $L^{-1}T$ respectively.\;\;It is also
known that the generators $J_a$ of rotations are dimensionless.\;\;Physically, $J_a$
correspond to the angular momenta and as said early $H$ corresponds to the Hamiltonian, $P_i$ correspond to the components of
linear momenta and $K_i$ correspond to the components of the static momenta.

 Furthermore, the relations
between these physical quantities $P_i, K_i$ and $H$ depend on the structure of the
Lie algebras and can be determined by symplectic realization methods of the
corresponding Lie groups \cite{nzo1}.\;\;The variables $p_i$ and $k_i$ yield the basic
canonical variables on the phase space via the
the coadjoint orbit method as it will be seen later in this thesis.\\

At this level, let us summarize the results obtained by Bacry and L\'evy-Leblond concerning the possible structures of groups of
transformations which relate two inertial systems under quite general physical hypotheses.
\subsection{Bacry and L\'evy-Leblond approach}
In \cite{5bacry}, Bacry and L\'evy-Leblond have classified the possible ten-parameters kinematical groups consisting of
the spacetime translations, spatial rotations and inertial transformations connecting different
inertial frames of reference.\;\;Bacry and L\'evy-Leblond have shown, under the assumption:
\begin{itemize}
 \item the space must be isotropic, meaning that the rotation group $SO(d-1)$ generated
by $J_a~,a=1,2,\dots,\frac{(d-2)(d-1)}{2}$, is a subgroup of
the kinematical group,
 \item the spacetime must be homogeneous, meaning that the space translations group generated
 by $P_i,i=1,2,\dots,d-1$ and the time translations
 group generated by $H$, are subgroups of the kinematical group,
\item the inertial transformations group generated by $K_i,~i=1,2,\dots,d-1$ is a noncompact subgroup of the kinematical group,
\item the parity ($\pi:H\rightarrow H,P_i\rightarrow -P_i,K_i\rightarrow-K_i,J_a\rightarrow J_a$) and the time-reversal
($\theta: H \rightarrow -H,P_i\rightarrow P_i,K_i\rightarrow-K_i,J_a\rightarrow J_a$) are automorphisms of the kinematical group,
\end{itemize} that there are eleven kinematical groups.\\

The corresponding kinematical algebras
are characterized by the fact that the inertial transformation generators and
the space translation generators behave as vectors under rotations while the time translation generator behaves as a scalar:
\begin{eqnarray}\label{rotations}
[J_j,J_k]=J_l\epsilon_{jk}^{l},~[J_j,K_k]=K_l\epsilon_{jk}^{l},~[J_j,P_k]=P_l\epsilon_{jk}^{l},~[J_j,H]=0
\end{eqnarray}
In the above relation, $\epsilon_{jk}^{l}$ are the three-dimensional totally antisymmetric Levi-Civita symbols and summation
convention for a repeated index up and down is implied (Einstein convention).

As each Lie bracket $[K_i,H], [K_i,P_j]$ and $ [P_i,H]$ is invariant under the parity and the time-reversal automorphisms as well as
 their product ($\Gamma= \pi \theta : H \rightarrow -H,P_i\rightarrow -P_i, K_i\rightarrow K_i,J_a\rightarrow J_a$), we also have that:
\begin{eqnarray}\label{boosts}
[K_j,K_k]=\mu J_l\epsilon_{jk}^{l},~[K_j,P_k] =\gamma\delta_{jk}H,~[K_j,H]=\lambda P_j
\end{eqnarray}
where the physical dimension of the parameters $\mu$ and $\gamma$ is $L^{-2}T^2$ while the parameter $\lambda$ is dimensionless and
 finally,
\begin{eqnarray}\label{translations}
 [P_j,P_k] =\alpha J_l\epsilon_{jk}^l,~[P_j,H]=\beta K_j
\end{eqnarray}
where the physical dimension of the parameter $\alpha$ is $L^{-2}$ while that of
the parameter $\beta$ is $T^{-2}$.\;\;Only three of the five parameters
$\alpha$, $\beta$, $\lambda$, $\mu$, $\gamma$ are independent.\;\;Effectively the Jacobi identities
\begin{eqnarray*}
[K_i,[P_j,P_k]]+[P_j,[P_k,K_i]]+[P_k,[K_i,P_j]]=0
\end{eqnarray*}
and
\begin{eqnarray*}
[K_i,[K_j,P_k]]+[K_j,[P_k,K_i]]+[P_k,[K_i,K_j]]=0~
\end{eqnarray*}
imply that
\begin{eqnarray*}
\alpha=\beta\gamma~,~\mu=-\lambda\gamma
\end{eqnarray*}
If we compute the adjoint representation of the generators $K_i$, we verify that the non compactness of the boost
transformations implies that $\mu \leq 0$, meaning that $\lambda$ and $\gamma$ are all positive or all negative when one of
them is not equal to zero.\;\;The brackets (\ref{boosts}) and (\ref{translations}) become
\begin{eqnarray*}\label{boosts1}
[K_j,K_k]=-\lambda\gamma J_l\epsilon_{jk}^{l},~[K_j,P_k] =\gamma\delta_{jk}H,~[K_j,H]=\lambda P_j
\end{eqnarray*}
and
\begin{eqnarray}\label{translations1}
 [P_j,P_k] =\beta\gamma J_l\epsilon_{jk}^l,~[P_j,H]=\beta K_j
\end{eqnarray}
We remain with three parameters $\beta$, $\gamma$ and $\lambda$ constrained by the fact that $\lambda$ and $\gamma$
are of the same sign when they are all different from zero.\;\;If each generator is multiplied by $-1$, the three parameters change
sign.\;\;As $\lambda$ is dimensionless, we can assume (after normalization ) that $\lambda=1$ or $\lambda=0$ and
then that $\gamma \geq 0$.

Let $\kappa$
denotes the inverse of the universe radius $r$ and let $\omega$ denotes the time curvature (frequency).\;\;Then, for each
of the two values of $\lambda$, $\gamma=\frac{1}{c^2}$ or
$\gamma=0$ where $c=\frac{\omega}{\kappa}$ is a velocity.\;\;Also for each of the two values of $\gamma$, there are three
Lie algebras corresponding to $\beta=\pm \omega^2$ and $\beta=0$.

Explicitly, we have:
\begin{itemize}
\item[A)] the kinematical algebras corresponding to $\lambda=1$ and defined by the brackets (\ref{rotations}),
(\ref{translations1}) and
\begin{eqnarray}\label{boosts2}
[K_j,K_k]=-\gamma J_l\epsilon_{jk}^{l},~[K_j,P_k] =\gamma\delta_{jk}H,~[K_j,H]= P_j
\end{eqnarray}
\subitem A1) Case $\gamma=\frac{1}{c^2}$ \\
In this case the Lie algebra is defined by (\ref{rotations}) and
\begin{eqnarray}\label{boosts3}
[K_j,K_k]=-\frac{1}{c^2} J_l\epsilon_{jk}^{l},~[K_j,P_k] =\frac{1}{c^2}\delta_{jk}H,~[K_j,H]= P_j
\end{eqnarray}
and
\subsubitem A11) Case $\beta=\omega^2$ (the de Sitter algebra $d{\cal{S}_+}$)
\begin{eqnarray}\label{translations31}
 [P_j,P_k] =\kappa^2 J_l\epsilon_{jk}^l,~[P_j,H]=\omega^2 K_j
\end{eqnarray}
\subsubitem A12) Case $\beta=0$ (the de Poincar\'e algebra ${\cal{P}}$)
\begin{eqnarray}\label{translations32}
 [P_j,P_k] =0,~[P_j,H]=0
 \end{eqnarray}
\subsubitem A13) Case $\beta=-\omega^2$ (the anti-de Sitter algebra $d{\cal{S}_-}$)
\begin{eqnarray}\label{translations33}
 [P_j,P_k] =-\kappa^2 J_l\epsilon_{jk}^l,~[P_j,H]=-\omega^2 K_j
\end{eqnarray}
\subitem A2) Case $\gamma=0$\\
In this case the Lie algebra is defined by (\ref{rotations}) and
\begin{eqnarray}\label{boosts4}
[K_j,K_k]=0,~[K_j,P_k] =0,~[K_j,H]= P_j
\end{eqnarray}
and
\subsubitem A21) Case $\beta=\omega^2$ (the expanding Newton-Hooke algebra ${\cal{NH}_+}$)
\begin{eqnarray}\label{translations41}
 [P_j,P_k] =0,~[P_j,H]=\omega^2 K_j
\end{eqnarray}
\subsubitem A22) Case $\beta=0$ (the Galilei algebra $\cal{G}$)
\begin{eqnarray}\label{translations42}
 [P_j,P_k] =0,~[P_j,H]=0
\end{eqnarray}
\subsubitem A23) Case $\beta=-\omega^2$ (the oscillating Newton-Hooke algebra ${\cal{NH}_-}$)
\begin{eqnarray}\label{translations43}
 [P_j,P_k] =0,~[P_j,H]=-\omega^2 K_j
\end{eqnarray}
\item[B)] the kinematical algebras corresponding to $\lambda=0$ and defined by the brackets (\ref{rotations}), (\ref{translations1})
 and
\begin{eqnarray}\label{boosts3}
[K_j,K_k]=0,~[K_j,P_k] =\gamma\delta_{jk}H,~[K_j,H]=0
\end{eqnarray}
\subitem B1) Case $\gamma=\frac{1}{c^2}$ \\
In this case the Lie algebra is defined by (\ref{rotations}),
\begin{eqnarray}\label{boosts5}
[K_j,K_k]=0,~[K_j,P_k] =\frac{1}{c^2}\delta_{jk}H,~[K_j,H]=0
\end{eqnarray}
and
\subsubitem B11) Case $\beta=\omega^2$ (the Para-Poincar\'e algebra ${\cal{P}_+}$)
\begin{eqnarray}\label{translations51}
 [P_j,P_k] =\kappa^2 J_l\epsilon_{jk}^l,~[P_j,H]=\omega^2 K_j
\end{eqnarray}
\subsubitem B12) Case $\beta=0$ (the Carroll algebra $\cal{C}$)
\begin{eqnarray}\label{translations52}
 [P_j,P_k] =0,~[P_j,H]=0
\end{eqnarray}
\subsubitem B13) Case $\beta=-\omega^2$ (the Para-Poincar\'e algebra ${\cal{P}_-}$)
\begin{eqnarray}\label{translations53}
 [P_j,P_k] =-\kappa^2 J_l\epsilon_{jk}^l,~[P_j,H]=-\omega^2 K_j
\end{eqnarray}
\subitem B2) Case $\gamma=0$\\
In this case we can assume that $\beta \geq 0$, the Lie algebra is then defined by (\ref{rotations}),
\begin{eqnarray}\label{boosts6}
[K_j,K_k]=0,~[K_j,P_k] =0,~[K_j,H]=0
\end{eqnarray}
\subsubitem B21) Case $\beta=\omega^2$ (the Para-Galilei algebra ${\cal{G}_+}$)\\
\begin{eqnarray}\label{translations42}
 [P_j,P_k] =0,~[P_j,H]=\omega^2 K_j
\end{eqnarray}
\subsubitem B22) Case $\beta=0$ (the Static algebra ${\cal{S}t}$)
\begin{eqnarray}\label{translations42}
 [P_j,P_k] =0,~[P_j,H]=0
\end{eqnarray}
 \end{itemize}
The possible planar kinematical algebras are then defined by the brackets (\ref{rotations}) reduced to:
\begin{eqnarray}\label{rotationsplanar}
 [J,K_k]=K_l\epsilon_{k}^{l},~[J,P_k]=P_l\epsilon_{k}^{l},~[J,H]=0
\end{eqnarray}
and to summarize the above discussion, we
distinguish the case where boosts do not commute with time translation (i.e $[K_i,H]=P_i)$ from the case where they commute \\
(i.e $[K_i,H]=0 $) as detailed below:
\subsection*{Boosts not commuting with time translations }
This case corresponds to $\lambda=1$.\;\;The planar kinematical algebras are then defined by the brackets (\ref{rotationsplanar}),
(\ref{translations1}) and (\ref{boosts2}).\;\;According to the values
of $\gamma$ and $\beta$ as detailed above, the possible planar kinematical algebras of this form are summarized
in the following
table where $d{\cal{S}}_+, ~ {\cal{P}}, ~d{\cal{S}}_-,~{\cal{NH}}_+,~{\cal{G}},~{\cal{NH}}_-$ stand respectively for the de Sitter,
Poincar\'e, the
anti de Sitter, the expanding Newton-Hooke, the Galilei and the oscillating Newton-Hooke Lie algebras and where we have omitted
the brackets of the form $[J,X_i]=X_j\epsilon^j_i$ and $[J,H]=0$.
\begin{table}[htbp]
\begin{center}
 \begin{tabular}{|c|c|c|}
\hline
$\gamma=\frac{1}{c^2}$&$~$&$[K_j,K_k]=-\frac{1}{c^2} J_l\epsilon_{jk}^{l},~[K_j,P_k] =\frac{1}{c^2}\delta_{jk}H,~[K_j,H]= P_j$\\
\hline
$d\cal{S}_+$&$\beta=\omega^2$&$[P_j,P_k] =\kappa^2 J_l\epsilon_{jk}^l,~~[P_j,H]=\omega^2 K_j$\\
\hline
$\cal{P}~~~~$&$\beta=0$&$[P_j,P_k] =0,~~~~~~~~~~~[P_j,H]=0~~~~~$\\
\hline
$d\cal{S}_-$&$\beta=-\omega^2$&$~~~[P_j,P_k] =-\kappa^2 J_l\epsilon_{jk}^l,~[P_j,H]=-\omega^2 K_j$\\
\hline
$\gamma=0$&$~$&$[K_j,K_k]=0$,$[K_j,P_k] =0$,$[K_j,H]= P_j$\\
\hline
$\cal{NH}_+$&$\beta=\omega^2$&$~~~[P_j,P_k] =0,~~~~~~~~~~~~~[P_j,H]=\omega^2 K_j$\\
\hline
$\cal{G}~~~$&$\beta=0$&$~~[P_j,P_k] =0,~~~~~~~~~~~~~~[P_j,H]=0$\\
\hline
$\cal{NH}_-$&$\beta=-\omega^2$&$~~~~~~~~~[P_j,P_k] =0,~~~~~~~~~~~[P_j,H]=-\omega^2 K_j$\\
\hline
 \end{tabular}
 \end{center}
 \caption{ \bf Kinematical algebras with $[K_i,H]=P_i$}
 \label{table1}
\end{table}
\subsection*{Boosts commuting with time translations }
This case corresponds to $\lambda=0$.\;\;The planar kinematical algebras are then defined by the brackets (\ref{rotationsplanar}),
(\ref{translations1}) and (\ref{boosts3}).\;\;In this case, according to the values of
$\gamma$ and $\beta$, we obtain the kinematical algebras summarized in the table below
 where ${\cal{P}}^{\prime}_+,~{\cal{P}}^{\prime}_-,~
{\cal{C}},~
{\cal{G}}^{\prime}_+,~{\cal{G}}^{\prime}_-$ and ${\cal{S}}$ stand respectively for the Para-Poincar\'e, the anti Para-Poincar\'e, the
 Carroll, the Para-Galilei, the
Anti-Para-Galilei and the Static Lie algebras and where we have omitted the brackets of the
form $[J,X_i]=X_j\epsilon^j_i$ and $[J,H]=0$.
\\
\\
\\
\\
\\
\\
\\
\\
\\
\\
\\
\\
\begin{table}[htbp]
\begin{center}
 \begin{tabular}{|c|c|c|}
\hline
$\gamma=\frac{1}{c^2}$&$~$&$[K_j,K_k]=0,~[K_j,P_k] =\frac{1}{c^2}\delta_{jk}H,~[K_j,H]= 0$\\
\hline
$\cal{P}^{\prime}_+$&$\beta=\omega^2$&$[P_j,P_k] =\kappa^2 J_l\epsilon_{jk}^l,~~[P_j,H]=\omega^2 K_j$\\
\hline
$\cal{C}~~$&$\beta=0$&$[P_j,P_k] =0,~~~~~~~~~~~[P_j,H]=0~~~~~$\\
\hline
${P}^{\prime}_- $&$\beta=-\omega^2$&$~~~[P_j,P_k] =-\kappa^2 J_l\epsilon_{jk}^l,~[P_j,H]=-\omega^2 K_j$\\
\hline
$\gamma=0$&$~$&$[K_j,K_k]=0~,~[K_j,P_k] =0,~~[K_j,H]= 0$\\
\hline
$\cal{G}^{\prime}_+$&$\beta=\omega^2$&$~~~[P_j,P_k] =0,~~~~~~~~~~~~~[P_j,H]=\omega^2 K_j$\\
\hline
$\cal{S}~~$&$\beta=0$&$~~[P_j,P_k] =0,~~~~~~~~~~~~~~[P_j,H]=0$\\
\hline
$\cal{G}^{\prime}_-$&$\beta=-\omega^2$&$~~~~~~~~~[P_j,P_k] =0,~~~~~~~~~~~~~[P_j,H]=-\omega^2 K_j$\\
\hline
 \end{tabular}
 \end{center}
  \caption{\bf Kinematical algebras with $[K_i,H]=0$}
 \label{table2}
\end{table}
\\
 Thus, under a natural assumption on the decomposition of the adjoint action into irreducible
(corresponding to isotropy of the space and homogeneity of the spacetime) and requiring that time and
space reversals are automorphisms, there are exactly
$11$ kinematical algebras \cite{5bacry}.\\

 In the discussion above, according to the link between the three parameters
$\beta$, $\gamma$ and $\lambda$, we have obtained an other Lie algebra: the Anti-Para-Galilei
Lie algebra denoted $\cal{G}^{\prime}_-$.\;\;This is consistent with the
contraction process.\;\;In fact, as the Para-Galilei algebra $\cal{G}^{\prime}_+$ is a velocity-space contraction of the
Para-Poincar\'e
algebra $\cal{P}^{\prime}_+$ or a velocity-time contraction of the Newton-Hooke group $NH_+$ \cite{inonu}, by the same process,
the anti Para-Galilei algebra $\cal{G}^{\prime}_-$ is a velocity-space contraction of the anti-Para-Poincar\'e
algebra $\cal{P}^{\prime}_-$ or a velocity-time contraction of the Newton-Hooke group $NH_-$, the latter
correspond to the value $\beta=-\omega^2$.\\

In conclusion, the Lie brackets for the possible planar kinematical algebras according to the classification in \cite {5bacry} are
summarized in the following table where the brackets of the form $[J,X_i]=X_j\epsilon^j_i$ and $[J,H]=0$
are omitted and where we have also considered the Anti-Para-Galilei Lie algebra $\cal{G}^{\prime}_-$.
\\
\\
\\
\\
\\
\\
\\
\\
\begin{table}[htbp]
\begin{center}
 \begin{tabular}{|c|c|c|c|c|c|c|}
\hline
Lie algebra& $[K_i,H]$&$[K_i,K_j]$&$[K_i,P_j]$&$[P_i,P_j]$& $[P_i,H]$\\
\hline
 $d\cal{S}_+$ & $P_i$ & $-\frac{1}{c^2}J_k\epsilon^k_{ij}$ & $\frac{1}{c^2}H\delta_{ij}$ & $\kappa^2J_k\epsilon^k_{ij}$& $\omega^2 K_i$\\
\hline
  $\cal{P}$ & $P_i$ & $-\frac{1}{c^2}J_k\epsilon^k_{ij}$ & $\frac{1}{c^2}H\delta_{ij}$ & $0$& $0$ \\
\hline
 $d\cal{S}_-$ & $P_i$ & $-\frac{1}{c^2}J_k\epsilon^k_{ij}$ & $\frac{1}{c^2}H\delta_{ij}$ & $-\kappa^2J_k\epsilon^k_{ij}$&
$-\omega^2 K_i$ \\
\hline
  $\cal{NH}_+$ & $P_i$ & $0$ & $0$ & $0$& $\omega^2 K_i$ \\
\hline
 $\cal{G}$& $P_i$ & $0$ & $0$ & $0$& $0$ \\
\hline
$\cal{NH}_-$ & $P_i$ & $0$ & $0$ & $0$& $-\omega^2 K_i$ \\
\hline
$\cal{P}^{\prime}_+$&  $0$ & $0$ & $\frac{1}{c^2}H\delta_{ij}$ & $\kappa^2J_k\epsilon^k_{ij}$& $\omega^2 K_i$ \\
\hline
  $\cal{C}$& $0$ & $0$ & $\frac{1}{c^2}H\delta_{ij}$ & $0$& $0$ \\
\hline
 $\cal{P}^{\prime}_-$ & $0$ & $0$ & $\frac{1}{c^2}H\delta_{ij}$ & $-\kappa^2J_k\epsilon^k_{ij}$& $-\omega^2 K_i$ \\
\hline
  $\cal{G}^{\prime}_+$&  $0$ & $0$ & $0$ & $0$& $\omega^2 K_i$ \\
\hline
$\cal{G}^{\prime}_-$&  $0$ & $0$ & $0$ & $0$& $-\omega^2 K_i$ \\
\hline
 $\cal{S}$ & $0$ & $0$ & $0$ & $0$& $0$ \\
 \hline
\end{tabular}
\end{center}
 \caption{\bf Lie brackets for the possible kinematical algebras }
\label{table3}
\end{table}
\\
Furthermore, through the classification in \cite{5bacry}, to each kinematical algebra is associated a corresponding Lie group.\;\;
Each of these associated groups is either the de Sitter or the anti-de Sitter or one of their contractions.\;\;Recall that there are
three fundamental types of contraction: velocity-space, velocity-time and space-time contraction corresponding respectively to
contracting to the subgroup generated by $H, ~P$ and $K$.\\

Geometrically, the velocity-space contraction (i.e making the
substitution $K\rightarrow \epsilon K,~P\rightarrow \epsilon P$ into the Lie algebra and calculating the
singular limit of the Lie brackets as $\epsilon \rightarrow 0$) means to describe spacetime near a timelike
geodesic equivalent to passing from relativistic to absolute time as stated in \cite{mcrae}.
\;\;So, $dS_+$ and $dS_-$ contract to $NH_{\pm}$ respectively, Poincar\'e
 group $P$ contracts to Galilei group $G$, Para-Poincar\'e groups $P^{\prime}_{\pm}$ contract to Para-Galilei
 groups $G^{\prime}_{\pm}$ and Carroll group $C$ contracts to Static group $S$ through a velocity-space contraction.
 \;\;In other words, the former are relativistic while the latter are absolute time groups. \\

 Similarly, the velocity-time contraction (i.e making the
substitution $K\rightarrow \epsilon K,~H\rightarrow \epsilon H$ into the Lie algebra and calculating the
singular limit of the Lie brackets as $\epsilon \rightarrow 0$) geometrically means to describe spacetime near a spacelike
geodesic equivalent to passing from relativistic to absolute space.\;\;So, $dS_+$ and $dS_-$ contract to $P^{\prime}_{\pm}$
respectively, $NH_{\pm}$ contract to $G^{\prime}_{\pm}$, Poincar\'e groups $P$ contract to Carroll $C$
 and the Galilei group $G$ contracts to Static group $S$ through a velocity-time contraction.\;\;In other words, the former
 are relativistic while the latter are absolute space groups. \\

 Finally, the space-time contraction (i.e making the
substitution $P\rightarrow \epsilon P,~H\rightarrow \epsilon H$ into the Lie algebra and calculating the
singular limit of the Lie brackets as $\epsilon \rightarrow 0$) geometrically means to describe spacetime near an event (physically
the spacelike and the timelike intervals are small but the boosts are not restricted).\;\;The corresponding group is called local group
as opposed to a cosmological group.\;\;
So, $dS_+$ and $dS_-$ contract to Poincar\'e group $P$, $NH_{\pm}$ contract to Galilei group $G$, Para-Galilei
 groups $G^{\prime}_{\pm}$ contracts to Static group $S$ and Para-Poincar\'e groups $P^{\prime}_{\pm}$ contract to Carroll group $C$
 through a space-time contraction.\;\;In other words, the former are cosmological groups while the
 latter are local groups. \\

The kinematical groups are then distributed according to the table below:

\begin{table}[htbp]
\begin{centering}
\begin{tabular}{|c|c|}
\hline
Relative time groups&de Sitter, Poincar\'e, Para-Poincar\'e, Carroll\\
\hline
Absolute time groups &Newton-Hooke, Galilei, Para-Galilei, Static\\
\hline
Relative space groups&de Sitter, Newton-Hooke, Poincar\'e, Galilei\\
\hline
Absolute space groups&Para-Poincar\'e, Para-Galilei, Carroll, Static\\
\hline
Cosmological groups&de Sitter, Newton-Hooke, Para-Poincar\'e, Para-Galilei\\
\hline
Local groups&Poincar\'e, Galilei, Carroll, Static\\
\hline
\end{tabular}
\end{centering}
 \caption{ \bf Kinematical groups classification according \cite{mcrae}}
\label{classification}
\end{table}
\subsection{Planar anisotropic kinematical algebras}
Let us consider the planar kinematical algebras of the table (\ref{table3}).\;\;Their corresponding planar
anisotropic Lie algebras are obtained by dropping the generators $J_i$ of rotations, meaning that
they are generated by $\{K_i,P_i,H\}$.\;\;This is only possible for the two \textbf{Newton-Hooke} Lie algebras, the \textbf{Galilei} and
the \textbf{Para-Galilei} Lie algebras, the \textbf{Carroll} Lie algebra and the \textbf{Static} Lie algebra where the rotation
generators do not appear in the right hand side of the brackets $[K_i,K_j]$ and $[P_i,P_j]$.\;\;The planar
anisotropic kinematical algebras are summarized in the following table (for~ $i,j=1,2)$:
\begin{table}[htbp]
\begin{center}
 \begin{tabular}{|c|c|c|c|c|c|c|}
\hline
Lie algebra& $[K_i,H]$&$[K_i,K_j]$&$[K_i,P_j]$&$[P_i,P_j]$& $[P_i,H]$\\
\hline
  $\cal{NH}_+$ & $P_i$ & $0$ & $0$ & $0$& $\omega^2 K_i$ \\
\hline
 $\cal{G}$& $P_i$ & $0$ & $0$ & $0$& $0$ \\
\hline
$\cal{NH}_-$ & $P_i$ & $0$ & $0$ & $0$& $-\omega^2 K_i$ \\
\hline
  $\cal{C}$& $0$ & $0$ & $\frac{1}{c^2}H\delta_{ij}$ & $0$& $0$ \\
\hline
  $\cal{G}^{\prime}_{\pm}$&  $0$ & $0$ & $0$ & $0$& $\pm\omega^2 K_i$ \\
\hline
 $\cal{S}$ & $0$ & $0$ & $0$ & $0$& $0$ \\
 \hline
\end{tabular}
\end{center}
 \caption{\bf Lie brackets for the planar anisotropic kinematical algebras}
\label{table5}
\end{table}
\\
Next to the above anisotropic kinematical algebras we consider in this thesis, also the absolute time Lie algebras which correspond
to the absolute time groups given in the table (\ref{classification})
with respect to the isotropy of the two-dimensional space.\;\;One simple reason we consider
the two types of kinematical algebras (namely anisotropic and absolute time) is that they both admit
central and noncentral abelian extensions.\;\;Hereafter, absolute (relative) time groups will be called nonrelativistic (relativistic)
kinematical groups.\;\;In this thesis, as a relative-time group \cite{mcrae, tybor}, the \textquotedblleft Carroll \textquotedblright
group is considered in a relativistic theory: i.e that incorporates the speed of light
as a parameter.\;\;In other
words, in its corresponding Lie algebra, the energy $H$ is not
considered as a central charge contrary to the assumption given in \cite{21} where the contracted Lie algebra contains
a Heisenberg sub-algebra in which energy $H$ appears as a central charge.\;\;In this context, all
possible anisotropic kinematical groups considered here are absolute time groups except
the \textquotedblleft Carroll \textquotedblright group.\\

In the following sections, we determine central and noncentral abelian extensions of the above planar kinematical algebras.
\section{Extensions of planar anisotropic kinematical algebras}
It is a well known fact in physics (particularly in quantum mechanics) that a projective representation of a group
is essentially equivalent to a regular representation of a central
extension of the group.\;\;For this reason, central extensions are of special importance in physics.\;\;In general, a kinematical group
should have or not a central extension in a certain dimension.\;\;However some of them
still have plenty of noncentral
extensions, some of which are interesting from the physical point of vue.\\

We restrict our study to the planar kinematical algebras which are six-dimensional and we are interested in the central extensions of
anisotropic kinematical algebras (without rotation parameters) and the noncentral abelian extensions of the absolute
time kinematical algebras (rotations included).\;\;Note also that
the passage from the extended Lie algebras to their corresponding extended Lie groups goes via the exponential function
as usually and the group multiplication laws are defined by the Baker-Campbell-Hausdorff formulas \cite{hall} as we will
see it later through our constructions.\\

To facilitate reading, let us first of all begin this section with some fundamental definitions.
\subsection{mathematical preliminaries: abelian extensions of Lie algebras }
\begin{dfn}
A Lie algebra ${\hat{\cal{G}}}$ is an extension of a Lie algebra ${\cal{G}}$ by an abelian Lie algebra ${\cal{A}}$ if there
is an exact sequence of Lie algebras \\
$$0\stackrel{i}\longrightarrow {\cal{A}}\longrightarrow {\hat{\cal{G}}}\stackrel{\pi}\longrightarrow {\cal{G}}\longrightarrow 0$$
i. e with continuous homomorphisms:
such that $\pi$ admit a continuous global section $s: {\cal{G}}\longrightarrow{\hat{\cal{G}}}$, $\pi\circ s= id_{{\cal{G}}}$
\end{dfn}
\begin{dfn}
Let ${\hat{\cal{G}}}_1$ and ${\hat{\cal{G}}}_2$ be two extensions of $\cal{G}$ by an abelian Lie algebra $\cal{A}$ i.e
 $$0\longrightarrow {\cal{A}}\longrightarrow {\hat{\cal{G}}}_1\stackrel{{\pi}_1}\longrightarrow {\cal{G}}\longrightarrow 0$$
and
$$0\longrightarrow {\cal{A}}\longrightarrow {\hat{\cal{G}}}_2\stackrel{{\pi}_2}\longrightarrow {\cal{G}}\longrightarrow 0$$
  ${\hat{\cal{G}}}_1$ and  ${\hat{\cal{G}}}_2$ are equivalent if there exists a Lie isomorphism $\Phi: {\hat{\cal{G}}}_1\rightarrow
{\hat{\cal{G}}}_2$ such that
\begin{eqnarray}
\pi_2\circ\Phi=\pi_1
\end{eqnarray}
\end{dfn}
There is a cocycle construction of abelian Lie algebra extensions.\;\;In order to define the
cohomology of ${\cal{G}}$ with coefficients in a $\varrho(\cal{G})$-module $\cal{A}$
(we shall take $\cal{A}$ here
to be the representation space of a finite-dimensional representation $\varrho$ of ${\cal G}$), the
coboundary operator is introduced:
\begin{eqnarray}\label{coboundaryonLiealgebra}\nonumber
  (\delta \alpha_n)(X_1,X_2,...,X_{n+1})=\sum^{n+1}_{i=1}(-1)^{i+1}\varrho(X_i)\alpha_n(X_1,X_2,...,\check{X}_i,...,X_{n+1})+\\
 \sum^{n+1}_{i,j=1,~i<j}\alpha_n([X_i,X_j],X_1,...,\check{X}_i,...,\check{X}_j,...,X_{n+1})
\end{eqnarray}
where $\alpha_n$ is a one $n$-cochain on ${\cal{G}}$ with values in $\cal{A}$ (i.e an alternate multilinear map),
 $X_1,X_2,\dots,X_{n+1} \in {\cal{G}}$ while the symbol \textquotedblleft ${\check{.}}$ \textquotedblright means that
 the variable under it has been deleted.

By using the fact that $\varrho$ is a Lie algebra homomorphism, we verify that $\delta^2=0$.\;\;This permits us
to obtain by the usual procedure the cohomology group
$$ H^{n}_{\varrho}({\cal{G}},\cal{A})= \frac{Z^{n}_{\varrho}({\cal{G}},\cal{A})}{B^{n}_{\varrho}({\cal{G}},\cal{A})}$$
of the Lie algebra ${\cal G}$ with values in $\cal{A}$ with respect to the representation $\varrho$ where
the group $Z^{n}_{\varrho}({\cal{G}},\cal{A})=Ker(\delta: C^n({\cal G}, \cal{A})\longrightarrow C^{n+1}({\cal G},\cal{A}))$ is
the group of $n$-cocycles and
$B^{n}_{\varrho}({\cal{G}},\cal{A})=Im(\delta: C^{n-1}({\cal G}, \cal{A})\longrightarrow C^n({\cal G}, \cal{A}))$ is the group of
$n$-coboundaries.\;\;It is well known that the
second cohomology group $ H^2( {\cal{G}}, \cal{A})$ is related to the extensions of ${\cal{G}}$
by $\cal{A}$ \cite{tuynmann}.

Note that the semi-direct sum of ${\cal{G}}$ and $\cal{A}$ consists of pairs $(X,A) \in {\cal{G}} \times \cal{A}$ with the commutator
\begin{eqnarray}
 [(X_i,A_a),(X_j,A_b)]=([X_i,X_j],\varrho(X_i)A_b-\varrho(X_j)A_a)
\end{eqnarray}
Let $\alpha_2 \in Z^2( {\cal{G}},\cal{A})$ be a two-cocycle.\;\;Define the following modified
commutator in the sum vector space $\cal{G}\oplus_{\alpha_2}\cal{A}$
\begin{eqnarray}\label{liebracketext}
 [(X_i,A_a),(X_j,A_b)]_{\alpha_2}=([X_i,X_j],\varrho(X_i)A_b-\varrho(X_j)A_a+\alpha_2(X_i,X_j))
\end{eqnarray}
\\
\begin{pro}
 The modified bracket $[.,.]_{\alpha_2}$ is a Lie bracket.
\end{pro}
\begin{itemize}
\item The bilinearity and antisymmetry of the above modified bracket follow from the definition of the two-cocycle
and from that of the Lie bracket of $\cal{G}$.
\item The Jacobi identity in ${\hat{\cal{G}}}$ is equivalent to the cocycle condition for $\alpha_2$:
\begin{eqnarray} \label{Cocycond}\nonumber
(\delta\alpha_2)(X_i,X_j,X_k)=\varrho(X_i)\alpha_2(X_j,X_k)-\varrho(X_j)\alpha_2(X_i,X_k)\\\nonumber
+\varrho(X_k)\alpha_2(X_i,X_j)+\alpha_2(X_i,[X_j,X_k])\\+\alpha_2(X_j,[X_k,X_i])+\alpha_2(X_k,[X_i,X_j])=0
\end{eqnarray}
\end{itemize}
Then $\cal{G}\oplus_{\alpha_2}\cal{A}$
endowed with the Lie bracket (\ref{liebracketext}) defines an abelian Lie algebra extension
$\hat{\cal{G}}=\cal{G}\oplus_{\alpha_2}\cal{A}$.

Let $\alpha_1$ and $\alpha_2$ $\in Z^2( {\cal{G}},\cal{A})$.\;\;The Lie algebras formed from these two-cocycles are isomorphic
through a map of the
type:
$$\Phi(X,A)=(X,A+\beta(X)), X\in \cal{G} , A\in \cal{A}$$
where $\beta \in H^{1}_{\varrho}({\cal{G}},{\cal{A}})$.
\begin{pro}
 The condition
\begin{eqnarray}
 [\Phi(X_i,A_a),\Phi(X_j,A_b)]_{{\hat{\cal{G}}}_2}=\Phi([(X_i,A_a),(X_j,A_b)]_{{\hat{\cal{G}}}_1})
\end{eqnarray}
is the same as
\begin{eqnarray}
\alpha_2-\alpha_2^{~\prime}=\delta\beta
\end{eqnarray}
\end{pro}
\begin{prof}
 $$[\Phi(X_i,A_a),\Phi(X_j,A_b)]_{{\hat{\cal{G}}}_2}=[(X_i,A_a+\beta(X_i)),(X_j,A_b+\beta(X_j))]_{{\hat{\cal{G}}}_2} $$
$$=([X_i,X_j],\varrho(X_i)(A_b+\beta(X_j))-\varrho(X_j)(A_a+\beta(X_i))+\alpha_2^{~\prime}(X_i,X_j))$$
$$=([X_i,X_j],\varrho(X_i)A_b+\varrho(X_i)\beta(X_j)-\varrho(X_j)A_a-\varrho(X_j)\beta(X_i)+\alpha_2^{~\prime}(X_i,X_j))(*)$$
$$\Phi([(X_i,A_a),(X_j,A_b)]_{{\hat{\cal{G}}}_1})=([X_i,X_j],\varrho(X_i)A_b-\varrho(X_j)A_a+\beta([X_i,X_j])
+\alpha_2(X_i,X_j))(**)$$
$$(*)=(**)\Longleftrightarrow \varrho(X_i)\beta(X_j)-\varrho(X_j)\beta(X_i)+\alpha_2^{\prime}(X_i,X_j))=\alpha_2(X_i,X_j))+\beta([X_i,X_j])$$
$$\alpha_2-\alpha_2^{~\prime}=\delta\beta$$
\end{prof}
\\
 i.e $\alpha_2$ and $\alpha_2^{\prime}$ are cohomologous.\;\;Thus, up to an isomorphism of
the above type, the Lie algebra extensions are parametrized by
elements of $H^{2}_{\varrho}({\cal{G}},{\cal{A}})$.\\
So, each $\alpha_2 \in H^{2}_{\varrho}({\cal{G}},{\cal{A}}) $ defines a new Lie algebra abelian extension.\;\;Then, the
second cohomology group $H^{2}_{\varrho}({\cal{G}},{\cal{A}})$ is related to the abelian extensions of ${\cal{G}}$
by an abelian Lie algebra ${\cal{A}}$ for a given representation $\varrho$.\\

The extension ${\hat{\cal{G}}}$ is called {\bf{central}} if ${\cal{A}}$ lies in the center of ${\hat{\cal{G}}}$, i.e $[{\cal{A}},
{\hat{\cal{G}}}]=0$, this
is the case when the representation $\varrho$ of ${\cal{G}}$ is trivial \cite{12}.\;\;In this case the
commutator (\ref{liebracketext}) is reduced to
\begin{eqnarray}\label{centralexttwococycle}
 [(X_i,A_a),(X_j,A_b)]_{\alpha_2}=([X_i,X_j], \alpha_2(X_i,X_j))
\end{eqnarray}
where the two-cocycle $\alpha_2$ satisfies:
\begin{eqnarray}\label{1}
 \alpha_{2}(X_i,X_j)=-\alpha_2(X_j,X_i)
\end{eqnarray}
and
\begin{eqnarray}\label{2}
 \alpha_{2}(X_i,[X_j,X_k])+\alpha_2(X_j,[X_k,X_i])+\alpha_2(X_k,[X_i,X_j])=0
\end{eqnarray}
from the properties of the Lie brackets in $\cal{G}$ and $\hat{\cal{G}}$.\\

As it has been argued in the introduction, we will consider also the general case: {\bf{ noncentral abelian extension}}
of ${\cal{G}}$ by an abelian Lie algebra
${\cal{A}}$ for a nontrivial action $\varrho$, the word noncentral meaning that $[{\cal{A}},{\hat{\cal{G}}}]\neq0$.\\

Explicitly, let us consider that ${\cal{G}}$ is a $n$-dimensional real Lie algebra generated by $X_i$ such that:
\begin{equation}
 [X_i,X_j]=X_kC_{ij}^k
\end{equation}
We know that the $n^3$ structure constants $C_{ij}^k$ satisfy the antisymmetry character:
\begin{eqnarray}\label{skew-sym1}
 C_{ij}^k=- C_{ji}^k
\end{eqnarray}
and the Jacobi identity:
\begin {eqnarray}\label{jacobi1}
 C_{ir}^l C_{jk}^r + C_{jr}^l C_{ki}^r + C_{kr}^l C_{ij}^r =0
\end {eqnarray}
Let us now consider that the Lie algebra ${\cal{A}}$ is generated by $Y_\alpha$ and $Z_\alpha, \alpha=1,\dots, dim \cal{A}$
such that the non trivial Lie brackets are:
\begin{eqnarray}\label{ext}
 [Y_\alpha,Y_\beta]=Z_aC_{\alpha\beta}^a
\end{eqnarray}
\begin{itemize}
\item A central abelian extension $ \hat{\cal{G}} $ of ${\cal{G}}$ by ${\cal{A}}$
is the Lie algebra generated by $X_i$ and $Y_\alpha $
and defined by the non trivial Lie brackets
\begin{eqnarray}\label{centralabelian}
 [X_i,X_j]&=&X_k C_{ij}^k +Y_{\alpha} d_{ij}^{\alpha}
\end{eqnarray}
or by (\ref{centralexttwococycle}) in terms of $2$-cocycle.\;\;It follows  that
\begin{eqnarray}\label{cextcocycl}
 \alpha_2(X_i,X_j)=Y_\alpha d_{ij}^{\alpha}
\end{eqnarray}
and we verify from the properties (\ref{1}) and (\ref{2}) of the two-cocycle that:
\begin{eqnarray}\label{skew-symm}
d_{ij}^\alpha=-d_{ji}^\alpha
\end{eqnarray}
and
\begin{eqnarray}\label{closurecond}
 d_{ir}^\alpha C_{jk}^r+ d_{jr}^\alpha C_{ki}^r+ d_{kr}^\alpha C_{ij}^r= 0
\end{eqnarray}
Then the structures constant $ d_{ij}^{\alpha}$ satisfy also the antisymmetry character (\ref{skew-symm}) and the closure
condition (\ref{closurecond}) in $\hat{\cal{G}}$.\\

Looking for solutions of the system (\ref{skew-symm})-(\ref{closurecond}), we obtain the maximal central
extension $\hat{\cal{G}}$ of the Lie algebra $\cal{G}$ by
the abelian Lie algebra $\cal{A}$.\;\;All central extensions of a Lie algebra by an abelian Lie algebra are obtained in this manner.
\item A noncentral nonabelian extension $ \hat{\cal{G}} $ of ${\cal{G}}$ by ${\cal{A}}$ is the Lie algebra
generated by $X_i,Y_\alpha,Z_a $ and $T_s$ and defined by the non trivial Lie brackets (\ref{ext}), (\ref{centralabelian}) and
\begin{eqnarray}\label{noncetralext}
[X_i,Y_{\beta}]&=&T_s C_{i\beta}^s
\end{eqnarray}
with Jacobi identities and the skew-symmetry character of the Lie bracket in $\hat{\cal{G}}$.\;\;We have the following relations
between the structure constants:
\begin{eqnarray}\label{systnoncgene}\nonumber
  d_{ij}^\alpha=-d_{ji}^\alpha  \\  \nonumber
 d_{ir}^\alpha C_{jk}^r+ d_{jr}^\alpha C_{ki}^r+ d_{kr}^\alpha C_{ij}^r= 0\\\nonumber
 C_{k\nu}^sd_{ij}^\nu+ C_{j\nu}^sd_{ki}^\nu+ C_{k\nu}^s d_{ij}^\nu= 0\\
C_{\alpha k}^s C_{ij}^k=0\\\nonumber
C_{\alpha\beta}^a d_{ij}^\beta =0\\\nonumber
\end{eqnarray}
 Particularly, the Lie algebra $\hat{\cal{G}}$ will be called noncentral abelian extension of ${\cal{G}}$ by ${\cal{A}}$ if $Z_a=0$
(i.e ${\cal{A}}$ is abelian).\;\;Its Lie structure is given by the system
of brackets (\ref{centralabelian}) and (\ref{noncetralext}).\;\;In this case the
 system (\ref{systnoncgene}) is reduced to
\begin{eqnarray}\label{systncpart}\nonumber
d_{ij}^\alpha=-d_{ji}^\alpha\\\nonumber
 d_{ir}^\alpha C_{jk}^r+ d_{jr}^\alpha C_{ki}^r+ d_{kr}^\alpha C_{ij}^r= 0\\
 C_{k\nu}^sd_{ij}^\nu+ C_{j\nu}^sd_{ki}^\nu+ C_{k\nu}^s d_{ij}^\nu= 0\\\nonumber
C_{\alpha k}^s C_{ij}^k=0\\\nonumber
\end{eqnarray}
\end{itemize}
Looking for solutions of the system (\ref{systncpart}), we obtain the maximal noncentral abelian
extension $\hat{\cal{G}}$ of the Lie algebra $\cal{G}$ by
the abelian Lie algebra $\cal{A}$.\;\;All noncentral abelian extensions of a Lie algebra by
an abelian Lie algebra are obtained in the same way.\\

Furthermore, central and noncentral abelian extensions of Lie algebras can also be found by looking at the second Lie algebra cohomology group
(determining
the two-cocycles which define extensions) but this method is not analyzed in this thesis.\;\;Note that
the facts cited above are standard knowledge and can be found in \cite{hekmati, tuynmann,  azcarraga, hamermesh}.\\
\subsection{Central extensions of planar anisotropic kinematical algebras}
In the following, we use the discussion detailed above to determine the central extensions of the possible planar anisotropic
kinematical algebras namely: the Carroll algebra, the Galilei algebra, the Para-Galilei algebras, the two Newton-Hooke
algebras and the Static algebra.
\subsubsection{i) Carroll algebra}
The Carroll algebra was first introduced in \cite{levy} as a velocity-time contraction
\cite{inonu} of the Poincar\'e algebra through a rescaling of the boosts and the time translations.\;\;Although appearing naturally
in the classification of kinematical groups, as an
alternative intermediate algebra in the contraction of the Poincar\'e group onto the Static group, and therefore as another
nonrelativistic limit (the other being the Galilei algebra), the Carroll algebra has played no
distinguish role in Kinematics.\;\;Recently it has been analyzed whether
the Carroll algebra (considered as a nonrelativistic limit of the Poincar\'e algebra: i.e the energy $H$ is considered as a central
extension parameter) constitutes
an object in the study of the problem of tachyon condensation in string theory \cite{21}.\\

In the classification by
Mcrae \cite{mcrae}, Carroll group appeared as a relative-time group and its Lie algebra can be considered therefore as a relativistic
kinematical algebra whose only nontrivial bracket: $[K_i,P_j]=\frac{1}{c^2}\delta_{ij} H$ is such that $H$ is not a central
extension parameter.\;\;Furthermore, in the classification detailed in \cite{tybor}, when the Lie bracket
$[K_i,P_j]\neq 0$, it is stated that the considered algebra is a relativistic
one.\;\;As pointed out previously, we consider
the \textquotedblleft Carroll \textquotedblright algebra at least relativistically
 and it is in this context where a possible cosmological interpretation and the noncommutative
phase spaces on the Carroll group recover some interest.\\

The only nontrivial Lie bracket of the anisotropic Carroll algebra is given by:
\begin{eqnarray}\label{carrolbracket}
[K_i,P_j]=\frac{1}{c^2}H\delta_{ij}~~~~~~~~~~~~~~~~~~~~~~~~~
\end{eqnarray}
By using standard methods \cite{ kostant, kirillov, 5bacry,  nzo1, hamermesh, stuart},
i.e looking for the solutions of the system (\ref{skew-symm})-(\ref{closurecond}), we obtain that the central extension of this
Lie algebra is defined by the following Lie brackets
\begin{eqnarray}\label{carrolbrackets}
 [K_i,K_j]=\frac{1}{c^2}S\epsilon_{ij}~~~~~~~~~~~~~~~~~~~~~~~~~~~~~~~\\\nonumber
[K_i,P_j]=\frac{1}{c^2}H\delta_{ij},~[P_i,P_j]=\kappa^2 S\epsilon_{ij}\\\nonumber
[K_i,H]=0,~~~~~~~~~~~~[P_i,H]=0~~~~~~~~~\nonumber
\end{eqnarray}
where $\kappa$ is the inverse of the universe radius $r$ while the dimensionless generator $S$ spans the center of the
extended Lie algebra.
\subsubsection{ii) Absolute time anisotropic Lie algebras}
The cohomological structure which determines the existence of central extensions of absolute time kinematical groups originates in
their invariant subgroup spanned by translations and boosts \cite{souriau}.

The planar nonrelativistic anisotropic Lie algebras are collectively defined by the following Lie structure :
\begin{eqnarray}\label{anisotropicnr}
[K_j,K_k]=0~~~~~~~~~~~~~~~~~~~~~~~~~~~~~~~~~~~~~~\\ \nonumber
[K_j,P_k] =0,~~~~[P_j,P_k] =0~~~~~~~~~~~~~\\\nonumber
[K_j,H]=\lambda P_j,~[P_j,H]=\beta K_j~~~~~~~ \nonumber
\end{eqnarray}
for $j,k=1,2$ and with $\lambda=1$ or zero, $\beta=\pm \omega^2$ or zero.

\begin{thm}[]
 The vector space of central extensions of the planar nonrelativistic anisotropic
 kinematical algebra (\ref{anisotropicnr}) is two-dimensional.
\end{thm}
\begin{prof}
Dimensional analysis permits us to set that a priori
the possible central extensions of (\ref{anisotropicnr}) are defined by the following Lie brackets:
\begin{eqnarray}\label{anisotropext}
[K_j,K_k]=\frac{\mu}{c^2}S\epsilon_{jk}~~~~~~~~~~~~~~~~~~~~~~~~~~~~~~~~~~\\ \nonumber
[K_j,P_k] =\gamma M\delta_{jk},~[P_j,P_k] =\kappa^2\alpha S\epsilon_{jk}\\\nonumber
[K_j,H]=\lambda P_j,~~~~~~~[P_j,H]=\beta K_j~~~~~~~ \nonumber
\end{eqnarray}
where $S$ , $\alpha$, $\gamma$ and $\mu$ are dimensionless while the dimension of $M$ is $L^{-2}T$
and that of the parameter $\beta$ is $L^{-2}$.\\

 The Lie brackets (\ref{anisotropext})
will form Lie algebras if every triplet satisfies the Jacobi identities
\begin{eqnarray*}
 [K_i,[P_j,H]]+[P_j,[H,K_i]]+[H,[K_i,P_j]]=0
\end{eqnarray*}
which imply that $\mu$ and $\alpha$ are related by the following relation:
\begin{eqnarray}\label{relatio}
 \frac{\mu\beta}{c^2}=-\kappa^2 \lambda\alpha
\end{eqnarray}
By using relation $c=\frac{\omega}{\kappa}$ in (\ref{relatio}) for the possible values of $\lambda$ and $\beta$ (in section $1.1$),
we get the following cases:
\begin{itemize}
 \item $\mu=\mp\alpha$ for $\lambda=1$ and $\beta=\pm\omega^2$
\item $\mu\in \Re$ while $\alpha =0$ when $\lambda=1$ and $\beta=0$
\item $\alpha\in\Re$ while $\mu=0$ when $\lambda=0$ and $\beta=\pm\omega^2$
\item$\mu,\alpha \in \Re$ when $\lambda=0$ and $\beta=0$
\end{itemize}
Thus, we are left with two independent central generators $M$ and $S$ and we
can assume after normalization ($\mu=1$ and $\gamma=1$)
that the central extended nonrelativistic planar anisotropic Lie algebras are given by the Lie brackets summarized in the following
 table (for ~$i,j=1,2$):
\begin{table}[ htbp]
\begin{center}
\begin{tabular}{|c|c|c|c|c|c|c|}
\hline
Lie algebra&$[P_i,H]$&$[K_i,H]$&$[P_i,P_j]$&$[K_i,K_j]$& $[K_i,P_j]$\\
\hline
 Extended ${\cal{NH}}_+$ & $\omega^2 K_{i}$& $P_i$ & $-\kappa^2 S\epsilon_{ij}$ & $\frac{1}{c^2}S\epsilon_{ij}$&$M\delta_{ij}$ \\
\hline
 Extended ${\cal{NH}}_-$& $-\omega^2 K_i$ & $P_i$ & $\kappa^2 S\epsilon_{ij}$ &$\frac{1}{c^2}S\epsilon_{ij}$ & $M\delta_{ij}$ \\
\hline
 Extended ${\cal{G}}~~~~$ & $0$ & $P_i$ & $0$ & $\frac{1}{c^2}S\epsilon_{ij}$ & $M\delta_{ij}$ \\
\hline
Extended ${\cal{G}}^{\prime}_{\pm}~$ & $\pm\omega^2 K_{i}$ & $0$ & $\kappa^2 S\epsilon_{ij}$  & $0$ & $M\delta_{ij}$ \\
\hline
Extended ${\cal{S}}~~~~ $& $0$ & $0$ &  $\kappa^2 S\epsilon_{ij}$ & $\frac{1}{c^2}S\epsilon_{ij}$ & $M\delta_{ij}$ \\
\hline
\end{tabular}
\end{center}
\caption{\bf Central extensions of nonrelativistic planar anisotropic kinematical algebras}
\label{tab:table6}
\end{table}
\\
The two-fold central extensions summarized in the above table (\ref{tab:table6})
are then the maximal extensions of the nonrelativistic
anisotropic Lie algebras that can exponentiate to the corresponding planar kinematical groups.
\end{prof}
\\

Furthermore,
these centrally extended anisotropic Lie algebras are new except for the Newton-Hooke groups case
 \cite{ancilla, zhang-horvathy, galajinski}.
\subsection{Noncentral abelian extensions of the absolute time planar kinematical algebras}
Many authors have had a big interest in studying
three absolute time Lie algebras namely Galilei algebra and the two Newton-Hooke
algebras as well as their centrally extended structures \cite{ bose, levyleblond2, Duval-horvathy, galajinski, brihaye, arratia,
zhang, zhang1}.\;\;In the
following, we revisit them and see three others namely Para-Galilei algebras and the Static one.\;\;But we are interested here in
their noncentrally abelian extended Lie structures.\\

The six planar absolute time Lie algebras are defined by the following nontrivial Lie brackets
\begin{eqnarray*}
 [J,K_j]=K_i\epsilon^i_j,~[J,P_j]=P_i\epsilon^i_j,~[K_i,H]=\lambda P_i,~[P_i,H]=\beta K_i,~i,j=1,2
\end{eqnarray*}
with $\lambda=1$ or zero, $\beta=\pm \omega^2$ or zero. \\

 We distinguish four cases :

\subsubsection{i) Galilei algebra}
In the discussion above (in section $1.1$), the Galilei algebra corresponds to
the case $\lambda=1$ and $\beta=0$.\;\;The corresponding ($2+1$)-Galilei group
$G$ is a six-parameter Lie group.\;\;It is the kinematical group of a classical, nonrelativistic
spacetime having two spatial and one time dimensions.\;\;It consists of translations of time and space, rotations in
the two-dimensional space and velocity boosts.\;\;Explicitly, the Galilei
group $G$ in two-dimensional space is defined by the multiplication law
\begin{eqnarray}\label{galileilaw}
(\theta,\vec{v},\vec{x},t)(\theta^{\prime},\vec{v}^{~\prime},\vec{x}^{~\prime},t^{\prime})=(\theta+\theta^{\prime},
R(\theta)\vec{v}^{~\prime}+\vec{v},R(\theta)\vec{x}^{~\prime}+\vec{v}t^{\prime}+\vec{x},t+t^{\prime})
\end{eqnarray}
where $\theta$ is an angle of rotations, $\vec{v}$ is a boost
vector, $\vec{x}$ is a space translation vector and $t$ is a time
translation parameter.\;\;Its Lie algebra $\cal{G}$ is then generated by
the left invariant vector fields
\begin{eqnarray}\label{vectorfieldsgalilei}
J=\frac{\partial}{\partial
\theta}~,~\vec{K}=R(-\theta)\frac{\partial}{\partial
\vec{v}}~~,~~\vec{P}=R(-\theta)\frac{\partial}{\partial
\vec{x}}~,~H=\frac{\partial}{\partial
t}+\vec{v}.\frac{\partial}{\partial \vec{x}}
\end{eqnarray}
that satisfy the nontrivial Lie brackets
\begin{eqnarray}\label{galileialgebra}
[J,K_j]=K_i\epsilon^i_j~,~[J,P_j]=P_i\epsilon^i_j~,~[K_i,H]=P_i~~;i,j=1,2
\end{eqnarray}
The Galilei algebra can also be obtained from the Poincar\'e algebra ${\cal{I}}so(1,3)$ through
the velocity-space contraction determined by
rescaling the boosts and the space translations \cite{inonu}.\;\;Furthermore, the planar Galilei group $G$ and its Lie algebra
${\cal{G}}$ have been defined in \cite{ancille1}.

More specifically, the Lie algebra ${\cal{G}}$ of the Galilei group in two-dimensional space is given by the Lie brackets:
\begin{eqnarray*}\nonumber
 [J,K_1]=K_2 ~~~~~~~~~~~~~~~~~~~~~~~~~~~~~~~~~~~~~~~~~~~~~~~~~~~~~~~~~~~~~~~~~~~~~~~~~~~~~~~~~~~~~~~~~~~~~\\\nonumber
 [J,K_2]=- K_1, ~ [K_1,K_2]=0~~~~~~~~~~~~~~~~~~~~~~~~~~~~~~~~~~~~~~~~~~~~~~~~~~~~~~~~~~~~~~~~~~\\\nonumber
 [J,P_1]= P_2, ~ [K_1,P_1]= 0, ~[K_2,P_1]=0~~~~~~~~~~~~~~~~~~~~~~~~~~~~~~~~~~~~~~~~~~~~~~~~\\\nonumber
 [J,P_2]= - P_1,~[K_1,P_2]= 0, ~ [K_2,P_2]=0, ~ [P_1,P_2]=0~~~~~~~~~~~~~~~~~~~~~~\\\nonumber
 [J,H]= 0,~ [K_1,H]=P_1, ~ [K_2,H]=P_2 , ~[P_1,H]=0, ~[P_2,H]= 0~~~~
\end{eqnarray*}
Note that the planar Galilei algebra admits three central extensions overall.\;\;This has been demonstrated by looking at the second
Lie algebra cohomology group $H^2(\cal{G},\Re)$.\\As noted in \cite{bose, grigore2}, $H^2(\cal{G},\Re)$ is three-dimensional
which in turn means
that the Galilei algebra admits three central extensions in total.\;\;However, only two of those extensions
containing the parameters $M$ and $S$ exponentiate to the corresponding groups.\;\;The third extension whose central parameter
extension measures the
noncommutativity of $[J,H]$ does not survive the exponentiation and is ignored.\;\;Here we are interested in the noncentral
extensions of the planar Galilei algebra (with the assumption that $[J,H]=0$ meaning that
rotation invariance is taking into account).\\

Using dimensional analysis and considering the extended Lie algebra which exponentiate in the Lie group and
then appear to have a clear physical interpretation,
we then find the following particular noncentral extended Lie algebra structure:
\begin{eqnarray}\label{galileinoncentral}
[J,K_j]=K_i\epsilon^i_j,~~[K_j,K_k]=\frac{1}{c^2}S\epsilon_{jk}~~~~~~~~~~~~~~~~~~~\\\nonumber
[J,P_j]=P_i\epsilon^i_j,~~~[K_j,P_k] = M\delta_{jk},~[P_j,P_k] =0\\\nonumber
[J,F_j]=F_i\epsilon^i_j,~~ [K_i,F_j]=0,~~~~~~~~~~[P_i,F_j]=0\\\nonumber
[J,H]=0,~~~~~~~~[K_j,H]= P_j,~~~~[P_j,H]=F_j\\\nonumber
\end{eqnarray}
where $F_i, M$ and $S$ are new generators such that $F_i$ behave as vectors under rotations while $M$ and $S$ behave as scalars.\;\;This
noncentrally abelian extended Galilei algebra has been used in \cite{ancille1}.
 \subsubsection{ii) Newton-Hooke algebras}
 As already argued, the two nonrelativistic Newton-Hooke cosmological groups were introduced by Bacry and L\'evy-Leblond in \cite{5bacry}
 who classified all kinematical groups in $(3+1)$-dimensional spacetime.\;\;The Newton-Hooke symmetries can be obtained
 through velocity-space contraction
 from the de Sitter and anti-de Sitter geometries and they describe respectively the nonrelativistic expanding
 (with the symmetry described by $NH_+$ algebra) and oscillating (with the symmetry described by $NH_-$ algebra)
 universes.\;\; They contract themselves through a space-time contraction in Galilei group.\\

 In the discussion detailed above (in section $1.1$),
the Newton-Hooke algebras correspond to the case $\lambda=1$ and $\beta=\pm \omega^2$.\;\;Here, we
determine the noncentral abelian extension of the Newton-Hooke algebras in the same manner as we did in
the discussion for the Galilei algebra.\;\;Dimensional analysis permits us to set that after normalization
the noncentral abelian extensions of $\cal{NH}_{\pm}$ coincides with their two-fold centrally extended Lie algebras
whose nontrivial Lie brackets are given by:
\begin{eqnarray}\label{anisotropicextncnh}
[J,K_j]=K_i\epsilon^i_j,~[J,P_j]=P_i\epsilon^i_j,~[K_j,K_k]=\frac{1}{c^2}S\epsilon_{jk},~[K_j,P_k] = M\delta_{jk}\\\nonumber
[P_j,P_k] =\kappa^2 S\epsilon_{jk}~,[K_j,H]= P_j,~[P_j,H]=\pm \omega^2 K_j~~~~~~~~~~~~~~~~~~~~~~~~
\end{eqnarray}
where $S$ and $M$ generate the center of these extended algebras, ${\kappa}$ is a constant whose dimension is inverse of that
of a length while $\omega$ is the time curvature (frequency).
\subsubsection{iii) Para-Galilei algebras}
The nonrelativistic Para-Galilei groups (i.e namely Para-Galilei and Anti-Para-Galilei) are obtained through a velocity-time
contraction of the Newton-Hooke groups or through a velocity-space
contraction of the Para-Poincar\'e groups (i.e namely Para-Poincar\'e and Anti-Para-Poincar\'e).\;\;They contract themselves by a
space-time contraction in the Static group \cite{5bacry, inonu}.\\

For the Para-Galilei algebras, which correspond to the case $\lambda=0$ and $\beta=\pm \omega^2$ in the discussion (of kinematical
algebras) encountered so far (in section $1.1$), we
proceed in a similar way to search for their associated noncentral abelian extensions.

We find that their corresponding noncentrally abelian extended Lie algebras has the following structure
 \begin{eqnarray}\label{anisotropicnoncentralparag}
[J,K_j]=K_i\epsilon^i_j,~[K_j,K_k]=0~~~~~~~~~~~~~~~~~~~~~~~~~~~~~~~~~~~~~~\\\nonumber
[J,P_j]=P_i\epsilon^i_j,~[K_j,P_k] = M\delta_{jk},~[P_j,P_k] =\kappa^2 S\epsilon_{jk}\\\nonumber
 [J,H]=0,~~~~~~[K_j,H]= \Pi_j,~~~~~~~[P_j,H]=\pm \omega^2 K_j \\\nonumber
[J,\Pi_j]=\Pi_i\epsilon^i_j~~~~~~~~~~~~~~~~~~~~~~~~~~~~~~~~~~~~~~~~~~~~~~~~~~~~~~~~~~~~~\\\nonumber
\end{eqnarray}
where $\vec{\Pi}$ behaves as a vector under rotation and has the same dimension as $\vec{P}$ while $S$ and $M$ commute with other
generators of this noncentrally abelian extended Lie algebras.\;\;This case corresponds
to the planar noncentrally abelian extended Para-Galilei algebras \cite{ancille1}.
\subsubsection{iv) Static algebra}
The planar Static algebra is an abelian Lie algebra generated by $J$ for rotation,
 $\vec{K}$ for boosts, $\vec{P}$ for space translations and $H$ for
time translations.\;\;The general element of the connected Static group can be written as $$g=\exp(\vec{v}\vec{K}+\vec{x}
\vec{P}+t H)\exp(\theta J)$$
where the parameter $\vec{v}$, $\vec{x}$ and $t$ are respectively the velocity parameter, the space and time translation parameters
while $\theta$ is an angle.\;\;As already argued,
we restrict our study to the planar anisotropic Static group. \\

In the discussion of section ($1.1$), we have seen that the Static algebra corresponds to the case $\lambda=0$ and
$\beta=0$.\;\;Similar computations as in
previous cases yield to the noncentral abelian extension of the Static algebra defined by the following Lie brackets:
 \begin{eqnarray}\label{anisotropicnoncentralstatic}
[J,K_j]=K_i\epsilon^i_j,~~[K_j,K_k]=0~~~~~~~~~~~~~~~~~~~~~~~~~~~~~~~~~~~~~\\\nonumber
[J,P_j]=P_i\epsilon^i_j,~[K_j,P_k] = M\delta_{jk},~[P_j,P_k] =0~~~~~~~~~\\\nonumber
[J,F_j]=F_i\epsilon^i_j,~[K_j,F_k]=B\delta_{jk},~[P_j,F_k]=\Lambda\delta_{jk}~~~~~\\\nonumber
[J,\Pi_j]=\Pi_i\epsilon^i_j,~[K_j,\Pi_k] = M^{\prime}\delta_{jk}, ~[P_j,\Pi_k] =B\delta_{jk}\\\nonumber
[J,H]=0,~~~~~~~[K_j,H]= \Pi_j,~~~~~~[P_j,H]=F_j~~~~~~~~\\\nonumber
\end{eqnarray}
This result is quite new and corresponds to the planar noncentrally abelian extended Static algebra.\\

Recall that the main focus for this work is the construction of noncommutative phase spaces on the above extended structures by
the coadjoint orbit
method in a two-dimensional space.\;\;As it will be established later in the text, each case treated here corresponds to a minimal
coupling in term of
symplectic structures.\;\;Thus, to be complete in this thesis, let us first of all
study the minimal couplings in detail own
to construct noncommutative phase spaces and then establish a relationship with the coadjoint orbit method (in Chapter $3$).

\chapter{NONCOMMUTATIVE PHASE SPACES BY MINIMAL COUPLINGS}
As it is well known \cite{arnold,marseden}, once the degrees of freedom of a physical system are identified, the dynamics is
determined by two
 elements: the Hamiltonian and the symplectic structure determined by the canonical Poisson brackets.\;\;It is also well known
 in symplectic mechanics \cite{souriau,  marseden, guillemin} that the interaction of a charged particle with a magnetic field can be
 described in a Hamiltonian formalism by means of a modified symplectic two-form
\begin{eqnarray}
 \sigma=\sigma_0+ e B
\end{eqnarray}
 where $\sigma_0$ is the canonical symplectic form
 \begin{eqnarray}\label{symplcanonical}
\sigma_0=dp_i\wedge dq^i,
\end{eqnarray}
 while $e$ is the particle's charge and the time-independent
 magnetic field $B$ is closed (i.e $dB=0$).\;\;With this symplectic
structure, the canonical momentum variables acquire nonvanishing Poisson brackets $\{p_i,p_j\}=-eB_{ij}$.

 Similarly, a closed two-form which interacts with the particle's dual charge $e^*$ may be
 introduced \cite{vanhecke11}.\;\;This provides
 a new modified symplectic two-form
 \begin{eqnarray}
 \sigma=\sigma_0+eB+ e^*B^*
 \end{eqnarray}
 With this, the canonical momentum and position variables
 acquire nonvanishing Poisson brackets: $\{p_i,p_j\}=-eB_{ij}$ and $\{q^i,q^j\}=-e^*B^*_{ij}$.\\

 Explicitly in noncommutative version, one generalizes the
symplectic structures by allowing further nonvanishing Poisson brackets among coordinates called
generalized Poisson brackets \cite {horvathy6, horvathy7, vanhecke11}:
\begin{eqnarray}\label{poissongeneralnc}
 \{q^i,q^j\}=G^{ij},~~~\{p_j,q^i\}=\delta^i_{j},~~~\{p_i,p_j\}=F_{ij}
\end{eqnarray}
where $\delta^{i}_j$ is a unit matrix, whereas $G^{ij}~(=-e^*B^*_{ij})$ and $F_{ij}~(=-eB_{ij})$ are functions of positions
and momenta.\;\;Note that the respective
physical dimensions of $G^{ij}$ and $F_{ij}$ are $M^{-1}T$ and $MT^{-1}$, $M$ representing a mass while $T$ represents a time.\\

 The fields $F$ and $G$ are responsible of the noncommutativity respectively of momenta and positions.\;\;This result is very general
in the sense that it takes into account the noncommutativity on the whole phase space.\;\;In this thesis, we will see in
Chapter $3$ that this can be performed by a group theoretical method (i.e the coadjoint orbit method) applied on kinematical groups
with the assumptions that $F$ is only momentum-dependent
while $G$ is only position-dependent, of course, the phase space coordinates (positions and momenta) are
determined through the group theoretical method itself.\;\;This method modifies the symplectic $2$-form, the
corresponding Hamiltonian and the equations of motion. \\

The aim of this Chapter is to introduce the construction of noncommutative spaces by using
different minimal couplings:
\begin{itemize}
 \item A coupling of momenta with magnetic potentials (the usual one)
\item A coupling of positions with dual magnetic potentials (the dual one )
\item A coupling with a magnetic field and with a dual field \cite{vanhecke11} (mixing the two couplings)
\end{itemize}
In other words, we study the planar mechanics in the following three situations:
\begin{itemize}
\item when a charged massive particle is in an electromagnetic field,
\item when a massless spring is in a dual magnetic field,
\item when a pendulum is in an electromagnetic field and in a dual one
\end{itemize}
 It is shown in the second section that under the presence of these fields
 \begin{itemize}
 \item charged massive particle acquires an oscillation state of motion with a certain frequency,
 \item massless spring acquires a mass,
\item pendulum looks like two synchronized oscillators.
 \end{itemize}
 The second and third results above are quite new and have been published in \cite{ancilla}.\;\;They
provide the formula of minimal couplings in symplectic terms as done by
Souriau \cite{souriau} for the first time (i.e in the case of minimal coupling of momentum with magnetic potential).\;\;Furthermore, as
already pointed out above, it will be seen in Chapter $3$ that each kind of minimal coupling
can be realized group theoretically by the coadjoint orbit method on a specific planar kinematical group.\\

 Note that the word \textquotedblleft dual \textquotedblright magnetic field has been also used
 in \cite{vanhecke11} and stands for
 a magnetic field which interacts with the particle's \textquotedblleft dual \textquotedblright charge $e^*$, $e$ being the
 particle's charge.\;\;Furthermore such a dual
 charge has been interpreted in \cite{plyushchary} as the anyon spin in a two-dimensional space ( $n=2$).\;\;But
 considering an arbitrary number of dimensions, no such interpretation of the dual charge is assumed.
\section{Noncommutative Phase Spaces}
In this section, we review Hamiltonian mechanics in both Darboux's coordinates and noncommutative
coordinates, the noncommutativity coming from the presence of two fields
$F_{ij}$ and $G^{ij}$.\;\;We will distinguish three cases of
noncommutative coordinates corresponding respectively to the presence of the magnetic field only,
of the dual magnetic field only and of both fields simultaneously.
 \subsection{Symplectic mechanics on Lie groups in commutative coordinates}
Symplectic geometry is the mathematical counterpart of the Hamiltonian formalism in classical mechanics.\;\;There are three main
sources of
symplectic manifold: cotangent bundles, complex algebraic manifolds and coadjoint orbits \cite{kirillovbook}.\;\;In our study, we are
interested
in this last source of symplectic manifolds as it will be seen later in this work. \\

As it has already said in the introduction of this Chapter, the main ingredients entering in the description of a classical
mechanical system are a symplectic manifold $(V,\sigma)$
(i.e a $2n-$ dimensional manifold equipped with a closed non degenerate $2-$ form $\sigma$)
called phase space and containing all the states of the system and a smooth function $H$ defined on  $V$ called the
Hamiltonian representing energy and determining the time-evolution of the system \cite{abraham}. \\

If $\sigma_{ab}$ are the matrix elements of the matrix representing the symplectic form $\sigma$ and if $\sigma^{ab}$ are solutions
 of $\sigma_{bc}\sigma^{ca}=\delta^a_b$, then a Poisson bracket of  $f,g \in C^{\infty}(V,{\mathbb R})$ is given by
\begin{eqnarray}\label{pb}
\{f,g\}=\sigma^{ab}\frac{\partial f}{\partial z^a}\frac{\partial g}{\partial z^b}
\end{eqnarray}
where $~a,b=1,2,\dots 2n$ and where we have introduced the definition:
\begin{eqnarray}
 \sigma^{ab}=\{x^a,x^b\}
\end{eqnarray}

Here and in what follows, we use summation over repeated indices.\\

The space  $C^{\infty}(V,{\mathbb R})$ endowed with the Poisson brackets given by (\ref{pb}) is an infinite Lie algebra
\cite{abraham}.\;\;If
$z^a=(p_i,q^i)$ are the canonical coordinates (Darboux's coordinates) on $V$,
the symplectic form on $V$ takes the form (\ref{symplcanonical})
 meaning that there is no coupling to a gauge field.\;\;These canonical coordinates
are defined up to canonical transformations preserving the basic form $\sigma$ called {\it symplectomorphisms or symplectic realizations}.\\

Moreover the Poisson brackets (\ref{pb}) becomes
\begin{eqnarray}\label{poissoncanonical}
\{f,g\}=\frac{\partial f}{\partial p_i}\frac{\partial
g}{\partial q^i}-\frac{\partial f}{\partial q^i}\frac{\partial
g}{\partial p_i}
\end{eqnarray}
and
\begin{eqnarray}\label{canonicalcoordinates}
\{p_k,p_i\}=0~~,~~\{p_k,q^i\}=\delta^i_k~,~\{q^k,q^i\}=0.
\end{eqnarray}
This means that momenta $p_i$ as well as the positions $q^i$ are commutative.\;\;It is also known that if $X_f$ is the
Hamiltonian vector field associated to $f$, then
$L_{X_f}(\sigma)=0$ where $L_X$ is the Lie derivative in the direction of $X$.\;\;This property
of $X_f$ suggests the following general definition of a vector field. \\

 A vector field $ X \in {\cal{F}}(V)$, the space
of vector fields defined on $V$, is said {\it locally Hamiltonian} if $i_X(\sigma)$ is closed.\;\;As $\sigma$
is closed, this definition is
equivalent to the following property: $X$ is locally Hamiltonian if and only if \\$$L_X(\sigma)=0$$\;\;In this case, the
vector field $X$ is also called {\it symplectic} because it defines an infinitesimal automorphism
of $(V,\sigma)$.\;\;In other
words, the flow generated on $V$ by a symplectic vector field consists of canonical transformations of $V$.\\

Recall that {\it all Hamiltonian vector fields are symplectic and if the  manifold $V$ is
simply connected, the converse
is also true} (for more details see \cite{kirillov1}).\\

Equivalently
$X_f(g)=\{f,g\}$ and the evolution equations under the flow $\Phi_{exp(sX_f)}$ on $V$ generated by $X_f$ are
\begin{eqnarray*}
\frac{dz^a}{ds}=X_f(z^a)
\end{eqnarray*}
which are exactly the usual Hamiltonian equations when $f$ is the energy.
\subsection{ Lie-Poisson structure}
Let us now turn to the algebraic approach and characterize a manifold $V$ by the algebra $C^{\infty} (V)$.\;\;On any
symplectic manifold we have
an additional algebraic operation on $C^{\infty} (V$).\;\;It is the so-called Poisson bracket $\{.,.\}$ which is
bilinear, skew-symmetric and
satisfies the Jacobi identity and the Leibniz rule.\;\;It is
defined in the following three equivalent ways:
\begin{eqnarray}\label{Poisson-symplectic}
 \{f_1,f_2\}= X_{f_1}(f_2)=-X_{f_2}(f_1)= \sigma(X_{f_1},X_{f_2})
\end{eqnarray}
In any canonical local coordinates we have relations (\ref{poissoncanonical}) yielding (\ref{canonicalcoordinates}).\;\;Note that
the functions $~1, q^i, p_i, ~1\leq i\leq n~$ satisfy the following commutator relations
$$\{p_i,p_j\}=\{q^i,q^j\}= \{p_i,1\}=\{q^i,1\}=0,~~~\{p_i,q^j\}=\delta^j_{i}$$
These functions constitute the basis of the ($2n+1$)-dimensional Heisenberg Lie algebra.\\

Later in the text, we will see that one can generalize the above canonical relations when the cotangent bundle of the
space is the dual of the Lie
algebra of a symmetry group assuming that the latter is simply connected.\\

Let $G$ be a Lie group whose the Lie algebra ${\cal{G}}$ is defined by the structure constants $C^k_{ij}$ satisfying (\ref{skew-sym1})
and (\ref{jacobi1}).\;\;Let $V$ be a differential manifold
whose the dimension is the same as that of $G$.\;\;In a basis $e^i$ of $V$, we associate to each $a \in V$ the matrix defined by
\begin{eqnarray}\label{kirillov01}
K_{ij}(a)=-a_kC^k_{ij}
\end{eqnarray}
\begin{pro}
 The bracket $\{.,.\}:C^{\infty}(V,\Re)\times C^{\infty}(V,\Re)\rightarrow C^{\infty}(V,\Re)$ defined by
\begin{eqnarray}\label{kirillov-poissonbracket}
\{f,g\}=K_{ij}(a)\frac{\partial f}{\partial a_i}\frac{\partial g}{\partial a_j}
\end{eqnarray}
 is a Poisson bracket.
\end{pro}
It suffices to demonstrate that the four properties of a Poissonian structure (bilinearity, skew-symmetry, Jacobi identity
and Leibniz rule)
are verified.\;\;For this, we use the
two properties of the structure constants given by relations (\ref{skew-sym1}) and (\ref{jacobi1}).\\

As this Poisson bracket is defined by the structure constants of the Lie algebra of $G$, we call it a {\it Poisson-Lie } bracket
also known as
the {\it Kirillov-Kostant-Souriau } bracket.\\

Thus, ($C^{\infty}(V,\Re),\{ ., .\}$) is a Lie algebra (or a Lie-Poisson algebra).\\

Equivalently, one can define a Poisson structure $\cal{P}$ on a manifold $V$ as a bi-vector field
$\cal{P}:\Lambda^2(T^*V)\rightarrow \Re$ so that
$\{f,g\}=\cal{P}(df,dg)$ is a Poisson bracket on $C^{\infty}(V)$.\;\;In that sense, the data $(V,P)$ define a Poisson manifold.\\

Furthermore,
the Poisson structure $\cal{P}$ can be used to identify the tangent and the cotangent spaces of manifold.\;\;Indeed, for
any one-form $\omega$,
$\cal{P}: T^*V\rightarrow TV,\\ \omega \mapsto \cal{P}(\omega,.)$  since $\cal{P}(\omega,.) \in TV$.\;\;Unfortunately no
assumption has been made on the
non-degeneracy of $\cal{P}$
and it cannot always be assumed invertible but when the manifold is symplectic the degeneracy for $\cal{P}$ is
guaranteed.\;\;Thus, every
symplectic manifold is also Poisson since the symplectic $2$-form defines a Poisson bracket via the relation
(\ref{Poisson-symplectic}).\\

 In this thesis, the definition of Poisson structure is given
in terms of a $2$-form $\sigma$ rather than a bivector $\cal{P}$.\\

Also, it has been shown that Poisson manifolds can be partitioned into symplectic leaves: Poisson sub-manifolds
which are equipped
with a symplectic structure.\;\;In fact, the symplectic leaves of a Poisson manifold $V$ are defined via the Hamiltonian
vector field since their
integral curves stay on the symplectic leaves (see \cite{marseden} for more details ).\\

When we replace $V$ by the dual ${\cal{G}}^*$ of the Lie algebra ${\cal{G}}$, then the $K_{ij}(a)$ in relation (\ref{kirillov01})
are elements of the Kirillov form \cite{levyleblond}.\;\;It defines a Poisson bracket on $C^{\infty}({\cal{G}}^*,\Re)$ of
the form (\ref{kirillov-poissonbracket}).\;\; As
the Kirillov form is generally degenerate, ${\cal{G}}^*$ is a presymplectic manifold.\;\;It is regular on the
coadjoint orbits.\;\;Thus, the
dual of a Lie algebra has a natural Poisson structure whose symplectic leaves are the coadjoint orbits.\;\;Depending
on the kinematical group, these orbits may provide noncommutative phase spaces. \\

 In this thesis, we will generalize this result by using the Lie algebra cohomology with central and noncentral abelian extensions of
kinematical algebras as detailed in Chapter $1$.\\

 Let us introduce now noncommutative coordinates by coupling the momentum $p_i$ with a
magnetic potential $A_i$, the position $q^i$ with a dual magnetic
potential $A^{*i}$ and by mixing the two couplings.
\subsection{Noncommutative coordinates}
Let us consider the change of coordinates $(p_i,q^i)\rightarrow (\pi_i,x^i)$ with
\begin{eqnarray}\label{newcoordinates}
\pi_i=p_i-\frac{1}{2}F_{ik}q^k~~,~~x^i=q^i+\frac{1}{2}p_kG^{ki}
\end{eqnarray}
The matrix form of (\ref{newcoordinates}) is
\begin{eqnarray*}
\left(\begin{array}{c}
\pi_i\\x^i
\end{array}
\right)=\left(
\begin{array}{cc}
\delta^k_i&-\frac{1}{2}F_{ik}\\\frac{1}{2}G^{ik}&\delta^i_k
\end{array}
\right)\left(\begin{array}{c}
p_k\\q^k
\end{array}
\right)
\end{eqnarray*}
As\\~\\
$\left(
\begin{array}{cc}
\delta^k_i&-\frac{1}{2}F_{ik}\\\frac{1}{2}G^{ik}&\delta^i_k
\end{array}
\right)$
\begin{eqnarray*}
=\left(
\begin{array}{cc}
\delta^j_i&-\frac{1}{2}F_{ij}\\0^{ij}&\delta^i_j
\end{array}
\right)\left(
\begin{array}{cc}
\delta^s_j-\frac{1}{4}F_{jm}G^{ms}&0_{js}\\0^{js}&\delta^j_s
\end{array}
\right)\left(
\begin{array}{cc}
\delta^k_s&0^{sk}\\-\frac{1}{2}G^{sk}&\delta^s_k
\end{array}
\right)
\end{eqnarray*}
the transformation (\ref{newcoordinates}) is a change of coordinates if $det(\delta^s_j-\frac{1}{4}F_{jm}G^{ms})\neq 0 $.\;\;It follows
 that (\cite{wei} equation (5))
\begin{eqnarray}\label{newpoisson}
\{\pi_i,\pi_k\}=F_{ik}~~,~~\{\pi_i,x^k\}=\delta^{k}_i~~,~~\{x^i,x^k\}=G^{ik}
\end{eqnarray}
i.e the new momenta as well as the new configuration coordinates are noncommutative.\\
By the Jacobi identity, $F_{ij}$ and $G^{ij}$ satisfy the following conditions
\begin{eqnarray*}\label{magneticfield}
\frac{\partial F_{jk}}{\partial x^i}+\frac{\partial F_{ki}}{\partial x^j}+\frac{\partial F_{ij}}{\partial x^k}=0~~,
~~\frac{\partial F_{ij}}{\partial \pi_k}=0
\end{eqnarray*}
\begin{eqnarray*}\label{exoticfield}
\frac{\partial G^{jk}}{\partial \pi_i}+\frac{\partial G^{ki}}{\partial \pi_j}+\frac{\partial G^{ij}}{\partial \pi_k}=0~~,
~~\frac{\partial G^{ij}}{\partial x^k}=0
\end{eqnarray*}
Thus the Jacobi identity implies that the $F_{ij}$'s depend only on positions, that the $G^{ij}$'s depend only on momenta and
that the two $2-$ forms $\sigma_1=F_{ij}(x)dx^i\wedge dx^j$ and\\
$\sigma_2=G^{ij}(\pi)d\pi_i\wedge d\pi_j$ are closed.\\
Let the Poisson brackets of two functions $f,g $ in the new coordinates be given by
\begin{eqnarray*}
\{f,g\}_{new}=\frac{\partial f}{\partial \pi_i}\frac{\partial
g}{\partial x^i}-\frac{\partial f}{\partial x^i}\frac{\partial
g}{\partial \pi_i}=Y_f(g)
\end{eqnarray*}
It follows that
\begin{eqnarray*}
Y_H=X_H+G^{ij}\frac{\partial H}{\partial q^i}\frac{\partial }{\partial
q^j}+F_{ij}\frac{\partial H}{\partial p_i}\frac{\partial }{\partial
p_j}
\end{eqnarray*}
The derivative of any function $f$ with respect to time $t$, in terms of $F$ and $G$, is then given by
\begin{eqnarray}\label{df}
\frac{df}{dt}=X_H(f)+G^{ij}\frac{\partial H}{\partial q^i}\frac{\partial f }{\partial
q^j}+F_{ij}\frac{\partial H}{\partial p_i}\frac{\partial f }{\partial
p_j}
\end{eqnarray}
and the equations of motion are then given by
\begin{eqnarray*}
\frac{dq^k}{dt}=\frac{\partial H}{\partial p_k}+G^{ki}\frac{\partial
H}{\partial q^i}~~,~~\frac{dp_k}{dt}=-\frac{\partial H}{\partial
q^k}+F_{ik}\frac{\partial H}{\partial p_i}
\end{eqnarray*}
 If for example
\begin{eqnarray*}
H=\frac{\delta^{ij}p_ip_j}{2m}+V_0
\end{eqnarray*}
is the Hamiltonian with the potential energy $V_0$
depending only on the configuration coordinates $q^i$, the equations of motion are then
\begin{eqnarray}\label{eq}
\frac{dq^k}{dt}=\frac{p^k}{m}+G^{ki}\frac{\partial V_0}{\partial
q^i}~~,~~\frac{dp_k}{dt}=-\frac{\partial V_0}{\partial
q^k}+F_{ik}\frac{p^i}{m}
\end{eqnarray}
They are equivalent to the modified second Newton's law \cite{romero, wei}
\begin{eqnarray*}
m\frac{d^2q^k}{dt^2}=-\frac{\partial V_0}{\partial
q^k}+F_{ik}\frac{p^i}{m}+m G^{ki}\frac{d}{dt}(\frac{\partial
V_0}{\partial q^i})
\end{eqnarray*}
In the absence of the potential $V_0$, the equations (\ref{eq}) become
\begin{eqnarray}\label{modifiedmotionequations2}
\frac{dq^k}{dt}=\frac{p^k}{m}~~,~~\frac{dp_k}{dt}=F_{ik}\frac{p^i}{m}
\end{eqnarray}
or equivalently
\begin{eqnarray}\label{modifiednewtonequations2}
m\frac{d^2q^k}{dt^2}=F_{ik}\frac{p^i}{m}
\end{eqnarray}
This means that the noncommutativity of the momenta (first equation of (\ref{newpoisson})) implies that the particle is accelerated
and is not free even if the potential $V_0$ vanishes identically.\;\;The particle will
be free if the momenta commute even if positions do not commute.
\section{Couplings in Planar Mechanics}
In this section we construct explicitly the noncommutative phase spaces by
introducing couplings.\;\;We start by the usual coupling of momentum
with a magnetic potential.\;\;Then, we introduce a new kind of coupling: of position with a dual magnetic potential and we
finish by a mixing model.
\subsection{Coupling of momentum with a magnetic Field}
\subsubsection{ i) commutative coordinates}
Consider a four-dimensional phase space (a cotangent space to a plane) equipped with the Darboux's
coordinates $(p_i,q^i$), $i=1,2$.\;\;This means that the momenta as well as the
positions commute.\;\;Consider also an electron with a mass $m$ and an
electric charge $e$, moving on a plane with the electromagnetic potential $A_{\mu}=(A_i=-\frac{1}{2}B\epsilon_{ik}q^k,~\phi=E_iq^i)$
 where a symmetric gauge has been chosen, $\vec{E}$ being an electric field while $\vec{B}$ is a magnetic field perpendicular to
the plane.\;\;It is known that the dynamics of the particle is governed by the
Hamiltonian
\begin{eqnarray}\label{hamiltonianelectron}
H=\frac{\vec{p}^{~2}}{2m}-e\phi
\end{eqnarray}
where the kinetic energy is minimally coupled to the electric
potential.\;\;It is also known that if we adapt the symmetric gauge,
the magnetic potential is given by
\begin{eqnarray}
A_i=-\frac{1}{2}B\epsilon_{ik}q^k
\end{eqnarray}
where $B$ is the magnetic field strength, source of the potential $A$ while
the electric potential is given by
\begin{eqnarray}
\phi=E_iq^i
\end{eqnarray}
the electric field $\vec{E}$ being the source of the potential $\phi$.\;\;The Hamiltonian is then
\begin{eqnarray}
H=\frac{\vec{p}^{~2}}{2m}-e\vec{E}.\vec{q}
\end{eqnarray}
and the equations of motion are
\begin{eqnarray}
\frac{ dq^{~i}}{dt}=\frac{p^i}{m}~~,~~\frac{ dp_i}{dt}=eE_i
\end{eqnarray}
or equivalently
\begin{eqnarray}
m\frac{d^2\vec{q}}{dt^2}=e\vec{E}
\end{eqnarray}
where the right hand side is the electric force.\;\;In this case we have commutativity of momenta
as well as positions.
\subsubsection{ii) Noncommutative Coordinates}
Let us construct noncommutative phase spaces through the minimal
coupling process.\;\;From the classical electromagnetism, it is known that the coupling of the momentum with the magnetic
potential is given by the relations
\begin{eqnarray}\label{coupling1}
 \pi_i=p_i-eA_i~~,~~x^i=q^i
\end{eqnarray}
or equivalently
\begin{eqnarray}\label{magneticoupling1}
 \pi_i=p_i+\frac{eB}{2} \epsilon_{ik}q^k~~,~~q^i=x^i
\end{eqnarray}
The coordinates $\pi_i$ and $x^i$ satisfy
\begin{eqnarray*}
\{x^i,x^k\}=0~,~\{\pi_i,x^k\}=\delta^k_i~,~\{\pi_i,\pi_k\}=-eB\epsilon_{ik}
\end{eqnarray*}
Then in the presence of an electromagnetic field, the momenta do not commute while the positions commute.\;\;Using the
equation (\ref{newpoisson}), we
have
\begin{eqnarray}\label{magnetics}
F_{ij}=-eB\epsilon_{ij}~~,~~G^{ij}=0^{ij}
\end{eqnarray}
Use of (\ref{hamiltonianelectron}) and (\ref{magnetics}) into (\ref{df}) gives rise the equations of motion
\begin{eqnarray}\label{newton1}
m\frac{d^2\vec{q}}{dt^2}=e(\vec{E}+ \frac{\vec{p}}{m}\times\vec{B}),
\end{eqnarray}
where the right hand side represents the Lorentz force.\;\;Moreover, in the noncommutative coordinates, the
Hamiltonian (\ref{hamiltonianelectron}) becomes
\begin{eqnarray}
H=\frac{\vec{\pi}^{~2}}{2m}-e\vec{E}.\vec{x}+\frac{m\omega^2\vec{x}^{~2}}{2}+\vec{\omega}.\vec{L}
\end{eqnarray}
where $\omega$ is the {\it cyclotron frequency},
$\vec{L}=\vec{x}\times \vec{p}~$ is the orbital angular momentum and
\begin{eqnarray*}
\vec{\omega}=\frac{eB}{2m}\vec{n}
\end{eqnarray*}
with $\vec{n}$ the unit vector in the direction perpendicular to
the plane.\;\;In the presence of a magnetic field, the massive
charged particle has became an oscillator with frequency $\omega$
given above and the equations of motion are
\begin{eqnarray}
\frac{d\vec{x}}{dt}=\frac{\vec{\pi}}{m}+\vec{\omega} \times
\vec{x}~~,~~\frac{d\vec{\pi}}{dt}=e\vec{E}+\vec{\omega} \times
\vec{\pi}-m\omega^2\vec{x}
\end{eqnarray}
or equivalently
\begin{eqnarray}\label{newton2}
m\frac{d^2{~\vec{x}}}{dt^2}=e(\vec{E}+ \frac{\vec{\pi}}{m}\times\vec{B}),
\end{eqnarray}
where we recognize again the Lorentz force
$\vec{f}_{Lorentz}=e\vec{E}+e\frac{\vec{\pi}}{m} \times \vec{B}$ and where $\vec{B}=B\vec{n}$.\;\;Note that
the relations (\ref{newton1}) and (\ref{newton2}) have the
same form.\;\;The Newton's equations are then covariant under the
coupling (\ref{magneticoupling1}).\\

In the next two subsections, we present quite new theories associated to an unusual coupling of
position with a dual magnetic \cite{vanhecke11} field.
\subsection{Coupling of position with a dual field}
\subsubsection{i) Commutative Coordinates}
 Consider a massless spring with $k$ as a Hooke's constant and a dual charge $e^*$ in a dual magnetic field $B^*$.\;\;Suppose that
the dynamics of the spring is governed by the Hamiltonian
\begin{eqnarray}\label{hamiltonian2}
H=k\frac{\vec{q}^{~2}}{2}-e^*\vec{p}.\vec{E^*}
\end{eqnarray}
where we have used the symmetric gauge
\begin{eqnarray*}
A^{*i}=-\frac{B^*}{2}p_k\epsilon^{ki}~~,~~\phi^*=p_iE^{*i}
\end{eqnarray*}
$B^*$ and $\vec{E}^*$ being respectively the sources of the dual magnetic
potential $A^*$ and $\phi^*$.\;\;The equations of motion are
\begin{eqnarray}\label{123}
\frac{d\vec{q}}{dt}=-e^*\vec{E^*}~~,~~\frac{d\vec{p}}{dt}=-k\vec{q}
\end{eqnarray}
where $e^*\vec{E}^*$ is a velocity.

It is well known that the first and the second derivative of position are respectively called velocity and acceleration but it is less
known that the third derivative is technically called jerk or jolt  \cite{gibbs, nzo2}.\;\;As its name suggests, jerk is applied
when evaluating the destructive effect of motion on a mechanism or discomfort caused to the passenger in a vehicle.\;\;It is
also known that momentum
equals mass times velocity and force equals mass times acceleration.\;\;Similarly, mass times jerk equals a quantity called yank,
$\vec{y}$.\;\;Equivalently, yank can be defined as
the second derivative of the momentum with respect to the time variable $t$
by the relations:
$$ \vec{y}=m \frac{d \vec{a}}{dt}= \frac{d}{dt}(\frac{d (m\vec{v})}{dt})= \frac{d^2 \vec{p}}{dt^2},~~ m =const$$ or the rate
of change of a force with time.\\

 The relations (\ref{123}) can take the form
\begin{eqnarray}\label{dualequation}
C\frac{d^2\vec{p}}{dt^2}=e^*\vec{E^*}
\end{eqnarray}
where $C=\frac{1}{\kappa}$ is the spring compliance while
$\frac{d^2\vec{p}}{dt^2}$ is a yank (as defined above).\;\;Furthermore, (\ref{dualequation}) is
 the analogue (dual) of the Newton's second equation for this case.
Note that the spring compliance plays, for the dual equations, a similar role as the mass for
the usual equations.
\subsubsection{ii) Noncommutative coordinates}
Let us consider the coupling of the position with the dual potential $A^{*i}$ depending on the momenta $p_i$:
\begin{eqnarray}\label{coupling2}
  \pi_i=p_i~~,~~q^i=x^i+\frac{e^*B^*}{2}p_k\epsilon^{ki}
\end{eqnarray}
In this case, the Poisson brackets become
\begin{eqnarray}
\{x^i,x^j\}=-e^*B^*\epsilon^{ij}~~,~~\{p_k,x^i\}=\delta^i_k~~,~~\{p_k,p_i\}=0
\end{eqnarray}
Therefore, in the presence of the dual field, positions do not commute while the momenta commute.\;\;It follows that
\begin{eqnarray}\label{fg}
F_{ij}=0_{ij}~~,~~G^{ij}=-e^*B^*\epsilon^{ij}
\end{eqnarray}
Use (\ref{hamiltonian2}) and (\ref{fg}) into (\ref{df}) gives rise to the Newton's analogue equation
\begin{eqnarray}\label{newton3}
C\frac{d^2\vec{p}}{dt^2}=e^*(\vec{E^*}+\frac{1}{C}{\vec{q}}\times \vec{B^*})
\end{eqnarray}
the right hand side being a velocity.\;\;In noncommutative coordinates, the Hamiltonian is
\begin{eqnarray}
H=\frac{\vec{x}^{~2}}{2C}-e^*\vec{\pi}.\vec{E^*}+\frac{\vec{\pi}^{~2}}{2m_s}-\vec{\omega}.\vec{L}
\end{eqnarray}
where the spring mass $m_s$ is defined by
\begin{eqnarray}
\frac{1}{m_s}=\frac{e^{*2}B^{*2}}{4C}
\end{eqnarray}
while the vector $\vec{\omega}$ is given by
\begin{eqnarray}
\vec{\omega}=\frac{e^*B^*}{2C}\vec{n}
\end{eqnarray}
The Hooke's compliance $C$ can be written as
\begin{eqnarray}\label{hookeconstant}
C=\frac{1}{m_s\omega^{2}}
\end{eqnarray}
In the presence of the dual field, the spring then acquires a mass
$m_s$ and the equation of motion is given by
\begin{eqnarray}\label{newton4}
C\frac{d^2\vec{\pi}}{dt^2}=e^*(\vec{E^*}+\frac{\vec{x}}{C}\times\vec{B^*})
\end{eqnarray}
The vector $\vec{f^*}=e^*(\vec{E^*}+\frac{\vec{x}}{C}\times\vec{B^*})$ can
be considered as a dual Lorentz force with the dimension of velocity.\;\;It represents for the spring
what the Lorentz force represents for a charged  particle.\;\;Here also the
coupling (\ref{coupling2}) preserves the covariance of the Newton's analogue equations.\;\;Comparing (\ref{newton3}) and (\ref{newton4})
and using (\ref{hookeconstant}) into (\ref{newton4}) we can conclude that
$e^*\omega^2(\vec{E^*}+\frac{\vec{x}}{C}\times\vec{B^*})$  is a kind of jerk \cite{nzo2}.
\subsection{Coupling with a magnetic field and with a dual magnetic field}
Now consider the case of a massive pendulum with mass $m$ and
Hooke's compliance $C$ under the action of an electromagnetic potential $A_{\mu}=(A_i,\phi)$ and a dual electromagnetic
potential $A^{*}_{\mu}=(A^{*}_i,\phi^*)$ with $A_i=-\frac{1}{2}B\epsilon_{ik}q^k$, $\phi=E_iq^i$,
$A^*_i=-\frac{1}{2}B^* p_k\epsilon_{ki}$ and  $\phi^*=p_iE^*_i$, where $\vec{E}$ is an electric field
and $\vec{E}^*$ its dual electric field while $\vec{B}$ is a magnetic field and $\vec{B}^*$ its corresponding dual magnetic field.\\

 The corresponding motion is governed by the Hamiltonian
 \begin{eqnarray}
H=\frac{\vec{p}^{~2}}{2m}+\frac{\vec{q}^{~2}}{2C}-e\phi-e^*\phi^*
\end{eqnarray}
where the extra term $e^*\phi^*$ is due to the presence of the dual magnetic \cite{vanhecke11} field in the model.
Let
\begin{eqnarray}\label{coupling5}
x^i=q^i+\frac{e^*B^*}{2}p_k\epsilon^{ki}~~,~~\pi_i=p_i+\frac{eB}{2}\epsilon_{ik}q^k
\end{eqnarray}
be the minimal coupling in the symmetric gauge;
that is \[G^{ij} = -e^*B^*\epsilon^{ij} ~~,~~ F_{ij} = -eB \epsilon_{ij}\]
We assume that the cyclotron frequency acquired by the massive charged particle is equal to the
frequency of the massless spring:
\begin{eqnarray*}
\frac{eB}{2}=m\omega ~,~ \frac{e^*B^*}{2}=\frac{1}{m_s\omega}
\end{eqnarray*}
where $m_s$ is the acquired mass by the spring while
\begin{eqnarray*}
\mu=\frac{m.m_s}{m+m_s}
\end{eqnarray*}
is the reduced mass of the two synchronized massive oscillators.\;\;It follows that
\begin{eqnarray}\label{noncommutativecoordinates}
\{x^i,x^j\}=-e^*B^*\epsilon^{ij}~~,~~\{\pi_k,x^i\}=\gamma\delta^i_k~~,~~\{\pi_k,\pi_i\}=-eB\epsilon_{ki}
\end{eqnarray}
with $\gamma=1+\frac{m}{m_s}$ and $m=\mu\gamma$.\\

In the presence of the two kind of fields, the positions as well as the momenta do not
commute.\;\;The Hamiltonian in noncommutative coordinates is written as
\begin{eqnarray}
H=\frac{\vec{\pi}^{~2}}{2\mu}+\frac{M
\omega^2\vec{x}^{~2}}{2}-e\phi -e^* \phi^*
\end{eqnarray}
where $M = m + m_s$ is the total mass, $ \phi=\vec{E}.\vec{x}+\vec{n}.\vec{E}\times
\frac{\vec{\pi}}{m_s\omega} $ \\ and $\phi^*=\vec{\pi}.\vec{E^*}+\vec{n}.m\omega
\vec{x}\times \vec{E^*}$.\;\;Note that $M=m_s\gamma$.

The equations of motion in noncommutative coordinates are then
\begin{eqnarray*}
\frac{d\vec{x}}{dt}=\frac{\vec{\pi}}{\mu}+e^*[\gamma\vec{E^*}+\vec{B^*}\times \frac{
\vec{x}}{C}-e\vec{B^*}\times \vec{E}]\\
\frac{d\vec{\pi}}{dt}=-\frac{\gamma\vec{x}}{C}+e[\gamma\vec{E}+\vec{B}\times
\frac{\vec{\pi}}{m}-e^*\vec{B}\times \vec{E^*}]
\end{eqnarray*}
where the Hooke's compliance $C$ is given by (\ref{hookeconstant}).\\

If the mass $m$ of the particle is much smaller than the mass $m_s$
acquired by the spring, i.e. $m<< m_s$, then $\gamma$ becomes $1$ and $\mu=m<<M=m_s$.\;\;In that limit, the brackets
(\ref{noncommutativecoordinates}) become
\begin{eqnarray}\label{p2}
\{x^i,x^j\}=e^*B^*\epsilon^{ij}~~,~~\{\pi_k,x^i\}=\delta^i_k~~,~~\{\pi_k,\pi_i\}=eB\epsilon_{ki}
\end{eqnarray}
and the Hamiltonian becomes
\begin{eqnarray*}
H=\frac{\vec{\pi}^{~2}}{2m}+\frac{\vec{x}^{~2}}{2C}-e[\vec{E}.\vec{x}+\vec{n}.\vec{E}\times
\omega C\vec{\pi}]-e^*[\vec{\pi}.\vec{E^*}+\vec{n}.m\omega
\vec{x}\times \vec{E^*}]
\end{eqnarray*}
and the equations of motion are given by
\begin{eqnarray*}
\frac{d\vec{x}}{dt}=\frac{\vec{\pi}}{m}+e^*[\vec{E^*}+\vec{B^*}\times\frac{
\vec{x}}{C}-e\vec{B^*}\times \vec{E}]\\
\frac{d\vec{\pi}}{dt}=-\frac{\vec{x}}{C}+e[\vec{E}+\vec{B}\times
\frac{\vec{\pi}}{m}-e^*\vec{B}\times \vec{E^*}]
\end{eqnarray*}
The velocity $ee^*\vec{B^*}\times \vec{E}$ and the force
$ee^*\vec{E^*}\times \vec{B}$ result from the coexistence of the two
fields.\;\;Thus, in the presence of the two kinds of fields, the positions do not
commute as well as the momenta (see equations (\ref{p2})). \\

In the following Chapter, we determine the maximal coadjoint orbits of the extended structures
determined in the first Chapter.\;\;These
orbits are physically interpreted as noncommutative
phase spaces of accelerated particles moving in the respective kinematical spacetimes.\;\;The noncommutativity is due
to the presence of
naturally introduced magnetic fields defining minimal couplings as already been reported.

\chapter{NONCOMMUTATIVE PHASE SPACES CONSTRUCTED GROUP THEORETICALLY}
One of the aims of this thesis is the study of classical dynamical systems associated with the planar kinematical
groups.\;\;This constitutes the main idea of this Chapter: to construct noncommutative phase spaces by coadjoint
orbit method and classify all the possible noncommutative symplectic structures in a two-dimensional
space when the symmetry groups are kinematical groups.\\

Historically, the {\it orbit method} was proposed in \cite{kirillov1} for the description of the unitary dual (i.e the set of
equivalence classes
of unitary irreducible representations) of nilpotent Lie groups.\;\;It turned out that the method not only solves this problem but
also gives simple
and visual solutions to all other principal questions in representation theory: topological structure of the unitary dual, the explicit
description
of the restriction and induction functors, the formulas for generalized and infinitesimal characters, the computation of the Plancherel
measure, \dots  \cite{kirillovbook}.\;\;The theory was extended later to the case of solvable groups in \cite{kostant}
and has also paved the way for the geometric quantization which constitutes its natural generalization.\\

Furthermore, while the idea that symmetry would determine the system which carries it goes back to Wigner \cite{wigner}, it has
been Souriau
\cite{souriau} who put it into precise form. As already mentioned in the introduction, his important theorem says that when a
symmetry group
$G$ acts transitively on the phase space then the latter is a coadjoint orbit of $G$ endowed with its canonical symplectic structure
(Kirillov-Kostant-Souriau). \\

The notion of {\it coadjoint orbits} (orbits of the action of a group on the dual of its Lie algebra) is the main
ingredient of the orbit method.\;\;Moreover it is the most important mathematical object that has
been brought into consideration in connection with the orbit method hence the title: {\it coadjoint orbit method}.\;\;Thus, the Souriau's
method relates the coadjoint orbit of a group $G$ to the phase space (where quantization yields the
irreducible unitary representation of that group \cite{kirillov}).  \\

In this Chapter, the construction of the maximal (i.e invariants are all nonvanishing) coadjoint orbits of centrally and noncentrally
extended kinematical groups and their modified symplectic structures is presented in detail own to describe and classify
noncommutative phase spaces.\\

More specifically, we realize in this Chapter the Poisson brackets (\ref{poissongeneralnc}) on the maximal
coadjoint orbits of a noncentrally abelian extended Aristotle group (i.e a subgroup of Galilei group),
of the  centrally extended planar nonrelativistic anisotropic kinematical groups (Newton-Hooke groups, Galilei  and Para-
Galilei groups, Static and Carroll groups) and of the
noncentrally abelian extended planar absolute time kinematical groups as classified in \cite {mcrae}.\;\;The main characteristic
features are the following:
\begin{itemize}
 \item there exists generators (namely momenta and boosts) which yield the basic canonical variables through the coadjoint
 orbit method,
 \item the Hamiltonian belongs to the Lie algebra itself (or to its central or noncentral abelian extension) and acts linearly on
 the canonical variables,
 \item the equations of motion are linear to the latter variables.
\end{itemize}
But before go on to determine the maximal coadjoint orbits of planar kinematical groups (or their corresponding group extensions) or
for a more pedagogical discussion of the applications
to construct noncommutative phase spaces, let us give here some formal definitions, keeping in mind that we can apply them to the
noncommutative algebraic structures as well as their physical interpretations we need in this thesis.\\

This Chapter is organized as follows.\\

In the next section, we give some formal definitions namely the coadjoint orbits and symplectic realization.\;\;However, it should
not be assumed that the following analysis is complete.\\

In section two, we study the classical dynamical systems associated with the Aristotle group.\;\;We obtain phase spaces which
do not commute in momentum sector due to the presence of a naturally introduced magnetic
field, i.e the obtained cases correspond to the minimal coupling of the momentum with a magnetic potential.\\

In section three and four, we construct noncommutative phase spaces as coadjoint orbits of centrally extended planar anisotropic
kinematical groups (oscillating and expanding Newton-Hooke, Galilei, Para-Galilei, Carroll and Static Lie groups) and
of noncentrally abelian extended absolute time planar kinematical groups (the same groups enumerated above when
rotation parameters are taken into account except the Carroll group) respectively.\;\;Through these constructions
the coordinates of the phase spaces do not commute due to the presence of naturally introduced fields
giving rise to minimal couplings.\;\;Finally in the section five, we conclude and
classify the obtained nonrelativistic models which are noncommutative phase spaces. \\

Some of the results of this Chapter have been published in \cite{ancille1, ancilla, ancille} and others in \cite{ancilla1}
(submitted to Journal of Mathematical Physics, August $2013$).

\section{Coadjoint orbit and symplectic realization methods}
\subsection{coadjoint orbit method}
Let $G$ be a Lie group and ${\cal{G}}$ its Lie algebra.  Let $Ad:G\rightarrow Aut({\cal{G}})$ be the adjoint representation
of $G$ on its Lie algebra ${\cal{G}}$ such that the automorphism $Ad_g$ associated to $g \in G$ is defined by
\begin{eqnarray*}
 Ad_g(X)=g Xg^{-1},~ X\in {\cal{G}}
\end{eqnarray*}
If ${\cal{G}^*}$ is the dual of ${\cal{G}}$, it is well known that the coadjoint action of $G$ on ${\cal{G}}^*$\\
$Ad^*:{\cal{G}}^*\times {\cal{G}} \rightarrow \Re$
is defined as the dual of the map $Ad$:
$$ \langle Ad_{g}^*\alpha,X\rangle=\langle \alpha,Ad_{g^{-1}}X\rangle, \alpha \in \cal{G}^*
$$
where we have denoted the pairing between $\cal{G}$ and $\cal{G}^*$ as $\langle.,.\rangle$.\;\;Then for a given element
$\alpha \in \cal{G}^*$, the coadjoint orbits through $\alpha$ are obtained as the images of the map:
$$g\rightarrow Ad_{g}^*\alpha$$
where the usual notation is
$$ \cal{O}_\alpha=\{Ad_g^{*}(\alpha), \alpha \in {\cal{G}}^*,g\in G\}$$
Consider now the derivative (push-forward) of $Ad^*$.\;\;This can be defined as the dual of the derivative of $Ad$.\;\;It is normally
denoted $ad$ and it defines a map :
$$ (Ad_g)_*\equiv ad_{X}: \cal{G}\rightarrow \cal{G}$$ such that $$ad_{X}Y=[X,Y]$$
where by $X$ we have denoting the Lie algebra element corresponding to $g\in G$.\;\;Similarly, we define
$ad^*$ as the derivative of the map $Ad^*$ i.e:
$$(Ad_{g}^*)_{*}=ad^{*}_X$$
 such that:
\begin{eqnarray*}
  \langle ad^*_X(\alpha),Y)\rangle=\langle\alpha,[X,Y]\rangle
\end{eqnarray*}
If $\alpha=\alpha_i\epsilon^i \in {\cal{G}}^*,~~X=e_iX^i,~~Y=e_iY^i \in \cal{G}~~$
then
\begin{eqnarray*}
  \langle ad^*_X(\alpha),Y)\rangle=K_{ij}(\alpha)X^iY^j
\end{eqnarray*}
where $ K_{ij}(\alpha)$ is of the form (\ref{kirillov01}), the Kirillov $2$-form \cite{levyleblond} on $\cal{G}^*$.\;\;The representation $\rho:\cal{G}\rightarrow {\cal{F}}({\cal{G}^*})$ of ${\cal{G}}$
on the space of vector fields on ${\cal{G}}^*$
defined by
 \begin{eqnarray*}
 \rho(X_i)=K_{ij}(\alpha)\frac{\partial }{\partial \alpha_j}
\end{eqnarray*}
is a Lie algebra homomorphism such that
$$ Ker(K(\alpha))=\{f \in C^{\infty}({\cal{G}}^*,\Re): \rho(X)f=0, X\in {\cal{G}}\}.$$
This means that $ Ker(K(\alpha))$ is the set of all invariants $f$ of $ {\cal{G}}$ in ${\cal{G}}^*$ satisfying the following relation:
\begin{eqnarray}\label{kirillovsystem}
  K_{ij}(\alpha)\frac{\partial f}{\partial \alpha_j}=0
 \end{eqnarray}
The quotient space $${\cal{O}}_{\alpha}=
{\cal{G}}^*/Ker(K(\alpha)),$$ called the coadjoint orbit of $G$ in ${\cal{G}}^*$, is a symplectic manifold \cite{giachetti}
whose symplectic
form $\sigma^{ij}$ is obtained from
$$ \Omega_{ij}\sigma^{jk}=\delta^k_{i}$$
where $\Omega_{ij}=K_{ij}|_{ {\cal{O}}_{\alpha}}$, i.e the restriction of the Kirillov form to the orbit.\;\;Explicitly, the
two-symplectic form is given by the following relation
 \begin{eqnarray}\label{symplectictwo-form}
 \sigma=(\Omega^{-1})^{ab}dx_b\wedge dx_a
 \end{eqnarray}
which takes the form $\sigma=dp_i\wedge dq^i$ in the canonical coordinates.\;\;The fact that $\alpha$ is
arbitrary means that one can choose the invariants
under the action of $G$ to label the orbit
(as we have silently done above).\;\;The invariants which labeled the orbit are generally the cohomology classes
as we will see it later in the Chapter $3$.\\

If $(p_i,q^i)$ are denoted collectively by $x_a$, the Poisson bracket implied by the Kirillov symplectic structure
\begin{eqnarray}\label{poissonbrackets}
\{H,f\}=-\Omega_{ab}\frac{\partial H}{\partial
x_a}\frac{\partial f}{\partial x_b}
\end{eqnarray}
leads to relations (\ref{canonicalcoordinates})
where $ p_i$, $q^i$ represent the generalized canonical coordinates and momenta of the system.\;\;Relations
(\ref{canonicalcoordinates}) mean that the momenta commute within themselves as well as the
positions.\\

Interesting consequences arise by considering central and noncentral abelian extensions of Lie algebras and this provides
a more general symplectic two-form (\ref{symplectictwo-form}) whose extended Poisson brackets of generalized
coordinates are defined by
\begin{eqnarray*}\label{poissonext}
 \{x_{a},x_{b}\}=\Theta_{ab}
\end{eqnarray*}
 where
\begin{eqnarray*}
\Theta=\left(\begin{array}{cccc}
0&G&1&0\\-G&0&0&1\\-1&0&0&F\\0&-1&-F&0
\end{array}
\right)
\end{eqnarray*}
is the inverse of the matrix of the symplectic form
\begin{eqnarray*}
\Omega=\frac{1}{1-GF}\left(\begin{array}{cccc}
0&F&-1&0\\-F&0&0&-1\\1&0&0&G\\0&1&-G&0
\end{array}
\right)
\end{eqnarray*}
The fields $F$ and $G$ are constant because they are coming from central or noncentral abelian extensions of Lie algebras.\;\;But cases
where they are not constant have been considered \cite{acatrinei}.\;\;Moreover the respective
physical dimensions of $G^{ij}$ and $F_{ij}$ are $M^{-1}T$ and $MT^{-1}$, $M$ representing a mass while $T$ represents a time.\\

The noncommutative phase space is then
 defined as a space on which variables satisfy the commutation relations (\ref{poissongeneralnc}).\;\;The equations of
motion corresponding to the above symplectic structure are given by:
\begin{eqnarray*}
 \frac{d x^i}{d t}=\{H,x^i\}=\Theta^{ij}\frac{\partial H}{\partial x^j}
\end{eqnarray*}
more explicitly:
\begin{eqnarray}\nonumber\label{motionequations0}
 \frac{dq^i}{dt}=\frac{\partial H}{\partial p_i}+G\epsilon^{ij}\frac{\partial H}{\partial x^j}\\
\frac{dp_i}{dt}=-\frac{\partial H}{\partial q^i}+F\epsilon_{ij}\frac{\partial H}{\partial p_j}
\end{eqnarray}
If $G=F=0$, then (\ref{motionequations0}) are the usual Hamiltonian equations.
\subsection{Symplectic realizations}
Let $(V,\sigma)$ be a symplectic manifold and let $G$ be a Lie group.\;\;The application\\$D:G\rightarrow Aut(V)$ is a
 symplectomorphism if
\begin{eqnarray}
 D_{g}^{*}\sigma =\sigma
\end{eqnarray}
It is also called a {\it symplectic realization } of $G$ on $(V,\sigma)$.
\begin{pro}
If $X$ is an element of ${\cal{G}}$ and if $ D_{\exp(tX)}$ is a symplectomorphism of $(V,\sigma)$, then $X$ is locally Hamiltonian
vector field.
\end{pro}
In fact, the fact that $D_{\exp(tX)}$ is a symplectomorphism of $(V,\sigma)$, means that \\$D_{\exp(tX)}^{*}\sigma=\sigma$.\;\;It
 follows that
$L_X\sigma=0$ hence $i_X(\sigma)$ is closed, which means effectively that $X$ is locally Hamiltonian.\;\;As a Hamiltonian vector field is
locally Hamiltonian, we can conclude that to a Hamiltonian vector field corresponds a symplectic
realization.\\

With these preliminaries out of the way, we can now proceed to tackle the problem of constructing and classifying noncommutative
phase spaces group theoretically on planar kinematical groups and compare the results obtained
with those found with others methods.\;\;Note that, in this thesis, the maximal coadjoint orbits are constructed
by considering only the evolution of time.
\section{Noncommutative phase space on Aristotle group}
Aristotle group is an intermediate group between the Euclidean and the Galilei groups dubbed by Souriau \cite{souriau1}.\;\;It
contains both Euclidean displacements and time translation but not boosts.\;\;In this section, we use the coadjoint orbit method
to construct phase spaces endowed with modified symplectic structure on this
intermediate group.\;\;This method
allows therefore the construction of the classical dynamical systems associated with this Lie group.\\

  More specifically, we demonstrate that such deformed objects can be generated in the framework of a
noncentrally abelian extended Aristotle algebra.\;\;We also realize
symplectically both the centrally  and noncentrally abelian extended
Aristotle Lie group.\;\;The obtained in such a way phase spaces (in the noncentrally abelian extended group case only) do not
commute in momentum sector
due to the presence of a naturally introduced magnetic field corresponding to the minimal coupling of the momentum
with a magnetic potential.\\
Thus, this case corresponds to the minimal coupling of the momentum with the magnetic potential.\;\;As the coadjoint orbit construction
has not been carried through this Lie group before, physical interpretations of new generators of the extended
corresponding Lie algebras are also given.
The results of this section have been published in \cite{ancille}.
\subsection{Aristotle group}
Let $V=M\times\Re$ be a ($n+1$)-dimensional nonrelativistic spacetime
where $\Re$ supports the absolute time coordinate $t$.\;\;
Let
\begin{eqnarray}
 x^{\mu}=\left(\begin{array}{c}x^{i}\\t
 \end{array}
 \right),~~i=1,....,n
\end{eqnarray}
be the coordinate of an arbitrary point of $V$. We use the Greek alphabet for the spacetime indices and the
Latin alphabet for the space indices.\;\;It is well known that the
displacement group with respect to the isotropy and
homogeneity of the spacetime $V$ together with the Galilean nonrelativistic principle of motion admits the following transformations:
\begin{itemize}
 \item spatial rotation characterized by three parameters:
 \begin{eqnarray}\label{ari1}
 x^{i~\prime }=a^{i}_{j}x^{j},~~~t^{\prime}=t
 \end{eqnarray}
 where $(a^{i}_{j})$ is an orthogonal matrix whose determinant is $+1$, that is an element of $SO(3)$.
 \item Galilean boosts characterized by the three component vector $v^i$:
 \begin{eqnarray}\label{ari2}
 x^{i~\prime }=x^{i}+ v^{i}t,~~t^{\prime}=t
 \end{eqnarray}
 \item spatial translations characterized by three parameters $a^i$:
 \begin{eqnarray}\label{ari3}
 x^{i~\prime }=x^{i}+ a^{i},~~t^{\prime}=t
 \end{eqnarray}
 \item temporal translations characterized by only one parameter:
 \begin{eqnarray}\label{ari4}
  x^{i~\prime }=x^{i},~~t^{\prime}=t+t_0
  \end{eqnarray}
\end{itemize}
 By composing the above transformations and under the usually composition of matrices, it has been verified that the above
 transformations form a group whose multiplication is:
\begin{eqnarray}\label{lawgalgen}
 (x^{i},t,v^{i}, a^{i}_{j})(x^{j~\prime },t^{\prime},v^{j~\prime },a^{j~\prime }_{k})=
 (x^{i}+ a^{i}_{j}x^{j~\prime }+v^{i}t^{\prime},t+t^{\prime},v^{i}+a^{i}_{j}v^{j~\prime },
 a^{i}_{j}a^{j~\prime }_{k})
\end{eqnarray}
This is the Galilei group multiplication law.\;\;When $n=2$, (\ref{lawgalgen}) becomes (\ref{galileilaw}).\\

Let us now define a subgroup of the Galilei group by dropping the Galilean boosts.\;\;We obtain the
Aristotle group.\;\;The latter is the
main group of the less known Aristotle mechanics and is
formed by the composition of the transformations (\ref{ari1}, (\ref{ari3}) and (\ref{ari4}).\;\;It
 has order $7$ expresses Aristotle's relativity principle.\;\;By composing the above transformations, we obtain
the Aristotle multiplication law given by:
\begin{eqnarray}\label{aristotegenlaw}
 (x^{i},t, a^{i}_{j})(x^{j~\prime },t^{\prime},a^{j~\prime }_{k})=
 (x^{i}+ a^{i}_{j}x^{j~\prime },t+t^{\prime},
 a^{i}_{j}a^{j~\prime }_{k})
\end{eqnarray}
Thus, the Aristotle group is
the group of both Euclidean displacements and time translations (there is no boosts for this group).\;\;In a two-dimensional
space, the multiplication law (\ref{aristotegenlaw} ) of this general Lie group is given by
\begin{eqnarray}\label{Aristotlelaw}
(\theta, \vec{x}, t)(\theta^{~\prime}, \vec{x}^{~\prime}, t^{~\prime})=(\theta+\theta^{~\prime}, R(\theta)\vec{x}^{~\prime}+\vec{x}
 , t+t^{~\prime})
\end{eqnarray}
where $\vec{x}$ is a space translation vector, $t$ is a time
translation parameter and $\theta$ is a rotation parameter. \\

 In the following subsections, we are going to prove that one can not construct noncommutative phase spaces by the coadjoint
 orbit method with the first central extension of the two-dimensional Aristotle group because the symplectic structure
obtained is canonical.\;\;This is due to the fact that in the centrally extended Aristotle Lie algebra, the generators of spatial
translations remain commutative.\;\;But
by considering a noncentral abelian extension of Aristotle group, we will realize a partially
noncommutative phase spaces (only momenta do not commute).\;\;Furthermore, the noncommutativity of momenta
is measured by a term which is associated to the naturally introduced magnetic field.\;\;Moreover, this case corresponds
to the minimal coupling of the momentum with the magnetic potential \cite{ancilla} as already argued previously.
 \subsection{Central extension of the Aristotle group and its maximal coadjoint orbit}
The Aristotle Lie algebra $\cal{A}$ is by definition generated by
the left invariant vector fields given by the formula
 \begin{eqnarray*}
 X_i=\frac{\partial{(gg^{~\prime}})^j}{\partial {\nu^{~\prime}_i}}
|_{g^{~\prime}=e}
(\frac{\partial}{\partial {\nu}_j})
 \end{eqnarray*}
where $X_i$ is the left invariant field corresponding to the parameter $\nu_i$, $e$
is the identity element of $G$ while $gg^{~\prime}$ is the group multiplication law.\\

Explicitly, the generators of the Aristotle Lie algebra are given by:
\begin{eqnarray*}
J=\frac{\partial}{\partial\theta},~~\vec{P}=R(-\theta )\frac{\partial}{\partial\vec{x}},~~H=\frac{\partial}{\partial t}
\end{eqnarray*}
such that the only nontrivial Lie brackets are
\begin{eqnarray}\label{Aristotleliealgebra}
[J,P_i]=P_j\epsilon^j_i,~~i,j=1,2.
\end{eqnarray}
The multiplication law (\ref{Aristotlelaw}) implies that the element $g$ of this group can be written as:
\begin{eqnarray}\nonumber
g=\left(
\begin{array}{cccc}
\cos\theta&-\sin\theta&0&x^1            \\
\sin\theta&\cos\theta&0&x^2\\
0&0&1&t\\
0&0&0&1\\
\end{array}
 \right)
\end{eqnarray}
or equivalently
\begin{eqnarray} \nonumber
g=\left(
\begin{array}{cccc}
\cos\theta&-\sin\theta&0&x^1\\
\sin\theta&\cos\theta&0&x^2\\
0&0&1&0\\
0&0&0&1\\
\end{array}
 \right)\times
\left(
\begin{array}{cccc}
1&0&0&0\\
0&1&0&0\\
0&0&1&t\\
0&0&0&1\\
\end{array}
 \right)
\end{eqnarray}
and then we can parametrize the element $g$ of the group in the following manner:
\begin{eqnarray}\label{elementg}
g=\exp(\vec{x}\vec{P}+t H)\exp(\theta J)
\end{eqnarray}
We are now going to employ the formulation we described in section ($3.1.1)$ to construct the
classical dynamical systems associated to the
Aristotle group.\;\;We construct indeed the maximal coadjoint orbits associated to the extensions of this Lie group when evolution of
time is considered.\;\;Following the Souriau prescription, these coadjoint orbits are
identified as elementary systems associated to the extended Aristotle
 groups.\;\;Here, the word elementary means that the action of the considered group should be transitive, at the quantum level, this
 means that the representation be irreducible \cite{Duval-horvathy}.\\

From the relation  $\exp(2\pi J)H \exp(-2\pi J)=H$ and by use of the standard methods,
we obtain the following nontrivial Lie brackets for the first central extension Lie algebra ${\hat{\cal{A}}}$ of ${\cal{A}}(2)$
\begin{eqnarray}\label{centralliebrackets}
[J,P_i]=P_j\epsilon^j_i,~~ [P_i,P_j]=\kappa^2 S\epsilon_{ij},~~i,j=1,2
\end{eqnarray}
where $S$ generates the center of  ${\hat{\cal{A}}}$ and is dimensionless.\;\;Dimensional analysis implies that the parameter
 $\kappa$ has dimension inverse of that of a length. \\

 Let us look now at the group associated to the above centrally extended Lie algebra.\\

Let $g$ be given by (\ref{elementg}) and ${\hat g}=\exp(\varphi S)g$ be the corresponding element in the connected
Lie group associated to
the extended Lie algebra ${\hat{\cal{A}}}$.\;\;By use of the Baker-Campbell-Hausdorff formulas \cite{hall} and by
identifying ${\hat g}$ with
$(\varphi,\theta,\vec{x},t)$, we find that the multiplication law of the connected extended Lie group $\hat{A}$ is:
\begin{eqnarray*}
(\varphi,\theta,\vec{x},t)(\varphi^{~\prime},\theta^{~\prime},\vec{x}^{~\prime},t^{~\prime})=(\varphi^{~\prime}+
\frac{1}{2}\kappa^2R(-\theta)\vec{x}\times \vec{x}^{~\prime}+\varphi,
\theta+\theta^{~\prime}, R(\theta)\vec{x}^{~\prime}+\vec{x},t+t^{~\prime})
\end{eqnarray*}
or equivalently
\begin{eqnarray*}
 (\alpha,g)(\alpha^{~\prime},g^{~\prime})=(\alpha+\alpha^{~\prime}+c(g,g^{~\prime}),g g^{~\prime})
\end{eqnarray*}
where $c(g,g^{~\prime})=\frac{1}{2}\kappa^2R(-\theta)\vec{x}\times \vec{x}^{~\prime}$
is the two-cocycle defining the central extension while $ g g^{~\prime}$ is the multiplication law (\ref{Aristotlelaw}).\\

The adjoint action $Ad_g(\delta{\hat{g}})=g( \delta{\hat{g}} )g^{-1}$ of $A$ on the Lie algebra ${\hat{\cal{A}}}$ is given by:
\begin{eqnarray}\label{ancillaa}\nonumber
 Ad_{(\theta,\vec{x},t)}(\delta \varphi,\delta \theta,\delta \vec{x},\delta t)=(\delta \varphi+\kappa^2 R(-\theta)\vec{x}\times
 \delta \vec{x}\\-
\frac{1}{2}\kappa^2 \vec{x}^{~2}\delta\theta,\delta \theta,R(\theta)\delta \vec{x} +\epsilon(\vec{x}) \delta\theta, \delta t)
\end{eqnarray}
with
\begin{eqnarray}\label{epsilon1}
\epsilon(\vec{x})=\left(
\begin{array}{cc}
0&x^{2}\\-x^{1}&0
\end{array}
\right)
\end{eqnarray}
If the duality between the centrally extended Lie algebra and its dual is given by the action
\begin{eqnarray}
 \langle(j,\vec{p},l,E),(\delta \theta,\delta \vec{x},\delta \varphi,\delta t)\rangle=j\delta \theta
 +\vec{p}.\delta \vec{x}+l\delta \varphi+E\delta t
\end{eqnarray}
where $j$ is an angular momentum, $\vec{p}$ is a linear momentum, $l$ is an action while $E$
is an energy, then the coadjoint action of the Aristotle Lie group is
\begin{eqnarray}\label{coadjointaction1}
Ad^*_{(\vec{x},t,\theta)}(j,\vec{p},l,E)=(j+\frac{m\omega}{2}(\vec{x}^{~2})+\vec{x} \times R(\theta)\vec{p},
R(\theta)\vec{p}-m\omega\epsilon(
{\vec{x}}),l,E)
\end{eqnarray}
where we have used the relation $l\omega=mc^2$: a relation remembering us the wave-particle duality, the left hand
side being an energy associated to a frequency $\omega$, the right hand side
being an energy associated to a mass,
and the relation $c=\frac{\omega}{\kappa}$ linking the velocity $c$, the
frequency $\omega$ and $\kappa$ whose dimension is the inverse of that of a length.\\

The parameters $l$ and $E$ in (\ref{coadjointaction1}) corresponding to the central generators remain
fixed (i.e are $Ad^*$-invariants): they are called trivial invariants
of the coadjoint action of the group on the dual of its centrally extended Lie algebra.\\

The Kirillov form in the basis
$(J,P_1,P_2,H,S)$ is
\begin{eqnarray}
K(a)=\left(
\begin{array}{ccccc}
0&p_2&-p_1&0&0\\
-p_2&0&m\omega&0&0\\
p_1&-m\omega&0&0&0\\
0&0&0&0&0\\
0&0&0&0&0\\
\end{array}
 \right)
\end{eqnarray}
The maximal coadjoint orbit of the centrally extended Lie group on the dual of its Lie algebra is characterized by the
two trivial invariants
 $l$ and $E$ and a nontrivial invariant, solution of the Kirillov system (\ref{kirillovsystem}) and given by:
\begin{eqnarray}\label{Aristotleinvariant10}
s=j+\frac{\vec{p}^{~2}}{2m\omega}.
\end{eqnarray}
interpreted as an internal angular momentum or spin with $j$, the angular momentum.\\

As already argued early, there must exist generators (namely momenta and boosts) which yield the basic canonical variables through
the coadjoint orbit method.\;\;Thus,
to define the canonical (Darboux's) coordinates on the orbit in this particular group case (where there is no boosts), let us
consider the following new variables:
\begin{eqnarray}\label{ancilla0}
 q=-\frac{p_2}{m \omega},  ~~~p=p_1
\end{eqnarray}
where dimensional analysis has been taking into account.\;\;Use of (\ref{ancilla0}) in (\ref{Aristotleinvariant10})
gives rise to the expression of the spin $s$ in function of the
canonical coordinates
\begin{eqnarray}
 s=j+\frac{{p}^{~2}}{2m\omega}+ m\omega\frac{q^2}{2}
\end{eqnarray}
Let us denote by  ${\cal{O}}_{(s,l,E)}$ the two-dimensional maximal coadjoint orbit of the Aristotle group $A(2)$ on the dual of its
central extended Lie algebra.\\

The restriction $\Omega=(\Omega_{ab})$ of the Kirillov form to the orbit is then
\begin{eqnarray*}
\Omega=\left(
\begin{array}{cc}
0&m\omega \\-m\omega&0\\
\end{array}
\right)
\end{eqnarray*}
 It follows that the symplectic form (\ref{symplectictwo-form}) is in this case given by  $\sigma=dp\wedge dq$ where we have used
relations (\ref{ancilla0} ).\\

The symplectic realization of the Aristotle Lie group is
\begin{eqnarray}
 D_{(\theta,\vec{x},t)}(p,q)=(\cos \theta ~p+m\omega q\sin\theta-m\omega x^2,-\frac{p}{m\omega}\sin\theta+q\cos\theta-x^1)
\end{eqnarray}
The Poisson bracket (\ref{poissonbrackets}) corresponding to this symplectic structure is then
the canonical one and the time translation subgroup acts trivially on the orbit (the position and
the linear momentum do not depend on time).\;\;To overcome this fact and to obtain a noncommutative
phase space, let us study the symplectic realization of a noncentral
extension of the Aristotle Lie group.\\

We prove, in the following paragraph, that noncommutative phase spaces can be obtained by
considering a noncentral abelian extension of the
two-dimensional Aristotle group.
\subsection{Noncentrally abelian extended group and its maximal coadjoint orbit}
In the previous section, we have found that one can not construct noncommutative phase spaces by coadjoint orbit method on the
the central extension of the Aristotle group because symplectic structure obtained is canonical which means that positions
commute as well as momenta.\;\;In the following subsection, we see that this construction is possible when we consider a noncentral
extension of this Lie group.\\

Consider the following noncentral abelian extension of the Aristotle Lie algebra defined by the nontrivial Lie brackets
\begin{eqnarray}\label{Aristotleextc}
[J,P_j]=P_i\epsilon^{i}_j,~~~[P_i,P_j]=\kappa^2 S\epsilon_{ij}~~~~~~~~~~~~~~~~~~~~\\\nonumber
[J,F_j]=F_i\epsilon^{i}_j~,~[P_i,H]= F_i,~~
[P_i,F_j]= K\delta_{ij}
 \end{eqnarray}
 where $F_i$ has the dimension of a force while $S$ and $K$ are dimensionless.

We recover the Lie algebra defined by (\ref{Aristotleliealgebra}) when $F_i=0$,~$K=0$ and $S=0$, the Lie algebra defined by
(\ref{centralliebrackets}) when
$F_i=0$,~$K=0$.\\

Consider now the general Lie algebra defined by (\ref{Aristotleextc}) and
let \\$\hat{g}=\exp(\varphi S+\gamma K)\exp(t H)\exp(\vec{\eta}\vec{F}+\vec{x}\vec{P})\exp(\theta J)$
be the general element of the corresponding connected extended Aristotle group.\;\;By identifying  $\hat{g}$ with $(\varphi,\gamma,t,
\vec{\eta},\vec{x},\theta)$ and by explicit calculation, we obtain the
multiplication law $\hat{g}^{~\prime\prime}=\hat{g}\hat{g}^{~\prime}$ in the extended Lie group.\;\;It is such that:
\begin{eqnarray*}
\varphi^{~\prime\prime}=\varphi^{~\prime}+\frac{1}{2}\kappa^2\vec{x}\times R(-\theta) \vec{x}^{~\prime}+\varphi,~~~~~~~~~~~~~
~~~~~~~~~~~~~~~~~~~~~~~~\\
\gamma^{~\prime\prime}=\gamma^{~\prime}+\frac{ 1}{2}\vec{x}. R(\theta)\vec{\eta}^{~\prime}-\frac{1}{2}(\vec{\eta}+\vec{x}~t^{~\prime})
.R(\theta)\vec{x}^{~\prime}+\gamma,~~~~~~~~~~\\
\vec{\eta}^{~\prime\prime}=R(\theta)\vec{\eta}^{~\prime}+\vec{\eta}+\vec{x}~t^{~\prime},~~~~~~~~~~~~~~~~~~~~~~~~~~~~~
~~~~~~~~~~~~~~~~~~~~~~\\
\vec{x}^{~\prime\prime}=R(\theta)\vec{x}^{~\prime}+\vec{x}~~~~~~~~~~~~~~~~~~~~~~~~~~~~~~~~~~~~~~~~~~~~~~~~~~~~~~~~~~~~~~~\\
\theta^{~\prime\prime}=\theta^{~\prime}+\theta,~~~~~~~~~~~~~~~~~~~~~~~~~~~~~~~~~~~~~~~~~~~~~~~~~~~~~~~~~~~~~~~~~~~~~~\\
t^{~\prime\prime}=t+t^{~\prime}~~~~~~~~~~~~~~~~~~~~~~~~~~~~~~~~~~~~~~~~~~~~~~~~~~~~~~~~~~~~~~~~~~~~~~~~~~
\end{eqnarray*}
The adjoint action $Ad_g(\delta{\hat{g}})=g( \delta{\hat{g}} )g^{-1}$ of the Aristotle group
$A$ on the noncentrally abelian extended Lie algebra is given by:
\begin{eqnarray}\label{ancillaa}\nonumber
 Ad_{( \vec{x}, \vec{\eta}, \theta, t)}(\delta \gamma, \delta \varphi,\delta \vec{x},
 \delta \vec{\eta},\delta \theta, \delta t)=(\delta \gamma^{\prime},\delta \varphi^{\prime}, \delta\vec{\eta}^{~\prime}, \delta \vec{x}^{~\prime},
 \delta \theta^{\prime}, \delta t^{\prime})
\end{eqnarray}
with
\begin{eqnarray*}
\delta \gamma^{\prime}=\delta \gamma+\vec{x} \times R(\theta)\delta \vec{\eta}-\vec{\eta}\times R(\theta)\delta \vec{x}-
\vec{\eta}\times\vec{x}\delta{\theta}
+\frac{1}{2}\vec{x}^{~2}\delta t~~~~~~~~\\
\delta\vec{\eta}^{~\prime}=R(\theta)\delta \vec{\eta}+\epsilon(\vec{\eta}-\vec{x}~t)\delta\theta-t
R(\theta)\delta \vec{x}+\vec{x}\delta t~~~~~~~~~~~~~~~~~~~~~~~~~~\\
\delta \varphi^{\prime}=\delta \varphi+\kappa^2 R(-\theta)\vec{x}\times \delta \vec{x}-
\frac{\vec{x}^{~2}}{2}\kappa^2\delta \theta~~~~~~~~~~~~~~~~~~~~~~~~~~~~~~~~~~~~~~~\\
\delta \vec{x}^{~\prime}=R(\theta)\delta \vec{x} +\epsilon(\vec{x})\delta \theta~~~~~~~~~~~~~~~~~~~~~~~~
~~~~~~~~~~~~~~~~~~~~~~~~~~~~~~~~~~~~~~~~~\\
\delta \theta^{\prime}=\delta{\theta}~~~~~~~~~~~~~~~~~~~~~~~~~~~~~~~~~~~~~~~~~~~~~~~~~~~~~~~~~~~~~~~~~~~~~~~~~~~~~~~~~~~~~~~~~\\
\delta t^{\prime}=\delta t~~~~~~~~~~~~~~~~~~~~~~~~~~~~~~~~~~~~~~~~~~~~~~~~~~~~~~~~~~~~~~~~~~~~~~~~~~~~~~~~~~~~~~~~~~~
\end{eqnarray*}
where $\epsilon (\vec{x})$ is given by the relation (\ref{epsilon1}).\\

If the duality between the noncentrally abelian extended Lie algebra and its dual Lie algebra gives rise to the action
\begin{eqnarray*}
 \langle(j,\vec{f},\vec{p},h,k, E),(\delta \theta, \delta \vec{\eta},\delta \vec{x},\delta \varphi,\delta \gamma,\delta t)\rangle=
 j\delta \theta+\vec{f}.\delta \vec{\eta}+\vec{p}.\delta \vec{x}\\+h\delta \varphi+E\delta t+k\delta\gamma
\end{eqnarray*}
then the coadjoint action is such that
\begin{eqnarray}\label{massaction0}
h^{\prime}=h,~~k^{\prime}=k
\end{eqnarray}
and
\begin{eqnarray}\label{Aristotlemomentum0}
\vec{p}^{~\prime}=R(\theta)\vec{p}+R(\theta)\vec{f}~t+k(\vec{\eta}-\vec{x}~t)+h\kappa^2\epsilon(\vec{x})~,\vec{f}^{~\prime}=
R(\theta)\vec{f}-k\vec{x}
\end{eqnarray}
\begin{eqnarray}\label{Aristotleangularmomentum0}
j^{~\prime}=j+\vec{x}\times R(\theta)\vec{p}+\vec{\eta}\times R(\theta)\vec{f}-\frac{h}{2}\kappa^2{\vec{x}~^2}~~~~~~~~~~~~~
~~~~~~~~~~~~~~~~
\end{eqnarray}
\begin{eqnarray}\label{Aristotleenergy0}
E^{~\prime}=E-\vec{x}.R(\theta)\vec{f}+\frac{1}{2} k\vec{x}^2~~~~~~~~~~~~~~~~~~~~~~~~~~~~~~~~~~~~~~~~~~~~~~~~~~~~~~
\end{eqnarray}
where $\vec{p}$ is a linear momentum, $h$ is an action, $\vec{f}$ is a force, $k$ is a Hooke's constant, $E$ is an energy while
$j$ is an angular momentum.\\

The coadjoint orbit is, in this case,
characterized by the two trivial invariants $h$ and $k$ (\ref{massaction0}), and by the
nontrivial invariants $s$ and $U$ given by:
\begin{eqnarray}\label{angular}
s=j-\vec{p}\times\vec{q}+\frac{1}{2}m\omega\vec{q}~^2,~~~U=E-\frac{1}{2}k\vec{q}~^2
\end{eqnarray}
where we have used the notation
\begin{eqnarray}\label{qfk}
 \vec{q}=-\frac{\vec{f}}{k}
\end{eqnarray}
which defines the basic position variables on the orbit.\\

Note that the quantities in (\ref{angular}) are interpreted as internal angular momentum and internal energy
respectively.\;\;By considering the coadjoint action, we see
 that the coadjoint orbit is $4-$dimensional.\;\;Let us denote it by
${\cal{O}}_{(h,k,s,U)}$.\\

The restriction $\Omega=(\Omega_{ab})$ of the Kirillov form to the orbit is then
\begin{eqnarray*}
\Omega=\left(
\begin{array}{cccc}
0&m\omega&k&0\\
-m\omega&0&0&k\\
-k&0&0&0\\
0&-k&0&0\\
\end{array}
\right)
\end{eqnarray*}
The modified symplectic form is explicitly given by
\begin{eqnarray}
\sigma=dp_i\wedge dq^i+\frac{1}{2}m\omega \epsilon^{ij}dq^i\wedge dq^j
\end{eqnarray}
where $\vec{q}$ is given by relation (\ref{qfk}).\\

If $(y_a)=(p_1,p_2,q^1,q^2)$, the Poisson brackets are then explicitly given by
\begin{eqnarray}
\{H,g\}=\frac{\partial H}{\partial p_i}\frac{\partial g}{\partial
q^i}-\frac{\partial H}{\partial q^i}\frac{\partial g}{\partial
p_i}+F_{ij}\frac{\partial H}{\partial p_i}\frac{\partial g}{\partial
p_j}
\end{eqnarray}
where
\begin{eqnarray}\label{fmomega}
F_{ij}=-m\omega{\epsilon_{ij}}
\end{eqnarray}
This implies that
\begin{eqnarray}
\{p_i,p_j\}=F_{ij}~,~\{p_i,q^j\}=\delta^{j}_i~,~\{q^i,q^j\}=0
\end{eqnarray}
Let us now use the symplectic realizations methods defined in section ($3.1.2$) to find the equations of motion in the dynamical
system constructed on this extension of Aristotle group.\\

 Let the symplectic realization of the extended Aristotle Lie group on its coadjoint orbit be given by
$(\vec{p}^{~\prime},\vec{q}^{~\prime})=L_{(\theta,\vec{\eta},\vec{x},t})(\vec{p},\vec{q})$.\;\;By using relations
(\ref{Aristotlemomentum0}), we have
\begin{eqnarray}
\vec{p}^{~\prime}=R(\theta)\vec{p}-k[(R(\theta)\vec{q}+\vec{x})t-\vec{\eta}~]+h\epsilon(\vec{x}),~\vec{q}^{~\prime}=
R(\theta)\vec{q}+\vec{x}
\end{eqnarray}
It follows that ($\vec{p}(t),\vec{q}(t))=D_{(0,0,0,0,0,t})(\vec{p},\vec{q})$ gives rise to
$$
\vec{p}(t)=\vec{p}-k\vec{q}~t,~~
\vec{q}(t)=\vec{q}
$$
The equations of motion are then
\begin{eqnarray}
\frac{d\vec{p}}{dt}=-k\vec{q},~~\frac{d\vec{q}}{dt}=0
\end{eqnarray}
From (\ref{angular}), the angular momentum is
\begin{eqnarray}
j=-\vec{q}\times \vec{p}+s-eB\frac{\vec{q}^{~2}}{2}
\end{eqnarray}
i.e. the sum of the orbital angular momentum
$L=\vec{q}\times \vec{p}$, the internal angular momentum $s$
and an extra term $-eB\frac{\vec{q}^{~2}}{2}$ associated to the
magnetic field $B$ \cite{horvathy}.\\

So with the above noncentral abelian extension of the two-dimensional Aristotle group, we have realized a phase space
where momenta do not commute and this noncommutativity is due to presence of a naturally introduced magnetic field given by
relation (\ref{fmomega})
or equivalently by
\begin{eqnarray}\label{magneticfield}
F_{ij}=-eB\epsilon_{ij}.
\end{eqnarray}
Let us proceed similarly to construct noncommutative phase spaces on the centrally extended anisotropic kinematical groups.\;\;They are
maximal coadjoint orbits identified as elementary systems associated with those extended Lie structures.
\section{Noncommutative phase spaces constructed on anisotropic kinematical groups}
Through this section, we follow the approach we have described above to construct dynamical systems and hence modified
symplectic structures (noncommutative phase spaces) associated to the extended groups corresponding to
the centrally extended Lie algebras described in section ($1.2.2$).\;\;As we have done for the
extended Aristotle groups, we proceed similarly for
the centrally extended anisotropic planar Newton-Hooke groups, Galilei group, Para-Galilei, Static and Carroll groups.\\

Three types of noncommutative phase spaces are obtained: noncommutative phase spaces in momenta and positions sectors in the
Newton-Hooke, Static and Carroll groups cases, noncommutative phase spaces in momenta sector only in the Para-Galilei groups case
and noncommutative phase spaces in positions sector only in the Galilei group case.
 \subsection{ Newton-Hooke noncommutative phase space}
Nonrelativistic particle models have been constructed following Souriau's method for the two-parameter centrally extended anisotropic
Newton-Hooke groups in a two-dimensional space \cite{ancilla} yielding similar results as in \cite{zhang-horvathy}
(for the oscillating case).

Indeed, by considering the planar oscillating Newton-Hooke group $NH_{-}$, the authors in
\cite{zhang-horvathy} have found a similar symmetry in the so-called Hill problem
(the latter is studied in celestial mechanics), which
is effectively an anisotropic harmonic oscillator in a magnetic field.\;\;The peculiarity is that this system has no rotational symmetry
 while translations and boosts still act as symmetries.\;\;Note also that, as already said in the introduction, the noncommutative
version of the Hill problem has been discussed in \cite{zhang-horvathy1}.\;\;To be complete in this work, let us review the
coadjoint orbit construction of the planar anisotropic Newton-Hooke groups $NH_{\pm}$.\\

The anisotropic
Newton-Hooke groups $ANH_{\pm}$ being Newton-Hooke groups $NH_{\pm}$ without the rotation parameters \cite{3derome}, their multiplication
laws are given by :
\begin{eqnarray}\label{newton-hookelawmoins}\nonumber
( x^i,v^i,t)(x^{i~\prime},v^{i~\prime},t^{~\prime})=(x^i~\cos\omega t^{~\prime}+\frac{v^i}{\omega}~\sin\omega t^{~\prime}+x^{i~\prime},\\
-\omega x^i~ \sin\omega t^{~\prime}+v^i~\cos \omega t^{~\prime}+v^{i~\prime},t+t^{~\prime})
\end{eqnarray}
 for $ANH_{-}$ ~~and
\begin{eqnarray}\label{newton-hookelawplus}\nonumber
( x^i,v^i,t)(x^{i~\prime},v^{i~\prime},t^{~\prime})=(x^i~\cosh\omega t^{~\prime}+\frac{v^i}{\omega}~\sinh\omega
t^{~\prime}+x^{i~\prime},\\
\omega x^i ~\sinh\omega t^{~\prime}+v^i~\cosh \omega t^{~\prime}+v^{i~\prime},t+t^{~\prime})
\end{eqnarray}
 for $ANH_{+}$~~ with ~~$i=1,2,3$.

 The multiplication law (\ref{newton-hookelawmoins}) implies that the general element $g$ of the group
$ANH_{-}$ can be written as:
\begin{eqnarray}\nonumber
g=\left(
\begin{array}{ccc}
\cos\omega t&-\omega\sin\omega t&0\\
\frac{1}{\omega}~\sin\omega t&\cos\omega t&0\\
x^i&v^i&1\\
\end{array}
 \right)
\end{eqnarray}
or equivalently
\begin{eqnarray} \nonumber
g=\left(
\begin{array}{ccc}
\cos\omega t&-\omega~\sin\omega t&0\\
\frac{1}{\omega}~\sin\omega t&\cos\omega t&0\\
0&0&1\\
\end{array}
 \right)\times
\left(
\begin{array}{ccc}
1&0&0\\
0&1&0\\
x^i&v^i&1\\
\end{array}
 \right)
\end{eqnarray}
Similarly, the multiplication law (\ref{newton-hookelawplus}) implies that
the general element $g$ of the group
$ANH_{+}$ can be written as:
\begin{eqnarray}\nonumber
g=\left(
\begin{array}{ccc}
\cosh\omega t&\omega~\sinh\omega t&0\\
\frac{1}{\omega}~\sinh\omega t&\cosh\omega t&0\\
x^i&v^i&1\\
\end{array}
 \right)
\end{eqnarray}
or equivalently
\begin{eqnarray} \nonumber
g=\left(
\begin{array}{ccc}
\cosh\omega t&\omega~\sinh\omega t&0\\
\frac{1}{\omega}~\sinh\omega t&\cosh\omega t&0\\
0&0&1\\
\end{array}
 \right)\times
\left(
\begin{array}{ccc}
1&0&0\\
0&1&0\\
x^i&v^i&1\\
\end{array}
 \right)
\end{eqnarray}
Then we can parametrize the element $g$ of the group $ANH_{\pm}$ in the following manner:
\begin{eqnarray}\label{elementgnh}
g=\exp(tH)\exp(\vec{x}\vec{P}+\vec{v}\vec{K})
\end{eqnarray}
where $ \vec{K}, \vec{P}, H$ are generators (i.e the left invariant vector fields) of the corresponding
Lie algebras ${\cal{ANH}}_{\pm}$ and are respectively given by:
\begin{eqnarray}
~\vec{K}=\frac{\partial}{\partial
\vec{v}}~~,~~\vec{P}=\frac{\partial}{\partial
\vec{x}}~,~H=\frac{\partial}{\partial
t}+\vec{v}.\frac{\partial}{\partial \vec{x}}\pm
\omega^2\vec{x}.\frac{\partial}{\partial \vec{v}}
\end{eqnarray}
In contrast to the Galilei case, the Newton-Hooke Lie algebras are not contraction of the Poincar\'e algebra, but can be obtained
directly from the de Sitter algebras by velocity-space contractions \cite{5bacry}.\;\;They contract themselves onto the Galilei
algebra through the space-time contractions.\\

Standard methods \cite{kostant, kirillov,  nzo1, hamermesh}
show that the structure of the central extensions of the Lie algebras
${\cal{ANH}}_{\pm}$ is
\begin{itemize}
\item in one-dimensional space\\
\begin{eqnarray*}
[K,H]=P~,~[P,H]=\pm \omega^2K~,~[K,P]=M
\end{eqnarray*}
\item  in two-dimensional spaces\\
\begin{eqnarray}
[K_i,K_j]&=&\frac{1}{c^2}J_3\epsilon_{ij}~~,~~[K_i,H]=P_i~~,~~[K_i,P_j]=M\delta_{ij} \crcr
[P_i,P_j]&=&\pm \kappa^2J_3\epsilon_{ij}~~,~~[P_i,H]=\pm
\omega^2K_i~~~~~~~~~~~~~~~~~~~~~~
\end{eqnarray}
\item in three-dimensional spaces\\
\begin{eqnarray}
[K_i,K_j]&=&\frac{1}{c^2}J_k\epsilon^k_{ij}~~,~~[K_i,H]=P_i~~,~~[K_i,P_j]=M\delta_{ij} \crcr
[P_i,P_j]&=&\pm\kappa^2J_k\epsilon^k_{ij}~~,~~[P_i,H]=\pm
\omega^2K_i~~~~~~~~~~~~~~~~~~~~~~
\end{eqnarray}
\end{itemize}
where $\kappa$ is a constant whose dimension is inverse of that of a length, $c$ is a
constant with the dimension of a speed while $J_{k}$ is an internal rotation parameter around the $k^{th}$ axis.\\

We are now going to employ the formulation we described in section ($3.1.1)$ to construct noncommutative phase space associated to the
one, two and three-dimensional anisotropic Newton-Hooke groups.\;\;We construct indeed the maximal coadjoint orbits associated to the
central extensions of these Lie groups when evolution of time is considered.\;\;Following the Souriau's
prescription, these coadjoint orbits are identified as elementary systems associated to the extended anisotropic
Newton-Hooke groups.\;\;Recall that the word elementary means that the action of the groups should be transitive and at
the quantum level, this means that the representation should be irreducible \cite{Duval-horvathy}.
 \subsubsection{i) One-dimensional space case}
Let $g$ be given by (\ref{elementgnh}) (i.e $ g=(x,v, t)$ for the short) and ${\hat g}=\exp(\varphi M)g$ be the
corresponding element in the connected Lie groups associated to
the extended Lie algebra ${\cal{ANH}}_{\pm}(1)$ in one-dimensional space.\;\;By use of the Baker-Campbell-Hausdorff
formulas \cite{hall} and by
identifying ${\hat g}$ with
$(\varphi,g)$, we find that the multiplication laws of the connected extended Lie groups in one- dimensional space are:
\begin{eqnarray*}
 (\varphi,g)(\varphi^{~\prime},g^{~\prime})=(\varphi+\varphi^{~\prime}+c(g,g^{~\prime}),g g^{~\prime})~~~~~~~~~~~~~~~~~~
\end{eqnarray*}
where
\begin{eqnarray}
 c(g,g^{~\prime})=-\frac{1}{2}(x~ \cos \omega t^{\prime}+\frac{v}{\omega}~\sin \omega t^{\prime})v^{\prime}
+\frac{1}{2}(v ~\cos \omega t^{\prime}-\omega x ~\sin \omega t^{\prime})x^{\prime}
\end{eqnarray}
for $ANH_{-}(1)$ ~~and
\begin{eqnarray}
c(g,g^{~\prime})=-\frac{1}{2}(x~ \cosh \omega t^{\prime}+\frac{v}{\omega}~\sinh \omega t^{\prime})v^{\prime}
+\frac{1}{2}(v~ \cosh \omega t^{\prime}+\omega x~ \sinh \omega t^{\prime})x^{\prime}
\end{eqnarray}
for $ANH_{+}(1)$\\
are the two-cocycles defining the central extensions of these groups while $ g g^{~\prime}$ is the multiplication law
(\ref{newton-hookelawmoins}) (for the group $ANH_{-}$) or (\ref{newton-hookelawplus} (for $ANH_{+}$) for ~~$i=1$.\\

The adjoint actions $Ad_g(\delta{\hat{g}})=g(\delta{\hat{g}} )g^{-1}$ of $ANH_{\pm}(1)$ on the Lie algebras ${\hat{\cal{ANH}}}_{\pm}(1)$
are given by:
\begin{eqnarray}\label{adjointnh1}\nonumber
 Ad_{(v,x,t)}(\delta \varphi,\delta x,\delta v, \delta t)=(\delta \varphi^{~\prime},\delta x^{~\prime},
 \delta v^{~\prime}, \delta t^{~\prime})
 \end{eqnarray}
 with
 \begin{eqnarray*}
\delta \varphi^{\prime}= \delta \varphi+v ~\delta x- x ~\delta v+\frac{1}{2}(v^2-\omega^2 x^2)~\delta t~~~~~~~~~~~~~~~
~~~~~~~~~~~~~\\\nonumber
\delta x^{\prime}= \cos \omega t~ \delta x-\frac{1}{\omega}~\sin \omega t~ \delta v+ (v ~\cos \omega t +\omega x ~\sin \omega t)~
\delta t~~~\\\nonumber
 \delta v^{\prime}=\cos \omega t ~\delta v+\omega~ \sin \omega t ~\delta x+\omega(v~ \sin \omega t- \omega x~ \cos \omega t)~
 \delta t\\\nonumber
 \delta t^{\prime}=\delta t~~~~~~~~~~~~~~~~~~~~~~~~~~~~~~~~~~~~~~~~~~~~~~~~~~~~~~~~~~~~~~~~~~~~~~~~~~~~~~~~~~~~~~\nonumber
\end{eqnarray*}
 in the case of $ANH_{-}(1)$ and
 \begin{eqnarray*}
\delta \varphi^{\prime}= \delta \varphi+v~ \delta x- x ~\delta v+\frac{1}{2}(v^2+\omega^2 x^2)~\delta t~~~~~~~~~~~~~~~~~~~~~~~~~~~~~~~~~
~~~~\\\nonumber
\delta x^{\prime}=  \cosh \omega t ~\delta x-\frac{1}{\omega}~\sinh \omega t~ \delta v-(v ~\cosh \omega t -\omega x ~\sinh \omega t)~
\delta t~~\\\nonumber
 \delta v^{\prime}=\cosh \omega t ~\delta v-\omega~ \sinh \omega t ~\delta x-\omega(v~ \sinh \omega t- \omega x ~\cosh \omega t)~
 \delta t~\\\nonumber
 \delta t^{\prime}=\delta t~~~~~~~~~~~~~~~~~~~~~~~~~~~~~~~~~~~~~~~~~~~~~~~~~~~~~~~~~~~~~~~~~~~~~~~~~~~~~~~~~~~~~~~~~~~~~~~\nonumber
\end{eqnarray*}
 in the case of $ANH_{+}(1)$.\\

If the duality between the extended Lie algebras and their duals is given by the action:
\begin{eqnarray}
 \langle(m,k,p,E),(\delta \varphi, \delta v,\delta x,\delta t)\rangle=m .\delta\varphi+k.\delta v+p.\delta x+E.\delta t
\end{eqnarray}
where $p$ is a linear momentum, $k$ is static momentum, $m$ is a mass while $E$
is an energy (in the dual Lie algebras), then the coadjoint actions of the anisotropic Newton-Hooke groups
on the dual of their centrally extended Lie algebras are
\begin{eqnarray*}\label{coadjointaction1}
Ad^*_{(x, v,t)}(m,p,k,E)=(m^{\prime}, p^{\prime}, k^{\prime}, E^{\prime})
\end{eqnarray*}
with
\begin{eqnarray}\label{coadjnh1moins}\nonumber
 m^{\prime}=m~~~~~~~~~~~~~~~~~~~~~~~~~~~~~~~~~~~~~~~~~~~~~~~~~~~~~~~~~~~~~~~~~~~~~~~~~~~~~~~\\\nonumber
 p^{\prime}=-\omega k ~\sin \omega t+p~\cos \omega t-m (v ~\cos \omega t+\omega x ~\sin \omega t) \\
 k^{\prime}=k ~\cos \omega+\frac{p}{\omega}~\sin\omega t+m( x ~\cos\omega t-\frac{v}{\omega}~ \sin \omega t)~~~~~~ \\\nonumber
 E^{\prime}=E-pv+\frac{mv^2}{2}+\frac{m\omega^2 x^2}{2}+\omega^2 k x~~~~~~~~~~~~~~~~~~~~~~~~~~~~\\\nonumber
\end{eqnarray}
in the $ANH_{-}(1)$ case and
\begin{eqnarray}\label{coadjnh1plus}\nonumber
 m^{\prime}=m~~~~~~~~~~~~~~~~~~~~~~~~~~~~~~~~~~~~~~~~~~~~~~~~~~~~~~~~~~~~~~~~~~~~~~~~~~~~~~~~~~~~~~\\\nonumber
 p^{\prime}=\omega k~ \sinh \omega t+p~\cosh \omega t-m (v~ \cosh \omega t- \omega x~ \sinh \omega t) \\
 k^{\prime}=k~ \cosh \omega t+\frac{p}{\omega}~\sinh \omega t-m (x~ \cosh\omega t+\frac{v}{\omega}~\sinh \omega t)~~~\\\nonumber
 E^{\prime}=E-pv+\frac{mv^2}{2}-\frac{m\omega^2 x^2}{2}-\omega^2 k x~~~~~~~~~~~~~~~~~~~~~~~~~~~~~~~~~~\\\nonumber
\end{eqnarray}
in the $ANH_{+}(1)$ case.\\

The parameter $m$ in (\ref{coadjnh1moins}) and (\ref{coadjnh1plus}) corresponding to the central generator remain fixed
(i.e is $Ad^*$-invariant): it is called trivial invariant
of the coadjoint actions of the anisotropic Newton-Hooke groups $ANH_{\pm}(1)$ on the dual of their centrally extended Lie algebras.\\

 The Kirillov form in the basis
$(K,P,H)$ is
 \begin{eqnarray}
 (K_{ij}(k,p,E))=\left(
 \begin{array}{cccc}
 0&m&p\\
 -m&0&\pm \omega^2k\\
 -p&\mp\omega^2 k&0\\
 \end{array}
 \right)
 \end{eqnarray}
 We verify that the other invariant, solution of the Kirillov's system (\ref{kirillovsystem}) is in this case,
\begin{eqnarray}\label{internalenergy}
U_{\pm}=E-\frac{p^2}{2m}\pm \frac{m\omega^2q^2}{2}
\end{eqnarray}
where $q=\frac{k}{m}$.
Note that (\ref{internalenergy}) can be written as:
\begin{eqnarray}
 E=\frac{p^2}{2m}\mp\frac{m\omega^2q^2}{2}+U_{\pm}
\end{eqnarray}
meaning that the total energy $E$ is the sum of the kinetic energy $\frac{p^2}{2m}$, the potential energy $\frac{m\omega^2q^2}{2}$
and the internal energy $U_{-}$ of the oscillating system in the case of $ANH_{-}(1)$ or $U_{+}$ in the case of the expanding system
with $ANH_{+}(1)$.\\

The inverse of the restriction of the Kirillov form to the orbit is in both cases:
 \begin{eqnarray}
 \Omega^{-1}=\frac{1}{m}\left(
 \begin{array}{cc}
 0&-1\\
 1&0\\
 \end{array}
 \right)
 \end{eqnarray}
 Then, the symplectic two-form (\ref{symplectictwo-form}) takes the form (\ref{symplcanonical}) in canonical
 coordinates (where we have used the notation $q=\frac{k}{m}$).\\

 Denote the two-dimensional orbit by
${\cal{O}}_{(m,U_{\pm})}$.\;\;It is a symplectic manifold endowed with the canonical symplectic two-form (\ref{symplcanonical}) and
therefore not interesting
for our study because there are one momentum and one position. \\

Note that the symplectic realizations of $ANH_{-}(1)$ and $ANH_{+}(1)$ are
 respectively given by
\begin{eqnarray}
 L_{(v,x,t)}(p,q)&=& (p~\cos \omega t - m\omega q~\sin \omega t - mv~\cos \omega t , \crcr
&&\frac{p}{m\omega}~\sin \omega t + (q + x) ~\cos \omega t  -\frac{v}{\omega}~\sin \omega t)
\end{eqnarray}
and
\begin{eqnarray}
L_{(v,x,t)}(p,q)= (p~\cosh \omega t + m\omega q~\sinh \omega t -m(v~\cosh \omega t \nonumber \\-\omega x~\sinh \omega t) ,
\frac{p}{m\omega}~\sinh \omega t + (q+x)~\cosh \omega  t -\frac{v}{\omega}~\sinh \omega t)
\end{eqnarray}
Let
 $(p(t),q(t))=L_{(0,0,t)}(p,q) $, it follows that
\begin{eqnarray}
 p(t) = p~\cos \omega t  -m\omega q~\sin \omega t~~ ,~~ q(t) =\frac{p}{m\omega}~ \sin \omega t + q~\cos \omega t
\end{eqnarray}
\mbox{for $ANH_{-}(1)$ and}
\begin{eqnarray}
 p(t) =  p~\cosh \omega t + m\omega q~\sinh \omega t ~~ ,~~ q(t) = \frac{p}{m\omega} ~\sinh \omega t+ q~\cosh \omega t
\end{eqnarray}
for $ANH_{+}(1)$. \\

The equations of motion are then given by
\begin{eqnarray}
 \frac{dp}{dt} = \pm m\omega^{2}q ~~,~~ \frac{dq}{dt} = \frac{p}{m}
\end{eqnarray}
or equivalently  $ \frac{d^{2}q}{dt^{2}} = \pm\omega^{2}q $; which is a second order differential equation whose
solutions are trigonometric functions for $ANH_{-}(1)$ case and
hyperbolic ones in $ANH_{+}(1)$ case.\;\;It is for this reason that
$ANH_{-}$ describes a universe in oscillation while $ANH_{+}$
describes a universe in expansion.

\subsubsection{ii) Two-dimensional spaces case}
 Note that the nontrivial
Lie brackets of the centrally extended Newton-Hooke Lie algebras $\cal{NH}_{\pm}$(rotation included) in two-dimensional
space are given by
\begin{eqnarray}
[J,K_i]=K_j\epsilon^j_i~,~[J,P_i]=P_j\epsilon^j_i~~~~~~~~~~~~~~~~~~~~~~~~~~~~~\crcr
[K_i,P_j]=M\delta_{ij}~,~[K_i,H]=P_i~,~[P_i,H]=\pm\omega^2 K_i
\end{eqnarray}
which means that the generators of space translations as well as pure Newton-Hooke transformations commute.\;\;One can not then
associate a noncommutative phase space to the Newton-Hooke group.\;\;It is then the absence of the symmetry rotations (anisotropy of the
plane) which guaranties the noncommutative phase space for the anisotropic Newton-Hooke group.\\

In this section, we prove that, with the anisotropic Newton-Hooke groups $ANH_{\pm}(2)$ in two-dimensional spaces, the phase spaces
obtained by the coadjoint
orbit method are completely noncommutative (i.e momenta as well as positions of the phase spaces do not commute).\;\;This is due
to the fact that both generators of the
pure kinematical (Newton-Hooke) transformations as well as generators of spatial transformations do not commute
in the centrally extended Lie algebras. \\

Indeed, let $g$ be given by (\ref{elementgnh}) (i.e $g=(\vec{x},\vec{v},t$) for the short) and \\
${\hat g}=\exp(\varphi M+ \psi S)g$ be the corresponding element in
the connected Lie groups associated to
the extended Lie algebra ${\cal{ANH}}_{\pm}(2)$ in two-dimensional space.\;\;By use of the Baker-Campbell-Hausdorff
formulas \cite{hall} and by identifying ${\hat g}$ with
$(\varphi+ \psi, g)$, we find that the multiplication laws of the connected extended Lie groups $ANH_{\pm}(2)$
 are respectively :
 \begin{eqnarray*}
 (\varphi+\psi, g)(\varphi^{~\prime}+\psi^{~\prime}, g^{~\prime})=
(\varphi+\varphi^{~\prime}+ \psi + \psi^{\prime}+c(g,g^{~\prime}),~ g g^{~\prime})
 \end{eqnarray*}
 where
\begin{eqnarray}\nonumber
  c(g,g^{~\prime})=\frac{1}{2}[\delta_{ij} (v^i ~\cos \omega t^{\prime}-\omega x^i~\sin \omega t^{\prime})x^{j\prime}
-(x^i ~\cos \omega t^{\prime}+\frac{v^i}{\omega }~ \sin \omega t^{\prime})v^{j\prime}]\\\nonumber+\frac{1}{2}\kappa^2\epsilon_{ij}(
x^i~ \cos \omega t^{\prime}+\frac{v^i}{\omega }~\sin \omega t^{\prime})x^{j\prime}+
\frac{1}{2c^2}\epsilon_{ij}(v^i~ \cos \omega t^{\prime}-\omega x^i~\sin \omega t^{\prime})v^{j\prime}
 \end{eqnarray}
 for $ANH_{-}(2)$
 and
 \begin{eqnarray}\nonumber
 c(g,g^{~\prime})=\frac{1}{2}[\delta_{ij} (v^i~ \cosh \omega t^{\prime}+\omega x^i~\sinh \omega t^{\prime})x^{j\prime}
-(x^i~ \cosh \omega t^{\prime}+\frac{v^i}{\omega }~ \sinh \omega t^{\prime})v^{j\prime}]\\\nonumber+\frac{1}{2}\kappa^2\epsilon_{ij}(
x^i~ \cosh \omega t^{\prime}+\frac{v^i}{\omega }~\sinh \omega t^{\prime})x^{j\prime}+
\frac{1}{2c^2}\epsilon_{ij}(v^i ~\cosh \omega t^{\prime}+\omega x^i~\sinh \omega t^{\prime})v^{j\prime}
 \end{eqnarray}
 for $ANH_{+}(2)$\\
are the two-cocycles defining the centrally extended anisotropic Newton-Hooke groups in two-dimensional space
while $ g g^{~\prime}$ is the multiplication law (\ref{newton-hookelawmoins}) in the $ANH_{-}(2)$ case or
 (\ref{newton-hookelawplus} in the $ANH_{+}(2)$ case respectively.\\

The adjoint actions $Ad_g(\delta{\hat{g}})=g( \delta{\hat{g}} )g^{-1}$ of $ANH_{\pm}(2)$ on the Lie algebras
${\hat{\cal{ANH}}}_{\pm}(2)$
are given by:
 \begin{eqnarray}\label{adjointnh2}\nonumber
 Ad_{(\vec{v},\vec{x},t)}(\delta \varphi,\delta \psi,\delta \vec{x},\delta \vec{v}, \delta t)=(\delta \varphi^{~\prime},
 \delta \vec{x}^{~\prime},\delta \vec{v}^{~\prime}, \delta t^{~\prime})
  \end{eqnarray}
with
 \begin{eqnarray}\nonumber
 \delta \varphi^{\prime}= \delta \varphi+ \delta_{ij}v^i~ \delta x^j- \delta _{ij}x^i~ \delta v^j ~~~~~~~~~~~~~~~~~~~~~~~~~~~~~~~~~~~~~~
 ~~~~~~~~~~~\\\nonumber
 \delta \psi ^{\prime}=\delta \psi+\frac{\epsilon_{ij}}{2c^2}v^i ~\delta v^j+\frac{\epsilon_{ij}}{2}\kappa^2x^i ~\delta x^j-\kappa^2
 \epsilon _{ij}v^i x^j~\delta t~~~~~~~~~~~~~~~~~~~~ \\
 \delta x^{i~\prime}= \cos \omega t~ \delta x^i- \frac{1}{\omega}~\sin \omega t ~\delta v^i-\omega(x^i ~\cos \omega t -\frac{v^i}{\omega}
 ~ \sin \omega t)~\delta t\\\nonumber
 \delta v^{i~\prime}=\cos \omega t~ \delta v^i+\omega~ \sin \omega t~ \delta x^i-\omega(v^i ~\cos \omega t+ \omega x^i~ \sin \omega t)
  ~\delta t\\\nonumber
  \delta t^{\prime}=\delta t~~~~~~~~~~~~~~~~~~~~~~~~~~~~~~~~~~~~~~~~~~~~~~~~~~~~~~~~~~~~~~~~~~~~~~~~~~~~~~~~~~~~~~~~~~~~~\nonumber
\end{eqnarray}
 in the case of $ANH_{-}(2)$ and
 \begin{eqnarray}\nonumber
 \delta \varphi^{\prime}= \delta \varphi+ \delta_{ij}v^i ~\delta x^j- \delta _{ij}x^i~ \delta v^j~~~~~~~~~~~~~~~~~~~
 ~~~~~~~~~~~~~~~~~~~~~~~~~~~~~~~~~~~~~~~~\\\nonumber
 \delta \psi ^{\prime}=\delta \psi+\frac{\epsilon_{ij}}{2c^2}v^i ~\delta v^j+\frac{\epsilon_{ij}}{2}\kappa^2x^i~ \delta x^j-\kappa^2
 \epsilon _{ij}v^i x^j~\delta t ~~~~~~~~~~~~~~~~~~~~~~~~~~~~~\\
 \delta x^{i~\prime}= \cosh \omega t~ \delta x^i- \frac{1}{\omega}~\sinh \omega t~ \delta v^i+\omega(x^i~ \cosh \omega t
 -\frac{v^i}{\omega}
  ~\sinh \omega t)~\delta t\\\nonumber
 \delta v^{i~\prime}=\cosh \omega t~ \delta v^i-\omega ~\sinh \omega t ~\delta x^i+\omega(v^i ~\cosh \omega t-
 \omega x^i~ \sinh \omega t)
 ~ \delta t\\\nonumber
  \delta t^{\prime}=\delta t~~~~~~~~~~~~~~~~~~~~~~~~~~~~~~~~~~~~~~~~~~~~~~~~~~~~~~~~~~~~~~~~~~~~~~~~~~~~~~~~~~~~~
  ~~~~~~~~~~~~~~~\nonumber
\end{eqnarray}
 in the case of $ANH_{+}(2)$.\\

 If the duality between the extended Lie algebras and their duals is given by the action \\
 $\vec{p}.\delta \vec{x}+\vec{k}.\delta \vec{v}+m .\delta\varphi+ h.\delta \psi +E.\delta t$,
 where the linear momentum $\vec{p}$, the static momentum $\vec{k}$, the mass $m$, the action $h$ and the energy $E$
 are elements of the dual Lie algebras, then the coadjoint actions of the anisotropic Newton-Hooke groups $ANH_{\pm}(2)$
 on the dual of their centrally extended Lie algebras are
\begin{eqnarray*}\label{coadjointaction2}
 Ad^*_{(x^i, v^i,t)}(m,h,p^i,k^i,E)=(m^{\prime},h^{\prime}, p^{i~\prime}, k^{i~\prime}, E^{\prime})
\end{eqnarray*}
 with
\begin{eqnarray}\label{coadjnh2moins}\nonumber
m^{\prime}=m~~~~~~~~~~~~~~~~~~~~~~~~~~~~~~~~~~~~~~~~~~~~~~~~~~~~~~~~~~~~~~~~~~~~~~~~~~~~~~~~~~~~~~~~~\\\nonumber
h^{\prime}=h~~~~~~~~~~~~~~~~~~~~~~~~~~~~~~~~~~~~~~~~~~~~~~~~~~~~~~~~~~~~~~~~~~~~~~~~~~~~~~~~~~~~~~~~~~~\\\nonumber
p^{\prime}_i=p_i~\cos \omega t-\omega k_i ~\sin \omega t-m \delta_{ij}(v^j~ \cos \omega t+\omega x^j~ \sin \omega t)+\\
 \frac{h\omega }{2c^2}\epsilon_{ij}(\omega x^j~\cos \omega t+v^j~\sin \omega t)~~~~~~~~~~~~~~~~~~~~~\\\nonumber
k^{\prime}_i=k_i~ \cos \omega t+\frac{p_i}{\omega}~\sin\omega t+m\delta_{ij}( x^j~ \cos\omega t-\frac{v^j}{\omega}~
\sin \omega t)+\\\nonumber
  \frac{h}{2}\epsilon_{ij}(\frac{1}{\omega}\kappa^2 x^j\sin \omega t-\frac{1}{c^2}v^j~\cos \omega t)~~~~~~~~~~~~\\\nonumber
E^{\prime}=E+\omega p_ix^i+\omega k_i v^i -h\kappa^2 \epsilon_{ij}v^ix^j~~~~~~~~~~~~~~~~~~~~~~~~~~~~~~~~~~~~~~~\nonumber
\end{eqnarray}
in the $ANH_{-}(2)$ case and
\begin{eqnarray}\label{coadjnh2plus}\nonumber
m^{\prime}=m~~~~~~~~~~~~~~~~~~~~~~~~~~~~~~~~~~~~~~~~~~~~~~~~~~~~~~~~~~~~~~~~~~~~~~~~~~~~~~~~~~~~~~~~~~~~~~~\\\nonumber
h^{\prime}=h~~~~~~~~~~~~~~~~~~~~~~~~~~~~~~~~~~~~~~~~~~~~~~~~~~~~~~~~~~~~~~~~~~~~~~~~~~~~~~~~~~~~~~~~~~~~~~~~\\\nonumber
p^{\prime}_i=p_i~\cosh \omega t+\omega k_i~ \sinh \omega t-m \delta_{ij}(v^j \cosh ~\omega t-\omega x^j~ \sinh \omega t)\\
-\frac{h\omega }{2c^2}\epsilon_{ij}(\omega x^j~\cosh \omega t+v^j~\sinh \omega t)\\\nonumber
k^{\prime}_i=k_i~ \cosh \omega t+\frac{p_i}{\omega}~\sinh \omega t+m\delta_{ij}( x^j~ \cosh\omega t-\frac{v^j}{\omega}~
\sinh \omega t)\\\nonumber-\frac{h}{c^2}\epsilon_{ij}(x^j~\sinh \omega t+\frac{1}{v^j}~\cosh \omega t)\\\nonumber
E^{\prime}=E-\omega p_ix^i-\omega k_i v^i +h\kappa^2 \epsilon_{ij}v^ix^j~~~~~~~~~~~~~~~~~~~~~~~~~~~~~~~~~~~~~~~\\\nonumber
\end{eqnarray}
in the $ANH_{+}(2)$ case.\\

The parameters $m$ and $h$ in (\ref{coadjnh2moins}) and (\ref{coadjnh2plus}) corresponding to the central generators
remain fixed.\;\;They are trivial invariants
under the coadjoint action of $ANH_{\pm}(2)$ on the dual of their centrally extended Lie algebras
in two-dimensional spaces. \\

The Kirillov form in the basis
$(K_i,P_i,H)$ is
 \begin{eqnarray}
 (K_{ij}(k_i,p_i,E,m,h))=\left(
 \begin{array}{ccccc}
 0&\frac{h}{c^2}&m&0&p_1\\
 -\frac{h}{c^2}&0&0&m&p_2\\
 -m&0&0&\pm h\kappa^2&\pm \omega^2 k_1\\
 0&-m&\mp h \kappa^2&0&\pm \omega^2k_2\\
 -p_1&-p_2&\mp \omega^2 k_1&\mp \omega^2k_2&0
 \end{array}
 \right)
 \end{eqnarray}
The other invariant, solution of the Kirillov's
system (\ref{kirillovsystem}) is explicitly given by
\begin{eqnarray}\label{anisotwoinvariant}
U=E-\frac{\vec{p}^{~2}}{2\mu_e}\pm\frac{\mu_e\omega^2\vec{q}^{~2}}{2}
\end{eqnarray}
with
\begin{eqnarray}\label{effectivemass}
\mu_e=m\pm \frac{h\kappa^2}{\omega }~~,~~\vec{q}=\frac{\vec{k}}{\mu_e}
\end{eqnarray}
where $h\omega_0=mc^2$ denotes the wave-particle duality, $\mu_e$ being an effective mass.\\

The inverse of the restriction of the Kirillov's matrix on the orbit is given by
\begin{eqnarray}\label{kirillovnewtonhooke2a}
\Omega^{-1}=\left(
\begin{array}{cccc}
0& \pm
\frac{\omega}{\mu_e}&-\frac{1}{\mu_e}&0\\\mp\frac{\omega}{\mu_e
}&0&0&-\frac{1}{\mu_e}\\\frac{1}{\mu_e}&0&0&\frac{1}{\mu_e\omega_0}\\0&\frac{1}{\mu_e}&-\frac{1}{\mu_e \omega_0}&0
\end{array}
\right)
\end{eqnarray}
where we have used the wave-particle duality and (\ref{effectivemass}).\\

The orbit is then equipped with the symplectic form
\begin{eqnarray}\label{symplnh}
\sigma=dp_i\wedge
dq^i+\frac{1}{\mu_e\omega_0}\epsilon^{ij}dp_i\wedge dp_j\pm
\mu_e\omega \epsilon_{ij} dq^i\wedge dq^j
\end{eqnarray}
The Poisson brackets corresponding to the above symplectic structure are given by:
\begin{eqnarray}
\{h,f\}=\frac{\partial h}{\partial p_i}\frac{\partial f}{\partial
q^i}-\frac{\partial h}{\partial q^i}\frac{\partial f}{\partial
p_i}+G^{ij}\frac{\partial h}{\partial q^i}\frac{\partial f}{\partial
q^j}+F_{ij}\frac{\partial h}{\partial p_i}\frac{\partial f}{\partial
p_j}~~;~~i,j=1,2
\end{eqnarray}
This implies that
\begin{eqnarray}
\{p_i,p_j\}=F_{ij}~,~\{p_i,q^j\}=\delta^j_i~,~\{q^i,q^j\}=G^{ij}
\end{eqnarray}
with
\begin{eqnarray}
G^{ij}=-\frac{\epsilon^{ij}}{m\omega_0}~~,~~ F_{ij}=-(m-\mu_{e})\omega\epsilon_{ij}
\end{eqnarray}
 It follows that the magnetic field $B$ and its dual field $B^*$ are such that
\begin{eqnarray}
e^*B^*=-\frac{1}{m\omega_0}~~,~~eB=(m-\mu_e)\omega
\end{eqnarray}
The effective mass is then given in function of the magnetic field
by
\begin{eqnarray}
\mu_e=m-\frac{eB}{\omega}
\end{eqnarray}
The Hamilton's equations are then
\begin{eqnarray}\label{hamiltonequationtwofields}
\frac{d\pi_i}{dt}=-\frac{\partial H}{\partial q^i}\pm
(m-\mu_e)\omega\epsilon_{ik}\frac{\partial H}{\partial
p_k}~~,~~\frac{dx^i}{dt}=\frac{\partial H}{\partial
p_i}+\frac{\epsilon^{ik}}{2m\omega_0}\frac{\partial H}{\partial q^k}
\end{eqnarray}
As already argued, similar results (for the oscillating case) have been found in \cite{zhang-horvathy} in the so-called Hill
problem, which is effectively an anisotropic harmonic oscillator in a magnetic field.\;\;This
 system has no rotational symmetry while translations and generalized boosts still act as symmetries.\;\;It also has a
two-parameter central extension and has been applied to the $3$-body problem and galactic dynamics.
Furthermore, the noncommutative version of the Hill problem was discussed in \cite{zhang-horvathy1}.
\subsubsection{iii) Three-dimensional spaces case}
Similarly, let $g$ be given by (\ref{elementgnh}) (i.e $g=(\vec{x},\vec{v},t$) for the short) and \\
${\hat g}=\exp(\varphi M+ \theta^i J_i)g$ be the corresponding element in
the connected Lie groups associated to
the extended Lie algebra ${\cal{ANH}}_{\pm}(3)$ in three-dimensional space.\;\;By use of the Baker-Campbell-Hausdorff
formulas \cite{hall} and by identifying ${\hat g}$ with
$(\varphi+ \theta^i, g)$, we find that the multiplication laws of the connected extended Lie groups $ANH_{\pm}(3)$
 are respectively :
 \begin{eqnarray*}
 (\varphi+\theta^i, g)(\varphi^{~\prime}+\theta^{i~\prime}, g^{~\prime})=
(\varphi+\varphi^{~\prime}+ \theta^i + \theta^{i~\prime}+c(g,g^{~\prime}), ~g g^{~\prime})
 \end{eqnarray*}
 where
\begin{eqnarray}\nonumber
  c(g,g^{~\prime})=\frac{\delta_{ij}}{2}[ (v^i~ \cos \omega t^{\prime}-\omega x^i~\sin \omega t^{\prime})x^{j~\prime}
-(x^i~ \cos \omega t^{\prime}+\frac{v^i}{\omega }~ \sin \omega t^{\prime})v^{j\prime}]\\\nonumber+\epsilon_{ij}^k[(
x^i~ \cos \omega t^{\prime}+\frac{v^i}{\omega }~\sin \omega t^{\prime})\frac{\kappa^2 x^{j~\prime}}{2}+
(v^i ~\cos \omega t^{\prime}-\omega x^i~\sin \omega t^{\prime})\frac{v^{j~\prime}}{2c^2}]
 \end{eqnarray}
 for $ANH_{-}(3)$\\
and
\begin{eqnarray}\nonumber
  c(g,g^{~\prime})=\frac{\delta_{ij}}{2}[ (v^i~ \cosh \omega t^{\prime}-\omega x^i~\sinh \omega t^{\prime})x^{j~\prime}
-(x^i~ \cosh \omega t^{\prime}+\frac{v^i}{\omega }~ \sinh \omega t^{\prime})v^{j~\prime}]\\\nonumber+\epsilon_{ij}^k[(
x^i~ \cosh \omega t^{\prime}+\frac{v^i}{\omega }~\sinh \omega t^{\prime})\frac{\kappa^2 x^{j~\prime}}{2}+
(v^i~ \cosh \omega t^{\prime}+\omega x^i~\sinh \omega t^{\prime})\frac{v^{j~\prime}}{2c^2}]
 \end{eqnarray}
 for $ANH_{+}(3)$\\
are the two-cocycles defining the centrally extended anisotropic Newton-Hooke groups in three-dimensional space
while $ g g^{~\prime}$ is the multiplication law (\ref{newton-hookelawmoins}) in the $ANH_{-}(3)$ case
or (\ref{newton-hookelawplus} in the $ANH_{+}(3)$ case respectively.\\

The adjoint actions $Ad_g(\delta{\hat{g}})=g( \delta{\hat{g}} )g^{-1}$ of $ANH_{\pm}(3)$ on the Lie algebras
${\hat{\cal{ANH}}}_{\pm}(3)$
are given by:
 \begin{eqnarray}\label{adjointnh3}\nonumber
 Ad_{(\vec{v},\vec{x},t)}(\delta \varphi,\delta \theta^i,\delta \vec{x},\delta \vec{v}, \delta t)=(\delta \varphi^{~\prime},
 \delta \vec{x}^{~\prime},\delta \vec{v}^{~\prime}, \delta t^{~\prime})
  \end{eqnarray}
with
 \begin{eqnarray}\nonumber
 \delta \varphi^{\prime}= \delta \varphi+ \delta_{ij}(v^i~ \delta x^j-x^i~ \delta v^j) ~~~~~~~~~~~~~~~~~~~~~~~~~~~~~~~~~~~~~~~~~
 ~~~~~~~\\\nonumber
 \delta \theta ^{i~\prime}=\delta \theta^i+\epsilon_{ij}^k(\frac{v^i}{2c^2} ~\delta v^j-\frac{\kappa^2}{2}{x^i} ~\delta x^j)+
 \epsilon _{ij}^k\kappa^2 v^i x^j~\delta t~~~~~~~~~~~~~~~~ \\
 \delta x^{i~\prime}= \cos \omega t~ \delta x^i- \frac{1}{\omega}\sin \omega t ~\delta v^i-\omega(x^i \cos \omega t -\frac{v^i}{\omega}
  \sin \omega t)~\delta t\\\nonumber
 \delta v^{i~\prime}=\cos \omega t~ \delta v^i+\omega \sin \omega t~ \delta x^i-\omega(v^i \cos \omega t+ \omega x^i \sin \omega t)
  ~\delta t\\\nonumber
  \delta t^{\prime}=\delta t~~~~~~~~~~~~~~~~~~~~~~~~~~~~~~~~~~~~~~~~~~~~~~~~~~~~~~~~~~~~~~~~~~~~~~~~~~~~~~~~~~~~~~~\nonumber
\end{eqnarray}
 in the case of $ANH_{-}(3)$ and
 \begin{eqnarray}\nonumber
 \delta \varphi^{\prime}= \delta \varphi+ \delta_{ij}(v^i~ \delta x^j-x^i~ \delta v^j) ~~~~~~~~~~~~~~~~~~~~~~~~~~~~~~~~~~~~~~~~~~~
 ~~~~~~~~~~~~~~~\\\nonumber
 \delta \theta ^{i~\prime}=\delta \theta^i+\epsilon_{ij}^k(\frac{v^i}{2c^2} ~\delta v^j+\frac{\kappa^2}{2}{x^i} ~\delta x^j)-
 \epsilon _{ij}^k\kappa^2 v^i x^j~\delta t~~~~~~~~~~~~~~~~~~~~~~~~~~~ \\
 \delta x^{i~\prime}= \cosh \omega t~ \delta x^i- \frac{1}{\omega}\sinh \omega t ~\delta v^i+(\omega x^i \cosh \omega t -
 v^i \sinh \omega t)~\delta t~~\\\nonumber
 \delta v^{i~\prime}=\cosh \omega t~ \delta v^i-\omega \sinh \omega t~ \delta x^i+\omega(v^i \cosh \omega t- \omega x^i \sinh \omega t)
  ~\delta t\\\nonumber
  \delta t^{\prime}=\delta t~~~~~~~~~~~~~~~~~~~~~~~~~~~~~~~~~~~~~~~~~~~~~~~~~~~~~~~~~~~~~~~~~~~~~~~~~~~~~~~~~~~~~~~~~~~~~~~~~\nonumber
\end{eqnarray}
 in the case of $ANH_{+}(3)$.\\

 If the duality between the extended Lie algebras and their duals is given by the action:
 \begin{eqnarray*}
  \langle(m,h_i,k_i,p_i,E),( \delta \varphi, \delta \theta^i,\delta v^i,\delta x^i,dt)\rangle=m .\delta\varphi+ h.\delta \theta^i+
 k_i.\delta v^i+p_i.\delta x^i+E.\delta t
  \end{eqnarray*}
 where the linear momentum $\vec{p}$, the static momentum $\vec{k}$, the mass $m$, the action $h$ and the energy $E$
 are elements of the dual Lie algebras, then the coadjoint actions of the anisotropic Newton-Hooke groups $ANH_{\pm}(3)$
 on the dual of their centrally extended Lie algebras are
\begin{eqnarray*}\label{coadjointaction3}
 Ad^*_{(x^i, v^i,t)}(m,h,p_i,k_i,E)=(m^{\prime},h^{\prime}, p_{i}^{\prime}, k_{i}^{\prime}, E^{\prime})
\end{eqnarray*}
 with
\begin{eqnarray}\label{coadjnh3moins}\nonumber
m^{\prime}=m~~~~~~~~~~~~~~~~~~~~~~~~~~~~~~~~~~~~~~~~~~~~~~~~~~~~~~~~~~~~~~~~~~~~~~~~~~~~~~~~~~~~~~~~~\\\nonumber
h^{\prime}=h~~~~~~~~~~~~~~~~~~~~~~~~~~~~~~~~~~~~~~~~~~~~~~~~~~~~~~~~~~~~~~~~~~~~~~~~~~~~~~~~~~~~~~~~~~~~\\\nonumber
p^{\prime}_i=p_i~\cos \omega t-\omega k_i ~\sin \omega t-m \delta_{ij}(v^j~ \cos \omega t+\omega x^j~ \sin \omega t)~~~~\\
 +\frac{h_k\omega }{2c^2}\epsilon_{ij}^k(\omega x^j~\cos \omega t+v^j~\sin \omega t)~~~~~~~~~~~~~~~~~~~~~\\\nonumber
k^{\prime}_i=k_i~ \cos \omega t+\frac{p_i}{\omega}~\sin\omega t+m\delta_{ij}( x^j~ \cos\omega t-\frac{v^j}{\omega}~
\sin \omega t)~~~~~\\\nonumber
  +\frac{h_k}{2}\epsilon_{ij}^k(\frac{1}{\omega}\kappa^2 x^j~\sin \omega t-\frac{1}{c^2}v^j~\cos \omega t)~~~~~~~~~\\\nonumber
E^{\prime}=E+\omega p_ix^i+\omega k_i v^i -\kappa^2 h_k\epsilon_{ij}^kv^ix^j~~~~~~~~~~~~~~~~~~~~~~~~~~~~~~~~~~~~\nonumber
\end{eqnarray}
in the $ANH_{-}(3)$ case and
\begin{eqnarray}\label{coadjnh3plus}\nonumber
m^{\prime}=m~~~~~~~~~~~~~~~~~~~~~~~~~~~~~~~~~~~~~~~~~~~~~~~~~~~~~~~~~~~~~~~~~~~~~~~~~~~~~~~~~~~~~~~~~~~~~~~~~\\\nonumber
h^{\prime}=h~~~~~~~~~~~~~~~~~~~~~~~~~~~~~~~~~~~~~~~~~~~~~~~~~~~~~~~~~~~~~~~~~~~~~~~~~~~~~~~~~~~~~~~~~~~~~~~~~~\\\nonumber
p^{\prime}_i=p_i~\cosh \omega t+\omega k_i~ \sinh \omega t-m \delta_{ij}(v^j ~\cosh \omega t-\omega x^j~ \sinh \omega t)~~~\\
-\frac{h_k\omega }{2c^2}\epsilon_{ij}^k(\omega x^j~\cosh \omega t+v^j~\sinh \omega t)\\\nonumber
k^{\prime}_i=k_i ~\cosh \omega t+\frac{p_i}{\omega}~\sinh \omega t+m\delta_{ij}( x^j ~\cosh\omega t-\frac{v^j}{\omega}~
\sinh \omega t)~~~\\\nonumber-\frac{h_k}{2c^2}\epsilon_{ij}^k(x^j~\sinh \omega t+\frac{1}{v^j}~\cosh \omega t)\\\nonumber
E^{\prime}=E-\omega p_ix^i-\omega k_i v^i +\kappa^2 h_k \epsilon_{ij}^kv^ix^j~~~~~~~~~~~~~~~~~~~~~~~~~~~~~~~~~~~~~~~~~\\\nonumber
\end{eqnarray}
in the $ANH_{+}(3)$ case.\\

The parameters $m$ and $h_k$ in (\ref{coadjnh3moins}) and (\ref{coadjnh3plus}) corresponding to the central
generators remain fixed.\;\;They are trivial invariants under the coadjoint
action of $ANH_{\pm}(3)$ on the dual of their centrally extended Lie algebras
in three-dimensional spaces. \\

The Kirillov form in the basis
$(K_i,P_i,H)$ is
 \begin{eqnarray}
 (K_{ij}(k_i,p_i,E,m,h_k))=\left(
 \begin{array}{ccc}
 \frac{h_k}{c^2}\epsilon_{ij}^k&m\delta_{ij}&p_i\\
-m\delta_{ij}&\pm\kappa^2 h_k \epsilon_{ij}^k&\pm \omega^2k_i\\
 -p_j&\mp \omega^2k_j&0\\
 \end{array}
 \right)
 \end{eqnarray}
 There is an other invariant.
Assuming that $\frac{\partial U}{\partial E}=1$, the latter is solution of the Kirillov's
system
\begin{eqnarray}\label{invsystemnh3}
\Big \{
\begin{array}{rl}
 A_{ij}\frac{\partial U}{\partial k_j}+\delta_{ij}\frac{\partial U}{\partial p_j}=-\frac{p_i}{m}~~~~~~~~~~~~~~~~~\\
-\delta_{ij}\frac{\partial U}{\partial k_j}+B_{ij}\frac{\partial U}{\partial p_j}=\mp\frac{\omega^2 k_i}{m}~~~~~~~~\\
\end{array}
\end{eqnarray}
where $$A_{ij}=\frac{h_k\epsilon_{ij}^k}{mc^2},~~~B_{ij}=\pm \kappa^2\frac{h_k\epsilon_{ij}^k}{m}$$
Using the relation $c=\frac{\omega}{\kappa}$ in the above relations, we obtain
$$ B_{ij}=\pm \omega^2 A_{ij}$$
Then the system (\ref{invsystemnh3}) takes the following form:
\begin{eqnarray}
\Big \{
\begin{array}{rl}
 A_{ij}\frac{\partial U}{\partial k_j}+\delta_{ij}\frac{\partial U}{\partial p_j}=-\frac{p_i}{m}~~~~~~~~~~~~~~~~~\\
-\delta_{ij}\frac{\partial U}{\partial k_j}\pm \omega^2 A_{ij}\frac{\partial U}{\partial p_j}=\mp\frac{\omega^2 k_i}{m}~~~~~~\\
\end{array}
\end{eqnarray}
or equivalently
\begin{eqnarray}
 \Big \{
\begin{array}{rl}
 A\frac{\partial U}{\partial \vec{k}}+\frac{\partial U}{\partial \vec{p}}=-\frac{\vec{p}}{m}~~~~~~~~~~~~~~~~~\\
-\frac{\partial U}{\partial \vec{k}}\pm \omega^2 A\frac{\partial U}{\partial \vec{p}}=\mp\frac{\omega^2 \vec{k}}{m}~~~~~~~~~~~~\\
\end{array}
\end{eqnarray}
Moreover,
\begin{eqnarray}
\Big \{
\begin{array}{rl}
 (I\pm \omega^2A^2)\frac{\partial U}{\partial \vec{p}}=-\frac{\vec{p}}{m}\mp A \omega^2 \frac{\vec{k}}
 {m}~~~~~~~~~~~~~~~~~\\
-(I\pm \omega^2A^2)\frac{\partial U}{\partial \vec{k}}=\mp\frac{\omega^2 \vec{k}}{m}\pm \omega^2 A \frac{\vec{p}}{m}~~~~~~~~~~~~~~~~\\
\end{array}
\end{eqnarray}
The solution of the above system is
\begin{eqnarray}
U=E-\frac{p_ip_j(\Phi_{\pm}^{-1})^{ij}}{2m}-\frac{m\omega^2q^iq^j(\Phi_{\pm}^{-1})_{ij}}{2}+\omega^2p_iq^j(\Phi_{\pm}^{-1}A)^i_j
\end{eqnarray}
where
\begin{eqnarray}
\Phi_{\pm}=I\pm \omega^2 A^2 \
\end{eqnarray}
and
\begin{eqnarray}\label{q^ik^i m}
 q_i=\frac{k_i}{m}
\end{eqnarray}
We see that $\Phi_{\pm}$ is a metric for ${\mathbb R}^3$.\;\;Note also that if $\Phi_{\pm}=I$ or equivalently if $A=0$, then
the nontrivial invariant takes the
form:
\begin{eqnarray}
 U=E-\frac{\vec{p}^{~2}}{2m}-\frac{m \omega^2\vec{q}^{~2}}{2}
\end{eqnarray}
 and is interpreted as the internal energy of a free oscillator. \\

 Let us denote the maximal coadjoint orbit by ${\cal{O}_{(m,\vec{h},U)}}$ . \\
The restriction of the Kirillov form on the orbit is then
\begin{eqnarray}
\Omega=m\left(
\begin{array}{cc}
A_{ij}&\delta_i^j\\-\delta^i_j&\pm\omega^2A_{ij}
\end{array}
\right)
\end{eqnarray}
and its inverse is
\begin{eqnarray}
\Omega^{-1}=\frac{1}{m}\left(
\begin{array}{cc}
\pm
\omega^2(A\Phi_{\pm}^{-1})_{ij}&-(\Phi_{\pm}^{-1})^j_i\\(\Phi_{\pm}^{-1})^i_j&(A\Phi_{\pm}^{-1})^{ij}
\end{array}
\right)
\end{eqnarray}
The maximal orbit is then equipped with the symplectic structure
\begin{eqnarray}
\sigma=(\Phi_{\pm}^{-1})^i_jdp_i\wedge
dq^j+\frac{1}{m}(A\Phi_{\pm}^{-1})^{ij}dp_i\wedge dp_j\pm
m\omega^2(A\Phi_{\pm}^{-1})_{ij}dq^i\wedge dq^j
\end{eqnarray}
and it follows that the Poisson brackets of two functions defined on
the orbit is then
\begin{eqnarray}
\{f,g\}=(\Phi_{\pm}^{-1})^j_i(\frac{\partial f}{\partial
p_i}\frac{\partial g}{\partial q^j}-\frac{\partial f}{\partial
q^i}\frac{\partial g}{\partial p_j})+F_{ij}\frac{\partial
f}{\partial p_i}\frac{\partial g}{\partial p_j}+G^{ij}\frac{\partial
f}{\partial q^i}\frac{\partial g}{\partial q^j}
\end{eqnarray}
This implies that
\begin{eqnarray}
\{p_i,p_j\}=F_{ij}~,~\{p_i,q^j\}=(\Phi_{\pm}^{-1})^j_i~,~\{q^i,q^j\}=G^{ij}
\end{eqnarray}
where the magnetic field $F_{ij}$ and the dual magnetic field
$G^{ij}$ are given by
\begin{eqnarray}
 F_{ij}=\frac{1}{C}(A\Phi_{\pm}^{-1})_{ij}\;\mbox{and}\;  G^{ij}=\frac{1}{m}(A\Phi_{\pm}^{-1})^{ij}
\end{eqnarray}
Moreover the Hamilton's equations are
\begin{eqnarray}
\frac{dp_k}{dt}&=&-(\Phi^{-1}_{\pm})^i_k\frac{\partial H}{\partial
q^i}+\frac{1}{C}(A\Phi_{\pm}^{-1})_{ik}\frac{\partial H}{\partial
p_i}\crcr\frac{dq^k}{dt}&=&(\Phi^{-1}_{\pm})^k_i\frac{\partial
H}{\partial p_i}+\frac{1}{m}(A\Phi^{-1}_{\pm})^{ik}\frac{\partial
H}{\partial q^i}
\end{eqnarray}
With the anisotropic Newton-Hooke groups $ANH_{\pm}$ in three-dimensional spaces, we also have realized phase
spaces where the momenta as well as the positions do not commute.\\
Note that in all these three cases, the fields are constant because they are coming from central extensions of Lie algebras.\\

Let us follow the similar fashion to construct noncommutative phase spaces on the other nonrelativistic anisotropic planar kinematical
groups as announced.\;\;But,
it is interesting to mention that we will not address explicit calculation at
some steps for the following kinematical groups cases as we have done previously.
\subsection{ Galilean noncommutative phase space}
The planar Galilei group admits a nontrivial two-parameter central extension leading to an exotic model \cite{Duval-horvathy}.\;\;In
this section, we prove that the planar anisotropic Galilei group which admits also a nontrivial central extension provides an exotic
structure (noncommutativity in position sector).

The Galilei group $G$ in two-dimensional space is defined
by the multiplication law (\ref{galileilaw}).\;\;Its Lie algebra $\cal{G}$ is then generated by
the left invariant vector fields $J,~\vec{K},~\vec{P}$ and $H$ defined by (\ref{vectorfieldsgalilei})
that satisfy the Lie brackets (\ref{galileialgebra}).\\

The central extension of the corresponding anisotropic Galilei Lie algebra is defined by the Lie brackets given in the table
(\ref{table5}).\;\;Explicitly, by standard methods, we verify that the
nontrivial brackets for the central extension ${\hat{\cal{G}}}$ are
(\ref{galileialgebra}) plus
\begin{eqnarray}
[K_i,K_j]=\frac{1}{c^2}S\epsilon_{ij}~,~[K_i,P_j]=M\delta_{ij}~~;i,j=1,2
\end{eqnarray}
where $M$ and $S$ generate the center of ${\hat{\cal{G}}}$, $c$
being a constant velocity.\\

Let
$k_iK^{*i}+p_iP^{*i}+EH^*+mM^*+hS^*$ be the general element of
the dual of the planar centrally extended  Lie algebra
where $\vec{k}$ is a kinematic momentum,
$\vec{p}$ is a linear momentum, $E$ is an energy, $m$ is a mass and
$h$ is an action.\;\;Then $m$ and $h$ are trivial invariants under the coadjoint action of the planar anisotropic Galilei group.\;\;The
other invariant, the solution of the Kirillov's system (\ref{kirillovsystem}), is explicitly given by:
\begin{eqnarray*}\label{galileaninvariants}
U=E-\frac{\vec{p}~^2}{2m}
\end{eqnarray*}
interpreted as the internal energy of the considered model.\\

The restriction $\Omega$ of the Kirillov's matrix on the orbit is given by
\begin{eqnarray*}
\Omega=\left(
\begin{array}{cccc}
0&\frac{h}{c^2}&m&0\\
-\frac{h}{c^2}&0&0&m\\
-m&0&0&0\\
0&-m&0&0\\
\end{array}
 \right)
\end{eqnarray*}
Then by using the relations (\ref{q^ik^i m}) and
\begin{eqnarray}\label{wave-partiduality}
 h\omega_0=mc^2
\end{eqnarray}
the latter remembering us the wave-particle duality, the left hand
side being an energy associated to a frequency $\omega_0$, the right hand side
being an energy associated to a mass, we obtain that, in the canonical coordinates (
Darboux coordinates), the Poisson bracket (\ref{poissonbrackets}) takes
the form:
\begin{eqnarray}\label{galileipoisson}
\{f,g\}=\frac{\partial f}{\partial p_i}\frac{\partial g}{\partial
q^i}-\frac{\partial f}{\partial q^i}\frac{\partial g}{\partial
p_i}+G^{ij}\frac{\partial f}{\partial q^i}\frac{\partial g}{\partial
q^j}
\end{eqnarray}
with
\begin{eqnarray}\label{g}
G^{ij}=-\frac{\epsilon^{ij}}{m\omega_0}
\end{eqnarray}
where $\omega_0$ is a frequency.

The inverse of $\Omega$ is given by
\begin{eqnarray*}
\Omega^{-1}=\left(
\begin{array}{cccc}
0&0&-\frac{1}{m}&0\\0&0&0&-\frac{1}{m}\\\frac{1}{m}&0&0&\frac{1}{m\omega_0}\\0&\frac{1}{m}&-\frac{1}{m\omega_0}&0
\end{array}
\right)
\end{eqnarray*}
Furthermore, the maximal coadjoint orbit denoted by
${\cal{O}}_{(m,h,U)}$
is equipped with the symplectic $2$-form:
\begin{eqnarray}\label{symplgal}
 \sigma=dp_i\wedge
dq^i+\frac{\epsilon^{ij}}{m\omega_0}dp_i\wedge dp_j
\end{eqnarray}
 It follows that the corresponding minimal coupling is
\begin{eqnarray}
\pi_i=p_i~~,~~x^i=q^i+\frac{p_k}{2m\omega_0}\epsilon^{ki}
\end{eqnarray}
and then that
\begin{eqnarray}
\{p_i,p_j\}=0~~,~~\{p_i,x^k\}=\delta^k_i~~,~~ \{x^i,x^j\}=G^{ij}
\end{eqnarray}
So following the coadjoint orbit method, we have constructed a nonrelativistic particle model for
the two-parameter centrally extended anisotropic Galilei group in a two-dimensional space, recovering the exotic model described in
 \cite{horvathy7}.\;\;It is a noncommutative phase space whose positions do not commute, i.e. where only the dual magnetic
field $B^*$ is present.
 It is such that
\begin{eqnarray}
e^*B^*=-\frac{1}{m\omega_0}
\end{eqnarray}
Moreover the equations of motion corresponding to the above symplectic structure are given by:
\begin{eqnarray}\label{garahamilton}
\frac{dp_i}{dt}=-\frac{\partial H}{\partial
q^i}~~,~~\frac{dx^i}{dt}=\frac{\partial H}{\partial
p_i}+\frac{\epsilon^{ki}}{2m\omega_0}\frac{\partial H}{\partial q^k}
\end{eqnarray}
i.e.
\begin{eqnarray}
\frac{dp_i}{dt}=-\frac{\partial H}{\partial
q^i}~~,~~\frac{dx^i}{dt}=\frac{\partial H}{\partial
p_i}-\frac{\epsilon^{ki}}{2m\omega_0}\frac{dp_k}{dt}
\end{eqnarray}
Comparing the results obtained for the anisotropic Newton-Hooke and Galilei groups cases in a two-dimensional space,
we observe that with the anisotropic Galiei group, the phase space obtained is
only partially noncommutative while the phase spaces obtained with the anisotropic Newton-Hooke groups (as detailed in the
previous section) are completely noncommutative.\;\;In fact, with the coadjoint orbit method applied to the centrally extended
Lie algebra of the Galilei group, the phase
space obtained is such that positions are noncommutative due to the noncommutativity of the generators of the
pure Galilei transformations.\;\;Although, in the anisotropic Newton-Hooke groups cases, generators of space translations and pure
Newton-Hooke transformations do not commute in centrally extended Lie algebras.\\

 Note that it has been proved in \cite {walczyk} that for Hamiltonian function of the form:
\begin{eqnarray}\label{Hamiltonian}
 H=\frac{1}{2m}(p_1^{2}+p_2^{2})+V(x^1,x^2),~~V(x^1,x^2)=\sum_iF_ix^i,~~F_i= \mbox{const.}
\end{eqnarray}
the corresponding Newton equation
\begin{eqnarray}\label{newtonlaw}
m \frac{d^2 x_i}{dt^2}=F_i
\end{eqnarray}
remains undeformed.\\

In the following section, we realize the Poisson brackets of the form :
\begin{eqnarray*}\label{un}
\{p_k,p_i\}=F_{ki}~,~\{p_k,q^i\}=\delta^i_k~,~\{q^k,q^i\}=0
\end{eqnarray*}
by the coadjoint orbit method on the planar anisotropic Para-Galilei groups and prove that the noncommutativity of momenta implies
the modification of the second Newton law (\ref{newtonlaw}) in the sense of \cite{romero, wei}.
\subsection{ Para-Galilean noncommutative phase spaces}
As already argued, the Para-Galilei non-relativistic algebra $\cal{G}^{\prime}_{+}$ was introduced in the framework of classification
of all kinematical groups \cite{5bacry}.\;\;
For radius parameter $r$ approaching infinity, the corresponding Para-Galilei group
contracts in the Static group acting on the standard (flat) nonrelativistic spacetime.\;\;In this thesis, we
have defined the other Para-Galilei algebra $\cal{G}^{\prime}_{-}$ whose Lie bracket $[P_i,H]=-\omega^2 K_i$.\;\;
The two Para-Galilei algebras are the obtained by a velocity-time
contaction of the
 nonrelativistic and cosmological Newton-Hooke algebras $\cal{NH}_{\pm}$.\\

 With the planar anisotropic Para-Galilei groups, we construct in this subsection,
noncommutative phase spaces whose momenta do not commute.\;\;This noncommutativity is due to the presence of naturally
introduced magnetic fields $B_{\pm}$.\\

The planar Para-Galilei groups and their Lie algebras have been defined in \cite{ancille1}.\;\;Explicitly, the Para-Galilei
groups $G^{\prime}_{\pm}$ in two-dimensional spaces are defined
by the multiplication law
\begin{eqnarray}\label{paragalileilaw}\nonumber
gg^{~\prime}=(\theta,\vec{v},\vec{x},t)(\theta^{~\prime},\vec{v}^{~\prime},\vec{x}^{~\prime},t^{~\prime})=(\theta+\theta^{~\prime},
R(\theta)\vec{v}^{~\prime}\\
+\vec{v}\pm\omega^2 \vec {x}t^{~\prime},R(\theta)\vec{x}^{~\prime}+\vec{x},t+t^{~\prime})
\end{eqnarray}
where $\theta$ is an angle of rotations, $\vec{v}$ is a boost
vector, $\vec{x}$ is a space translation vector and $t$ is a time
translation parameter.\\

Their Lie algebras $\cal{G}^{\prime}_{\pm}$ are then generated by the left invariant vector fields
\begin{eqnarray}\label{vectorfieldsparagalilei}
J=\frac{\partial}{\partial
\theta}~,~\vec{K}=R(-\theta)\frac{\partial}{\partial
\vec{v}}~~,~~\vec{P}=R(-\theta)\frac{\partial}{\partial
\vec{x}}~~~~H=\frac{\partial}{\partial
t}\pm\omega^2 \vec{x}\frac{\partial}{\partial
\vec{v}}~~
\end{eqnarray}
satisfying the nontrivial Lie brackets
\begin{eqnarray}\label{para-galiliealgebra}
[J,K_j]=K_i\epsilon^i_j~,~[J,P_j]=P_i\epsilon^i_j~,~[P_i,H]=\pm\omega^2 K_i~
\end{eqnarray}
Their corresponding anisotropic Lie algebras generated by ${K_i,P_i,H}$ are defined in the
table (\ref{table5}) by the nontrivial  Lie brackets:
\begin{eqnarray}
 [P_i,H]=\pm \omega^2 K_i
\end{eqnarray}
The central extensions of the anisotropic Para-Galilei Lie algebras are given in the table (\ref{tab:table6})
where $M$ and $S$ are the central generators of the extended Lie algebras.\\

Let $k_iK^{*i}+p_iP^{*i}+EH^*+mM^*+hS^*$ be a general element of
the dual of the centrally extended anisotropic Para-Galilei Lie algebras where
$\vec{k}$ is a kinematic momentum,
$\vec{p}$ is a linear momentum, $E$ is an energy, $m$ is a mass and
$h$ is an action.\\

Then the Kirillov form (\ref{kirillov01}) in the basis ($K_1,K_2, P_1,P_2,H$), is in this case given by:
\begin{eqnarray}\label{kirillovPara-Gal1}
(K_{ij})=\left(
\begin{array}{ccccc}
0&0&m&0&0\\
0&0&0&m&0\\
-m&0&0&\kappa^2h&\pm\omega^2k_1\\
0&-m&-\kappa^2h&0&\pm\omega^2k_2\\
0&0&\mp\omega^2k_1&\mp\omega^2k_2&0\\
\end{array}
 \right)
\end{eqnarray}
The coadjoint orbits of the centrally extended anisotropic Para-Galilei Lie groups on the dual of their Lie algebras
are then characterized by the two trivial invariants $m$ and $h$, and a
nontrivial invariant $U_{\pm}$, solution of the system (\ref{kirillovsystem}) interpreted as internal energy and given by:
\begin{eqnarray}\label{para-galileaninvariants}
U_{\pm}=E\pm\frac{\vec{q}~^2}{2C}
\end{eqnarray}
where we have used relation $c=\frac{\omega}{\kappa}$, (\ref{q^ik^i m}) and (\ref{wave-partiduality})
where
\begin{eqnarray}\label{compliance}
C=\pm \frac{1}{m \omega^2}
\end{eqnarray}
 Let us denote by
${\cal{O}}_{(m,h,U_{\pm})}$ the coadjoint orbits.\\

The restriction $\Omega=(\Omega_{ab})$ of the Kirillov form (\ref{kirillovPara-Gal1}) on ${\cal{O}}_{(m,h,U_{\pm})}$ is then
\begin{eqnarray*}
\Omega=\left(
\begin{array}{cccc}
0&0&m&0\\
0&0&0&m\\
-m&0&0&\kappa^2h\\
0&-m&-\kappa^2h&0\\
\end{array}
\right)
\end{eqnarray*}
It follows that
\begin{eqnarray*}
\Omega^{-1}=\left(
\begin{array}{cccc}
0&\pm\frac{\omega^2}{m\omega_0}&-\frac{1}{m}&0\\\mp\frac{\omega^2}{m\omega_0}&0&0&-\frac{1}{m}\\\frac{1}{m}&0&0&0\\0&\frac{1}{m}&0&0
\end{array}
\right)
\end{eqnarray*}
where we have used the wave-particle duality (\ref{wave-partiduality}) and the relation $c=\frac{\omega}{\kappa}$.\\

The Poisson bracket (\ref{poissonbrackets}) is in this case given by:
\begin{eqnarray}\label{para-galileipoisson}
\{H,f\}=\frac{\partial H}{\partial p_i}\frac{\partial f}{\partial
q^i}-\frac{\partial H}{\partial q^i}\frac{\partial f}{\partial
p_i}+F_{ij}\frac{\partial H}{\partial p_i}\frac{\partial f}{\partial
p_j}
\end{eqnarray}
with
\begin{eqnarray}
F_{ij}=\mp\frac{\epsilon^{ij} m\omega^2}{\omega_0}
\end{eqnarray}
where we have used (\ref{q^ik^i m}) and the wave-particle duality.

Moreover, the symplectic form (\ref{symplectictwo-form}) takes the form:
\begin{eqnarray} \label{symplecticpara-1}
\sigma=dp_i\wedge
dq^i\pm\frac{\epsilon^{ij}m\omega^2}{\omega_0}dq^i\wedge dq^j
\end{eqnarray}
 If the frequency of the charged particle still unchanged: that is $\omega_0=\omega$, then the symplectic
 structure (\ref{symplecticpara-1})
 becomes equivalent to the one obtained in
 \cite{zhang-horvathy} in the case of Para-Galilei group $G_+^{\prime}$.\\

 The corresponding minimal coupling \cite{ancilla} is
\begin{eqnarray}\label{magneticcoupling1}
 \pi_i=p_i\pm\frac{\epsilon^{ij} m \omega^2}{2 \omega_0} q^k~~,~~q^i=x^i
\end{eqnarray}
The coordinates $\pi_i$ and $x^i$ are such that
\begin{eqnarray}
\{x^i,x^k\}=0~~~,~~~\{\pi_i,x^k\}=\delta^k_i~~,~~~\{\pi_i,\pi_k\}= F_{ik}
\end{eqnarray}
So with the planar anisotropic Para-Galilei groups, we have
obtained noncommutative phase spaces in momenta sector.\;\;This noncommutativity is due to the presence of naturally
introduced magnetic fields $B_{\pm}$ given by the relation:
\begin{eqnarray}\label{magneticparagal}
e B_{\pm}\epsilon^{ij}=F_{ij}
\end{eqnarray}
The corresponding equations of motion take the form:
\begin{eqnarray}
\frac{dp_i}{dt}=-\frac{\partial H}{\partial
q^i}+F_{ij}\frac{\partial H}{\partial p_i}\delta^i{_j}~~,~~\frac{dx^i}{dt}=\frac{\partial H}{\partial
p_i}
\end{eqnarray}
i.e.
\begin{eqnarray}
\frac{dp_i}{dt}=-\frac{\partial V}{\partial
q^i}\mp\frac{\epsilon^{ij} m \omega^2}{\omega_0}\frac{dx^j}{dt}~~~,~~\frac{dx^i}{dt}=\frac{\partial H}{\partial
p_i}
\end{eqnarray}
or equivalently
\begin{eqnarray}\label{secondnewtonlaw}
 m\frac{d^2 x^i}{dt^2}=-\frac{\partial V}{\partial{x^i}}+eB_{\pm} \epsilon^{ij}\frac{p_j}{m}
\end{eqnarray}
for Hamiltonian of the form (\ref{Hamiltonian}) and where we have used (\ref{magneticparagal}).
We interpret the equation (\ref{secondnewtonlaw}) as the modified Newton's second law \cite{romero, wei}.\;\;The second term
in this equation is a correction due to
the noncommutativity of momenta.\;\;It is a damping force which depends on the space through the factor of noncommutativity $F_{ij}$.
For $F_{ij}=0$ or $B=0$, equation (\ref{secondnewtonlaw}) leads to the usual Newton's second law.

\subsection{ Static noncommutative phase spaces}
We are now going to employ the formulation we described in the above sections to construct noncommutative phase spaces
on anisotropic planar Static group.\\

Let us consider the central extension of the planar anisotropic Static Lie algebra whose Lie algebra structure is given in the
table (\ref{table5}) by:
 \begin{eqnarray}\label{anisotropcentstatic}
[K_j,K_k]=\frac{1}{c^2}S\epsilon_{jk}~~~~~~~~~~~~~~~~~~~~~~~~~~~~~~~~~~\\ \nonumber
[K_j,P_k] =M\delta_{jk},~[P_j,P_k] =\kappa^2 S\epsilon_{jk}~~~~~\\\nonumber
[K_j,H]=0,~~~~~~~[P_j,H]=0~~~~~~~~~~~~~~~~~ \nonumber
\end{eqnarray}

Let
$hS^*+k_iK^{*i}+p_iP^{*i}+EH^*+mM^*$ be the general element of
the dual of the planar anisotropic centrally extended Static Lie algebra.\;\;Then $E$, $m$ and $h$ are
trivial invariants under the coadjoint action of the planar anisotropic Static Lie group. \\

The restriction of the Kirillov's matrix on the orbit is given by
\begin{eqnarray}
\Omega=\left(
\begin{array}{cccc}
0&\frac{h}{c^2}&m&0\\-\frac{h}{c^2}&0&0&m\\-m&0&0&\kappa^2h\\0&-m&-\kappa^2h&0
\end{array}
\right)
\end{eqnarray}
By using the wave-particle duality (\ref{wave-partiduality}) and the equality $c=\frac{\omega}{\kappa}$, where $\kappa$
is a constant whose dimension is the inverse of that of a length, we obtain that
the Poisson bracket of two functions $f$ and $g$ implied by the Kirillov symplectic structure is given by
\begin{eqnarray}
\{f,g\}=\frac{\partial f}{\partial p_i}\frac{\partial g}{\partial
q^i}-\frac{\partial f}{\partial q^i}\frac{\partial g}{\partial
p_i}+G^{ij}\frac{\partial f}{\partial q^i}\frac{\partial g}{\partial
q^j}+F_{ij}\frac{\partial f}{\partial p_i}\frac{\partial g}{\partial
p_j}~~;~~i,j=1,2
\end{eqnarray}
with
\begin{eqnarray}
G^{ij}=-\frac{\epsilon^{ij}}{m\omega_0}~~,~~ F_{ij}=-(m-\mu_{e})\omega\epsilon_{ij}
\end{eqnarray}
and where
\begin{eqnarray}\label{effectivemass0}
\mu_e=m-\frac{\kappa^2 h}{\omega},~~~\vec{q}=\frac{\vec{k}}{\mu_e}
\end{eqnarray}
 $\mu_e$ being an effective mass.\;\;It follows that the magnetic fields $B$ and $B^*$ are such that
\begin{eqnarray}\label{magneticfields}
e^*B^*=-\frac{1}{m\omega_0}~~,~~eB=(m-\mu_e)\omega
\end{eqnarray}
The effective mass is then given in function of the magnetic field
by
\begin{eqnarray}
\mu_e=m-\frac{eB}{\omega}
\end{eqnarray}
The Hamilton's equations are then
\begin{eqnarray}\label{hamiltonequationtwofields}
\frac{d\pi_i}{dt}=-\frac{\partial H}{\partial q^i}-
(m-\mu_e)\omega\epsilon_{ik}\frac{\partial H}{\partial
p_k}~~,~~\frac{dx^i}{dt}=\frac{\partial H}{\partial
p_i}+\frac{\epsilon^{ik}}{2m\omega_0}\frac{\partial H}{\partial q^k}
\end{eqnarray}
The inverse of $\Omega$ is
\begin{eqnarray}\label{kirillovnewtonhooke2a}
\Omega^{-1}=\left(
\begin{array}{cccc}
0& -
\frac{\omega}{\mu_e}&-\frac{1}{\mu_e}&0\\\frac{\omega}{\mu_e
}&0&0&-\frac{1}{\mu_e}\\\frac{1}{\mu_e}&0&0&\frac{1}{\mu_e\omega_0}\\0&\frac{1}{\mu_e}&-\frac{1}{\mu_e \omega_0}&0
\end{array}
\right)
\end{eqnarray}
where we have used the wave-particle duality and (\ref{effectivemass0}).\\

Finally the orbit is equipped with the symplectic form
\begin{eqnarray}\label{symplstatic}
\sigma=dp_i\wedge
dq^i+\frac{1}{\mu_e\omega_0}\epsilon^{ij}dp_i\wedge dp_j-
\mu_e\omega \epsilon_{ij} dq^i\wedge dq^j
\end{eqnarray}

We observe that with the planar anisotropic Static group, the phase space obtained is completely noncommutative.\;\;In fact, with
the coadjoint orbit method applied to the centrally extended anisotropic Static Lie algebra, the phase space
obtained is such that positions as well as momenta are noncommutative due to the noncommutativity of both the generators of the
pure Static transformations and the generators of space transformations.\;\;This noncommutativity is measured by two naturally
introduced magnetic fields expressed by relations (\ref{magneticfields}).\;\;The same results
have been obtained in the planar anisotropic oscillating Newton-Hooke group case \cite{ancilla}.
\subsection{ Carroll noncommutative phase spaces}
The planar anisotropic Carroll group is the Carroll group $C(2)$ without the rotations parameters \cite{3derome}.\;\;Its Lie algebra has
the only nontrivial Lie bracket given by (\ref{carrolbracket})
where the left invariant vector fields are given by:
$$\vec{K}=\frac{\partial}{\partial \vec{v}},~~\vec{P}=\frac{\vec{v}}{c^2}\frac{\partial}{\partial t}+\frac{\partial}{\partial \vec{x}},~
H=\frac{\partial}{\partial t}$$
By using standard methods, we have obtained that the central extension of the planar anisotropic Carroll algebra
is given by (\ref{carrolbrackets}).\\

Let
$hS^*+k_iK^{*i}+p_iP^{*i}+EH^*$ be the general element of
the dual of the centrally extended planar Carroll Lie algebra.\;\;Then $E$ and $h$ are
trivial invariants under the coadjoint action. \\
 The restriction of the Kirillov form on the orbit in the basis ( $K_1,K_2,P_1,P_2,H$) is in this case
\begin{eqnarray}\label{kirillovcarroll}
\Omega=\left(
\begin{array}{cccc}
0&\frac{h}{c^2}&\frac{E}{c^2}&0\\-\frac{h}{c^2}&0&0&\frac{E}{c^2}\\-\frac{E}{c^2}&0&0&\kappa^2 h\\0&-\frac{E}{c^2}&-\kappa^2 h&0
\end{array}
\right)
\end{eqnarray}
By using relations $h\omega_0=mc^2=E$, $c=\frac{\omega}{\kappa}$ and (\ref{effectivemass0}), $\kappa$ being
a constant whose dimension is the inverse of that of a length,
 we obtain that the Poisson brackets of two functions defined on the orbit are given by
\begin{eqnarray}\label{carrolleanpoissons}
\{h,f\}=\frac{\partial h}{\partial p_i}\frac{\partial f}{\partial
q^i}-\frac{\partial h}{\partial q^i}\frac{\partial f}{\partial
p_i}+G^{ij}\frac{\partial h}{\partial q^i}\frac{\partial f}{\partial
q^j}+F_{ij}\frac{\partial h}{\partial p_i}\frac{\partial f}{\partial
p_j}~~;~~i,j=1,2
\end{eqnarray}
with
\begin{eqnarray}
G^{ij}=-\frac{\epsilon^{ij}}{m\omega_0}~~,~~ F_{ij}=-(m-\mu_{e})\omega\epsilon_{ij}
\end{eqnarray}
and where we have used the relations (\ref{effectivemass0}), $\mu_e$ being an effective mass.\;\;It follows that the magnetic
 field $B$ and its dual field $B^*$ are such that
\begin{eqnarray}
e^*B^*=-\frac{1}{m\omega_0}~~,~~eB=(m-\mu_e)\omega
\end{eqnarray}
The effective mass is then given in function of the magnetic field
by
\begin{eqnarray}
\mu_e=m-\frac{eB}{\omega}
\end{eqnarray}
The Hamilton's equations are then
\begin{eqnarray}\label{hamiltonequationtwofields}
\frac{d\pi_i}{dt}=-\frac{\partial H}{\partial q^i}-
(m-\mu_e)\omega\epsilon_{ik}\frac{\partial H}{\partial
p_k}~~,~~\frac{dx^i}{dt}=\frac{\partial H}{\partial
p_i}+\frac{\epsilon^{ik}}{2m\omega_0}\frac{\partial H}{\partial q^k}
\end{eqnarray}
The inverse of $\Omega$ is
\begin{eqnarray}\label{kirillovnewtonhooke2a}
\Omega^{-1}=\left(
\begin{array}{cccc}
0& -
\frac{\omega}{\mu_e}&-\frac{1}{\mu_e}&0\\\frac{\omega}{\mu_e
}&0&0&-\frac{1}{\mu_e}\\\frac{1}{\mu_e}&0&0&\frac{1}{\mu_e\omega_0}\\0&\frac{1}{\mu_e}&-\frac{1}{\mu_e \omega_0}&0
\end{array}
\right)
\end{eqnarray}
where we have used the duality wave-particle and (\ref{effectivemass0}).\;\;Finally the orbit is equipped with the symplectic form
\begin{eqnarray}\label{sympcarroll}
\sigma=dp_i\wedge
dq^i+\frac{1}{\mu_e\omega_0}\epsilon^{ij}dp_i\wedge dp_j-
\mu_e\omega \epsilon_{ij} dq^i\wedge dq^j
\end{eqnarray}
We observe that with the anisotropic Carroll group in two-dimensional spaces, we obtain similar results as in
the Static group case.\;\;Thus, by applying coadjoint orbit
method to the extended planar anisotropic Carroll group, the structure of the phase space obtained is the
same as in the Static group case: positions as well as momenta do not
commute due to the noncommutativity of both the generators of the pure Carroll transformations and the generators of
space transformations.\\

If we apply the coadjoint orbit method to the centrally extended anisotropic kinematical algebras, we obtain three kinds
of noncommutative phase spaces:
\begin{itemize}
 \item A phase space whose only positions do not commute, i.e: generators of pure kinematical group transformations are
noncommutative in extended Lie algebras.\;\;This kind of noncommutative phase spaces is realized on the Galilei algebra ${\cal G}$.
\item A phase space whose only momenta are noncommutative due to the noncommutativity of generators of space translations in
centrally extended Lie algebras.\;\;This case is realized on the Para-Galilei algebras ${\cal G}^{\prime}$.\;\;It arrives also
for the Aristotle Lie algebra.
\item phase spaces which are completely noncommutative: meaning that both generators of pure kinematical group transformations
and generators of space translations do not commute.\;\; These are obtained by considering the case of the Newton-Hooke Lie
algebras ${\cal N}_{\pm}$, the Static Lie algebra and the Carroll Lie algebra.
\end{itemize}
In the following section, we realize noncommutative phase spaces on the planar absolute time groups by considering their noncentral
extensions.
\section{ Noncommutative phase spaces constructed on absolute time groups}
For the planar absolute time groups, the noncommutative phase spaces are obtained by
working with noncentral abelian extensions which have been determined in the Chapter $1$.\;\;In this section, we see in detail that
noncommutative phase spaces can be also constructed in the framework of noncentrally abelian extended kinematical groups (when rotations are
taking into account) by the coadjoint orbit method.\;\;It will be highlighted that the deformed structures obtained in such way are
algebraically more general than those obtained in the anisotropic case.
\subsection{Galilei noncommutative phase spaces}
With the coadjoint orbit method applied to the noncentrally abelian extended Galilei group, the phase space obtained is a six-dimensional
phase space where one coordinates commute with the others and where the positions do not commute due to the noncommutativity
of the generators
of pure Galilei transformations.\;\;This noncommutativity is measured by the presence of the dual magnetic filed $B^*$.\;\;With the
Galilei group, same results have been obtained in \cite{horvathy} but by using the two-
centrally extended Lie algebra of this group as it has been done also in the section ($3.3.3$).\\

Let us consider the Galilei group in two-dimensional space as described previously.\;\;Its Lie algebra $\cal{G}$ is generated by
the left invariant vector fields given by relation (\ref{vectorfieldsgalilei}) and
satisfy the nontrivial Lie brackets (\ref{galileialgebra}) as already pointed out.
\subsubsection{i) Noncentrally abelian extended group and its maximal coadjoint orbit}
Let ${\hat{\cal{G}}}$ be the noncentrally abelian extended Galilei Lie algebra satisfying relations (\ref{galileialgebra}) and
\begin{eqnarray}\label{galileialgebraextended}
[J,F_i]=F_k\epsilon^k_i~,~[P_i,H]=F_i~,~[K_i,K_j]=\frac{1}{c^2}S\epsilon_{ij}~,~[K_i,P_j]=M\delta_{ij}~~
\end{eqnarray}
where $c$ is a constant whose dimension is a velocity, the
 extending generators $M$ and $F_i$  have $L^{-2}T$ and $L^{-1}T^{-1}$ as physical dimensions respectively \cite{ kostant,
kirillov, nzo1,
hamermesh} while $S$ is dimensionless.\;\;Note that $S$ and $M$ generate
the center $ Z({\hat{\cal{G}}})$ of ${\hat{\cal{G}}}$. \\

Let $$\hat{g}=exp(\varphi S+\xi M)exp(\eta^iF_i)exp(x^iP_i+tH)exp(v^iK_i)exp(\theta J)$$
be the general element of the connected extended Galilei group $\hat{G}$.\;\;Assume also that
\\$\hat{g}=(\beta,\vec{\eta},g)\in \hat{G}$
 with $\beta=(\varphi,\xi)$ and $g$ the general element of the Galilei group.\;\;By using relations (\ref{galileialgebra}),
(\ref{galileialgebraextended}) and the Baker-Campbell-Hausdorff formulas \cite{hall}, we obtain that the multiplication
law of $\hat{G}$ is given by
\begin{eqnarray}
 (\beta,\vec{\eta},g)(\beta^{~\prime},\vec{\eta}^{~\prime},g^{~\prime})=(\beta+\beta^{~\prime}+c(g,g^{~\prime}),R(\theta)
\vec{\eta}^{~\prime}+
\vec{\eta}+\vec{c}~(g,g^{~\prime}),gg^{~\prime})
\end{eqnarray}
with $gg^{\prime}$ given by (\ref{galileilaw}) and where
\begin{eqnarray}
c(g,g^{~\prime})=(\frac{1}{2c^2}R(-\theta)\vec{v}\times \vec{v}^{~\prime},R(-\theta)\vec{v}.\vec{x}^{~\prime}+
\frac{\vec{v}^{~2}}{2}t^{~\prime})
\end{eqnarray}
and
\begin{eqnarray}
\vec{c}~(g,g^{~\prime})=\frac{1}{2}[(\vec{x}-\vec{v}t)t^{~\prime}-tR(\theta)\vec{x}^{~\prime}].
\end{eqnarray}
are two $2-$ cocycles defining the extended structure.\\

It follows that the adjoint action
 $Ad_g(\delta{\hat{g}})=g( \delta{\hat{g}} )g^{-1}$ of the quotient group \\$Q=\hat{G}/Z({\hat{G}})$ on
 the Lie algebra ${\hat{\cal{G}}}$ is given by:
\begin{eqnarray}\label{ancillag}\nonumber
 Ad_{( \vec{x}, \vec{v}, \theta, t)}(\delta \theta, \delta \varphi,\delta \vec{x},
 \delta \vec{v},\delta \vec{\eta},\delta \xi, \delta t)=(\delta \theta^{\prime},\delta \varphi^{\prime},
 \delta \vec{x}^{~\prime},\delta \vec{v}^{~\prime}, \delta\vec{\eta}^{~\prime}, \delta \xi^{\prime},
  \delta t^{\prime})
\end{eqnarray}
with

$\delta \theta^{~\prime}=\delta{\theta},~~\delta t^{~\prime}=\delta t$,~~$\delta \vec{v}^{~\prime}=R(\theta)\delta \vec{v} +
\epsilon(\vec{v})\delta \theta,~~~~~~~~~~~~~~~~~~~~~$\\
$\delta \varphi^{~\prime}=\delta \varphi+\frac{1}{c^2} R(-\theta)\vec{v}\times \delta \vec{v}-\frac{\vec{v}~^2}{2c^2}\delta \theta$,\\
$\delta\vec{x}^{~\prime}=R(\theta)\delta \vec{x}+\vec{v}\delta t+\epsilon(\vec{x}-\vec{v}t)\delta \theta-tR(\theta)\delta \vec{v}$,\\
$\delta \xi^{~\prime}=\delta \xi+R(-\theta)\vec{v}.\delta \vec{x}-R(-\theta)\vec{x}.\delta \vec{v}+\frac{\vec{v}~^2}{2}\delta t+
\vec{v}\times \vec{x}~ \delta \theta$,\\
$\delta \vec{\eta}^{~\prime}=R(\theta)\delta \vec{\eta}-t R(\theta)\delta \vec{x}+(\vec{x}-\vec{v}t)\delta t+\epsilon(\vec{\eta}-
\frac{\vec{x}t}{2}+\frac{\vec{v}t^2}{2})\delta \theta+\frac{t^2}{2}R(\theta)\delta \vec{v}$\\
with
\begin{eqnarray}\label{epsilo}
\epsilon(\vec{v})=\left(
\begin{array}{c}
v^{2}\\-v^{1}
\end{array}
\right)
\end{eqnarray}
If the duality between the extended Lie algebra ${\hat{\cal{G}}}$ and its dual ${\hat{\cal{G}}}^*$ is defined by the action
$j\delta \theta+\vec{k}.\delta \vec{v}+\vec{p}.\delta \vec{x}+E\delta t+\vec{f}.\delta {\vec{\eta}}+m\delta \xi+h\delta \varphi$,
then the coadjoint action of the quotient group $Q$ on ${\hat{\cal{G}}}^*$ is
\begin{eqnarray}
(h^{~\prime},m^{~\prime},\vec{f}^{~\prime},j^{~\prime},\vec{k}^{~\prime},\vec{p}^{~\prime},E^{~\prime})=
Ad^*_{(\vec{\eta},g)}(h,m,\vec{f},j,\vec{k},\vec{p},E)
\end{eqnarray}
 with
\begin{eqnarray}\label{masactionforcemom}
m^{\prime}=m,~~h^{\prime}=h,~~\vec{f}^{~\prime}=R(\theta)\vec{f},~\vec{p}^{~\prime}=R(\theta)\vec{p}+tR(\theta)\vec{f}-m\vec{v}~~~~
~~~~~~~~~~~
\end{eqnarray}
\begin{eqnarray}\label{galileanpassage}
\vec{k}^{~\prime}=R(\theta)\vec{k}+tR(\theta)\vec{p}+\frac{t^2}{2}R(\theta)\vec{f} +m(\vec{x}-\vec{v}t)+\frac{h}{c^2}\epsilon(\vec{v})~~
~~~~~~~~~~
~~~~~~~~~~
\end{eqnarray}
\begin{eqnarray}\label{galileanenergy}
E^{\prime}=E-\vec{v}.R(\theta)\vec{p}-\vec{x}.R(\theta)\vec{f}+\frac{m\vec{v}~^2}{2}~~~~~~~~~~~~~~~~~~~~~~~~~~~~~~~~~~~~~
~~~~~~~~~~~~~~~~~~
\end{eqnarray}
\begin{eqnarray}\label{galileianangularmomentum}
j^{\prime}=j+\vec{x}\times R(\theta)\vec{p}+\vec{v}\times R(\theta)\vec{k}+{\vec{\eta}}\times R(\theta)\vec{f}+
\frac{\vec{x}t}{2}\times R(\theta)\vec{f}+m\vec{v}\times \vec{x}
\end{eqnarray}
The observables $j, \vec{k}, \vec{p},E, \vec{f}, m$  and $h$ have respectively the physical dimensions of an angular momentum, a
static momentum,
 a linear momentum, an energy, a force, a mass and an action. \\

 Let the position vector $\vec{q}$ and the dual magnetic field $B^*$ \cite{ancilla} be defined by
 \begin{eqnarray}\label{dualmagnetic}
 \vec{q}=\frac{\vec{k}}{m},~~B^*=\frac{1}{e^*m\omega},
 \end{eqnarray}
where $e^*$ is a dual charge.\\

The maximal coadjoint orbit of $Q$ on ${\hat{\cal{G}}^*}$ denoted by ${\cal{O}}_{(m,B^*,f,U)}$ is then characterized by
the two trivial
invariants $m$ and $h$ and by two nontrivial invariants: the force intensity $f$ and the (internal) energy $U$ given respectively by
\begin{eqnarray}
f=||\vec{f}||,~~U=E-\frac{\vec{p}~^2}{2m}+\vec{f}.\vec{q}+e^*B^*\vec{f}\times\vec{p}
\end{eqnarray}
where the relation $h\omega_0=mc^2$ has been used. \\

 In the basis
 $(J,F_1,K_1,P_1,K_2,P_2,F_2,H,M,S)$ of the noncentrally abelian extended Galilei Lie algebra, the restriction of
the Kirillov's matrix on the
orbit is
\begin{eqnarray}
\Omega=\left (\begin{array}{cccccc}
0&f sin\alpha&mq^2&p_2&-mq^1&-p_1\\
-f sin\alpha&0&0&0&0&0\\-mq^2&0&0&m&e^*m^2B^*&0\\
-p_2&0&-m&0&0&0\\mq^1&0&-e^*m^2B^*&0&0&m\\p_1&0&0&0&-m&0
\end{array}
\right)
\end{eqnarray}
where $f_1=f cos\alpha~,~f_2=f sin\alpha$.\\

Its inverse is
\begin{eqnarray}
\Omega^{-1}=\frac{1}{f~ sin\alpha}\left (\begin{array}{cccccc}
0&-1&0&0&0&0\\1&0&-\frac{p_2}{m}&q^2-e^*B^*p_1&\frac{p_1}{m}&-q^1-e^*B^*p_2\\
0&\frac{p_2}{m}&0&-\frac{f~ sin\alpha}{m}&0&0\\0&-q^2+e^*B^*p_1&\frac{f~ sin\alpha}{m}&0&0&e^*B^*f~\sin \alpha\\
0&-\frac{p_1}{m}&0&0&0&-\frac{f~ sin\alpha}{m}\\0&q^1+e^*B^*p_2&0&-e^*B^*f~\sin\alpha&\frac{f~ sin\alpha}{m}&0
\end{array}
\right)
\end{eqnarray}
where $\vec{A^*}=\frac{1}{2}\vec{B^*}\times \vec{p} ~$ is the dual magnetic potential \cite{ancilla} while
 $\vec{B^*}=B^*\vec{n}$ with $\vec{n}$ the unit vector perpendicular to the plane.\\

The orbit is then equipped with the symplectic form
\begin{eqnarray}\label{symplecnc}
\sigma_1=\sigma_0+G^{ij}dp_i\wedge dp_j
\end{eqnarray}
where $\sigma_0=ds\wedge d\alpha+dp_i\wedge dq^i$, $G^{ij}=e^*B^*\epsilon^{ij}$ and $s=j-\vec{p}\times (\vec{q}-e^*\vec{A}^*)$ is
the sum of the orbital momentum $\vec{L}=\vec{q}\times \vec{p}$, the angular momentum $j$ and an extra term $\vec{p}\times e^*\vec{A}^*$
associated to the dual magnetic field $B^*$ \cite{horvathy6}.\\

The Poisson brackets (of two functions $H$ and $f$ defined on the orbit)\\
corresponding to the symplectic structure (\ref{symplecnc}), are given by
\begin{eqnarray}
\{H,f\}=\frac{\partial H}{\partial s}\frac{\partial f}{\partial \alpha}-\frac{\partial H}{\partial \alpha}\frac{\partial f}
{\partial s}+
\frac{\partial H}{\partial p_i}\frac{\partial f}{\partial q^i}-\frac{\partial H}{\partial q^i}\frac{\partial f}{\partial p_i}-
e^*B^*\epsilon^{ij}\frac{\partial H}{\partial q^i}\frac{\partial f}{\partial q^j}
\end{eqnarray}
Then
\begin{eqnarray}\label{poissongalilei}\label{01}\nonumber
\{s,p_j\}=p_i\epsilon^i_j~,~\{s,q^i\}=\epsilon^i_j(q^j-e^*A^{*j})\\
\{q^i,q^j\}=-e^*B^*\epsilon^{ij}~,~\{p_i,q^j\}=\delta^j_i~~~~~~~~~
\end{eqnarray}
are the Poisson brackets within the coordinates on the orbit.\;\;We observe that $\alpha$ commute with all the other coordinates, that
the momenta commute and form a vector,
that positions do not commute and do not form a vector due to the presence of the dual magnetic field $B^*$.\\

Note also that $(s,p_i,\alpha,\tilde{q}~^i=q^i-e^*A^{*i})$ are canonical phase coordinates on the coadjoint orbit and
 that
\begin{eqnarray}
\{s,A^*_j\}=A^*_i\epsilon^i_j~,~~\{q^i,A^*_j\}=\frac{B^*}{2}\epsilon^i_j.
\end{eqnarray}
\subsubsection{ii) Symplectic realization and equations of motion}
 Let  $(s^{\prime},\vec{p}^{~\prime},\alpha^{\prime},\vec{q}^{~\prime})=D_{(\vec{\eta},\theta,\vec{v},\vec{x},t)}
(s,\vec{p},\alpha,\vec{q}~)$ be
the symplectic realization of the extended Galilei group on its coadjoint orbit.\;\;Use of the relations
(\ref{masactionforcemom}) to (\ref{galileianangularmomentum}) gives rise to
\begin{eqnarray}
\alpha^{\prime}=\alpha+\theta~,~\vec{q}^{~\prime}=R(\theta)\vec{q}+\frac{1}{m}(R(\theta)\vec{p}-m\vec{v})t+
\frac{R(\theta)\vec{f}}{m}\frac{t^2}{2}+\vec{x}-e^*m\vec{v}\times \vec{B}^*
\end{eqnarray}
\begin{eqnarray}
s^{\prime}=s+\frac{K^*}{m}t+\frac{N^*}{m}\frac{t^2}{2}+(\vec{\eta}-\frac{\vec{x}t}{2}+\frac{\vec{v}t^2}{2})\times
R(\theta)\vec{f}-\frac{e^*B^*m^2\vec{v}^2}{2}~~~~~~~~~~~~~~~~~~~~~~~~~
\end{eqnarray}
\begin{eqnarray}
\vec{p}^{~\prime}=R(\theta)\vec{p}+R(\theta)\vec{f}t-m\vec{v}~~~~~~~~~~~~~~~~~~~~~~~~~~~~~~~~~~~~~~~~~~~~~~~~~~~~~~~~~~~~~~~~~~~~~~~~~
~~~~~~~~~~
\end{eqnarray}
where $\frac{K^*}{m}$ and $\frac{N^*}{m}$ are respectively an energy and a power with $K^*$ and $N^*$ given by:
\begin{eqnarray}\label{energypower}
K^*=m\vec{q}\times \vec{f}+e^*mB^*\vec{p}.\vec{f}~,~N^*=\vec{f}\times \vec{p}-e^*mB^*\frac{\vec{f}~^2}{2}
\end{eqnarray}
It follows that the evolution with respect to the time $t$ is given by
\begin{eqnarray}\label{configurationevolution1}
\alpha(t)=\alpha,~\vec{q}~(t)=\vec{q}+\frac{\vec{p}}{m}t+\frac{\vec{f}}{m}\frac{t^2}{2}
\end{eqnarray}
and
\begin{eqnarray}\label{momentaevolution1}
s(t)=s+\frac{K^*}{m}t+\frac{N^*}{m}\frac{t^2}{2},
~\vec{p}~(t)=\vec{p}+\vec{f}t
\end{eqnarray}
The corresponding Hamiltonian vector field is
\begin{eqnarray}
X_H=\frac{\vec{p}}{m}.\frac{\partial}{\partial \vec{q}}+\vec{f}.\frac{\partial}{\partial \vec{p}}+\frac{K^*}{m}\frac{\partial}
{\partial s}
\end{eqnarray}
 and then the Hamiltonian function, given by
\begin{eqnarray}
H=\frac{\vec{p}^{~2}}{2m}+V_0(\vec{q},\alpha)+e^*\vec{p}.(\vec{f}\times \vec{B}^*),
\end{eqnarray}
is the sum of a kinetic energy $T(\vec{p})=\frac{\vec{p}^{~2}}{2m}$, of a potential energy \\$V_0(\vec{q},\alpha)=-\vec{f}.\vec{q}-
\frac{K^*}{m}\alpha$ depending on the position-angle variables $(\vec{q},\alpha)$ and of an exotic energy $E^*_{exotic}=e^*\vec{p}.
(\vec{f}\times \vec{B}^*)$ depending on the dual magnetic field.\\

The equations of motion are then given by
\begin{eqnarray}\label{forcemomentum}
\left(
\begin{array}{c}
\vec{p}(t)\\K^*(t)
\end{array}
\right)=m\frac{d}{dt}\left(
\begin{array}{c}
\vec{q}(t)\\ s(t)
\end{array}
\right)~,~
\frac{d}{dt}\left(
\begin{array}{c}
\vec{p}(t)\\ \alpha(t)
\end{array}
\right)= \left(
\begin{array}{c}
\vec{f}\\0
\end{array}
\right)
\end{eqnarray}
The equations (\ref{forcemomentum}) define the linear momentum $\vec{p}~(t)=m\frac{d\vec{q}~(t)}{dt}$, the force \\
$\vec{f}=\frac{d\vec{p}}{dt}$ and the quantity
$K^*(t)=m\frac{ds}{dt}$ whose dimension is that of a linear momentum squared.\;\;Moreover
\begin{eqnarray}\label{accelleration}
\frac{d^2}{dt^2}\left(
\begin{array}{c}
\vec{p}\\\alpha
\end{array}
\right)=\left(
\begin{array}{c}
\vec{0}\\0
\end{array}
\right)
,
\frac{d^2}{dt^2}\left(
\begin{array}{c}
\vec{q}\\s
\end{array}
\right)=\frac{1}{m}\left(
\begin{array}{c}
\vec{f}\\N^*
\end{array}
\right)
\end{eqnarray}
We interpret the equations (\ref{accelleration}) as the Euler-Newton's second law \cite{wei} associated to the above
noncentral extended Galilei group.\\

So with the noncentrally abelian extended Galilei group, we obtain also an exotic model \cite{horvathy} i.e with
noncommutativity in positions sector.\\

 In the following paragraph, we study the case of the absolute time Para-Galilei groups and as in the anisotropic case, we show that the
 noncommutativity of the
momenta induces some modification of the Newton's second law \cite{romero, wei}.\;\; Comparative summary between results obtained on
both the absolute time Galilei (see previous section) and Para-Galilei groups will be given.
\subsection{Para-Galilei noncommutative phase spaces}
In this case, the phase space obtained is a six-dimensional phase space where one coordinates commute
with the others and where the momenta do not commute due to the noncommutativity of the generators of space translations
in the extended Para-Galilei groups.\;\;This noncommutativity
is measured by the presence of a magnetic field $B$.\\

More specifically, let us consider the planar Para-Galilei groups as described in the previous
section.\;\;Their Lie algebras ${\cal{G}}^{\prime}_{\pm}$ are generated by
the left invariant vector fields given by relation (\ref{vectorfieldsparagalilei}) and
satisfy the nontrivial Lie brackets (\ref{para-galiliealgebra}).
\subsubsection{i)Noncentrally abelian extended groups and their maximal coadjoint orbits}
 The noncentrally abelian extended Para-Galilei Lie algebras ${\hat{\cal{G}}^{\prime}}_{\pm}$ satisfy the nontrivial Lie brackets
(\ref{para-galiliealgebra}) and
\begin{eqnarray}\label{extendedparagalileialgebra}
[J,\Pi_i]=\Pi_k\epsilon^k_i~,~[K_i,H]=\Pi_i~,~[P_i,P_j]=\kappa^2 S\epsilon_{ij}~,~[K_i,P_j]=M\delta_{ij}~~
\end{eqnarray}
where $\kappa$ is a constant whose dimension is the inverse of that of a length.\;\;As in
the Galilei case, $S$ and $M$ generate the center
of ${\hat{\cal{G}}^{\prime}}_{\pm}$ and $\Pi_i$ are new generators which do not commute with $J$,
hence the term noncentral abelian extension.\\

 Let $$\hat{g}=exp(\varphi S+\xi M)exp({\cal{l}}^i\Pi_i)exp(v^iK_i+tH)exp(x^iP_i)exp(\theta J)$$
be the general element of the connected extended
Para-Galilei groups ${\hat{G}}^{\prime}_{\pm}$.\;\;As in the case of the Galilei group, the
relations (\ref{para-galiliealgebra}), (\ref{extendedparagalileialgebra}) and
the Baker-Campbell-Hausdorff
formulas give rise to the group multiplication laws
\begin{eqnarray}
(\beta,{\vec{l}},g)(\beta^{\prime},{\vec{l}}^{~\prime},g^{\prime})=(\beta+\beta^{\prime}+c(g,g^{\prime}),
{\vec{l}}+R(\theta){\vec{l}}^{~\prime}+\vec{c}~(g,g^{\prime}),gg^{\prime})
\end{eqnarray}
with $gg^{\prime}$ given by (\ref{paragalileilaw}) and
where
\begin{eqnarray}
c(g,g^{\prime})=(\frac{1}{2}\kappa^2 R(-\theta)\vec{x}\times \vec{x}^{~\prime},-R(-\theta)\vec{x}.\vec{v}^{~\prime}
\mp\frac{\omega^2\vec{x}~^2}{2}
t^{\prime})
\end{eqnarray}
and
\begin{eqnarray}
\vec{c}~(g,g^{\prime})=\frac{1}{2}[(\vec{v}\mp \omega^2\vec{x}t)t^{\prime}-tR(\theta)\vec{v}^{~\prime}].
\end{eqnarray}
are two $2$-cocycles.\\

It follows that the adjoint action $Ad_g(\delta{\hat{g}})=g( \delta{\hat{g}} )g^{-1}$
of the quotient groups \\$Q^{\prime}_{\pm}=\hat{G}^{\prime}_{\pm}/Z({\hat{G}}^{\prime}_{\pm})$
on the extended Para-Galilei Lie algebras
${\hat{\cal{G}}^{\prime}}_{\pm}$ is
\begin{eqnarray}\label{ancillag}\nonumber
 Ad_{( \vec{x}, \vec{v}, \theta, t)}(\delta \theta, \delta \varphi,\delta \vec{x},
 \delta \vec{v},\delta \vec{l},\delta \xi, \delta t)=(\delta \theta^{\prime},\delta \varphi^{\prime},
 \delta \vec{x}^{~\prime},\delta \vec{v}^{~\prime}, \delta\vec{l}^{~\prime}, \delta \xi^{\prime},
  \delta t^{\prime})
\end{eqnarray}
with\\
$\delta \theta^{~\prime}=\delta{\theta}~,~\delta t^{\prime}=\delta t~,~\delta\vec{x}^{~\prime}=R(\theta)\delta \vec{x}+
\epsilon(\vec{x})\delta \theta~~~~~~~~~~~~~~~~~~~~~~~~~~~~~~~~~~~~~~~~~~~~~~~~~~~$\\
$\delta \vec{v}^{~\prime}=R(\theta)\delta \vec{v}\pm \omega^2\vec{x}\delta t+
\epsilon(\vec{v}\mp \omega^2\vec{x}t)\delta \theta \mp \omega^2 t R(\theta)\delta \vec{x}~~~~~~~~~~~~~~~~~~~~~~~~~~~~$\\
$\delta {\vec{l}}^{~\prime}=R(\theta){\vec{l}}-t R(\theta)\delta \vec{v}\pm \frac{\omega^2t^2}{2}R(\theta)\delta \vec{x}+
(\vec{v}\mp\omega^2\vec{x}t)\delta t+\epsilon({\vec{l}}-\frac{t}{2}(\vec{v}\mp \omega^2\vec{x}t))\delta \theta$\\
$\delta \varphi^{\prime}=\delta \varphi+\kappa^2 R(-\theta)\vec{x}\times \delta \vec{x}-\frac{\vec{x}~^2}{2}\kappa^2\delta \theta~
~~~~~~~~~~~~~~~~~~~~~~~~~~~~~~~~~~~~~~~~~~~~~~~~~~~~~~~$\\
$~\delta \xi^{\prime}=\delta \xi-R(-\theta)\vec{x}.\delta \vec{v}+R(-\theta)\vec{v}.\delta \vec{x}\mp\frac
{\omega^2\vec{x}^2}{2}\delta t+\vec{v}\times \vec{x}~ \delta \theta~~~~~~~~~~~~~~~~~~~~~~~~~~~~~~~~~$\\

where $\epsilon(\vec{v})$ is given by (\ref{epsilo}).

If the duality between the extended Lie algebras ${\hat{\cal{G}}^{\prime}}_{\pm}$ and their duals ${\hat{\cal{G}}^{*{\prime}}}_{\pm}$
gives rise
to the action $j\delta \theta+\vec{k}.\delta \vec{v}+
\vec{I}.\delta \vec{x}+E\delta t+\vec{p}.\delta {\vec{l}}+m\delta \xi+h\delta \varphi$, then the coadjoint actions are such that
\begin{eqnarray}\label{massaction}
m^{\prime}=m~,~h^{\prime}=h~~~~~~~~~~~~~~~~~~~~~~~~~~~~~~~~~~~~~~~~~~~~~~~~~~~~~~~~~~~~~~~~~~~~~~~~~
\end{eqnarray}
\begin{eqnarray}\label{passage1}
\vec{p}^{~\prime}=R(\theta)\vec{p}~,~~\vec{k}^{~\prime}=R(\theta)\vec{k}+tR(\theta)\vec{p}+m\vec{x}~~~~~~~~~~~~~~~~~~~~~~~~~~~~~~~~
\end{eqnarray}
\begin{eqnarray}\label{linearmomentum}
\vec{I}^{~\prime}=R(\theta)\vec{I}\pm\omega^2tR(\theta)\vec{k}\pm\frac{\omega^2t^2}{2}R(\theta)\vec{p} +
m\omega\epsilon(\vec{x})-m(\vec{v}\mp\omega^2\vec{x}t)
\end{eqnarray}
\begin{eqnarray}\label{energy}
E^{\prime}=E\mp \omega^2\vec{x}.R(\theta)\vec{k}-\vec{v}.R(\theta)\vec{p}-\frac{\vec{x}~^2}{2C}~~~~~~~~~~~~~~~~~~~~~~~~~~~~~~~~~~~~~~~~~~
\end{eqnarray}
where $C$ is given by relation (\ref{compliance})
\begin{eqnarray}\label{angularmomentum1}
j^{~\prime}=j+\vec{x}\times R(\theta)\vec{I}+\vec{v}\times R(\theta)\vec{k}+{\vec{l}}\times R(\theta)\vec{p}+
\frac{\vec{v}t}{2}\times R(\theta)\vec{p}+m\vec{v}\times \vec{x}
\end{eqnarray}
Define the vector $\vec{q}~$ by (\ref{dualmagnetic}) and the magnetic field $B$ by
\begin{eqnarray}
B=\frac{m\omega}{e}
\end{eqnarray}
where $e$ is the electric charge.\;\;The coadjoint orbits denoted by ${\cal{O}}_{(m,B,p,U_{\pm})}$ are characterized by two
trivial invariants $m$ and
$B$ and by two nontrivial invariants $p$ and $U_{\pm}$ respectively given by:
\begin{eqnarray}
p=||\vec{~p}||,~~U_{\pm}=E+\frac{\vec{q}~^2}{2C}-\vec{p}.\frac{\vec{I}}{m
}+eB\vec{p}\times\vec{q}
\end{eqnarray}
where we have used the relation $h\omega_0=mc^2$ and (\ref{compliance}).\\

In the basis $(J,P_1,K_1,\Pi_1,K_2,\Pi_2,P_2,H,M,S)$ of the extended Para-Galilei Lie algebras, the restriction of the Kirillov
form on the coadjoint orbit is then
\begin{eqnarray}
\Omega=\left (\begin{array}{cccccc}
0&p~ sin\alpha&mq^2&I_2&-mq^1&-I_1\\-p~ sin\alpha&0&0&0&0&0\\-mq^2&0&0&m&0&0\\-p_2&0&-m&0&0&eB\\mq^1&0&0&0&0&m\\I_1&0&0&-eB&-m&0
\end{array}
\right)
\end{eqnarray}
where $p_1=p~ cos\alpha~,~p_2=p~ sin\alpha$. \\

The inverse of $\Omega$ is
\begin{eqnarray}
\Omega^{-1}=\frac{1}{p~ sin\alpha}\left (\begin{array}{cccccc}
0&-1&0&0&0&0\\1&0&-\frac{I_2}{m}-\frac{eBq^1}{m}&q^2&\frac{I_1}{m}-\frac{eBq^2}{m}&-q^1\\0&\frac{I_2}{m}+\frac{eBq^1}{m}&0&-
\frac{p~ sin\alpha}{m}&\frac{eBp~ sin\alpha}{m^2}&0\\0&-q^2&\frac{p ~sin\alpha}{m}&0&0&0\\0&-\frac{I_1}{m}+\frac{eBq^2}{m}&-
\frac{eBp~ sin\alpha}{m^2}&0&0&-\frac{p~ sin\alpha}{m}\\0&q^1&0&0&\frac{p~ sin\alpha}{m}&0
\end{array}
\right)
\end{eqnarray}
We then verify that the symplectic form on the orbit ${\cal{O}}_{(m,B,p,U_{\pm})}$ is
\begin{eqnarray}\label{symplecticformnc}
\sigma_{1}^{\prime}=\sigma_0^{\prime}+F_{ij}dq^i\wedge dq^j
\end{eqnarray}
where $\sigma_0^{\prime}=ds\wedge d\alpha+dI_i\wedge dq^i$, $F_{ij}=eB\epsilon_{ij}$ and $s =j-\vec{q}\times (\vec{I}+e\vec{A})$ is
 the sum of the angular momentum $j$, the orbital angular momentum $\vec{L}=\vec{q}\times \vec{I}$
and an extra term $eB\frac{\vec{q}~^2}{2}$ associated to the magnetic field $B$, $\vec{A}=\frac{1}{2}\vec{B}\times \vec{q}~$ being the
magnetic potential \cite{ancilla} with $\vec{B}=B\vec{n}$.\\

The Poisson brackets of two functions $H$ and $f$ on the orbit corresponding to the symplectic form
(\ref{symplecticformnc}) is
\begin{eqnarray}
\{H,f\}=\frac{\partial H}{\partial s}\frac{\partial f}{\partial \alpha}-\frac{\partial H}{\partial \alpha}\frac{\partial f}{\partial s}+
\frac{\partial H}{\partial I_i}\frac{\partial f}{\partial q^i}-\frac{\partial H}{\partial q^i}\frac{\partial f}{\partial I_i}-
eB\epsilon_{ij}\frac{\partial H}{\partial I_i}\frac{\partial f}{\partial I_j}
\end{eqnarray}
and the nontrivial Poisson brackets within the coordinates are
\begin{eqnarray}\label{possonparagalilei}
\{s,I_j\}=(I_i-eA_i)\epsilon^i_j~,~\{s,q^i\}=\epsilon^i_jq^j~,~\{I_i,I_j\}=-eB\epsilon_{ij}~,~\{I_i,q^j\}=\delta^j_i~
\end{eqnarray}
This means that $\alpha$ commute with all the other coordinates, that the positions coordinates commute and form a vector
while momenta do not commute and do not form a vector due to the presence of the magnetic field $B$.\\

Note that $(s,\tilde{I}_i=I_i-eA_i,\alpha,q^i)$ are the canonical phase coordinates on the coadjoint orbit and that
we have moreover these Poisson brackets
\begin{eqnarray}
\{s,A_j\}=-A_i\epsilon^i_j~,~\{I_i,A_j\}=\frac{B}{2}\epsilon_{ij}
\end{eqnarray}
\subsubsection{ii) Symplectic realizations and equations of motion}
 Let the symplectic realizations of the extended Para-Galilei groups on their coadjoint orbits be given by
$(s^{\prime},\vec{p}^{~\prime},\alpha^{\prime},\vec{q}^{~\prime})=D_{({\vec{l}},\theta,\vec{v},\vec{x},t)}
(s,\vec{p},\alpha,\vec{q})$.\;\;
By using relations (\ref{passage1}) to (\ref{angularmomentum1}), we obtain
\begin{eqnarray}
\alpha^{\prime}=\alpha+\theta~,~\vec{q}^{~\prime}=R(\theta)\vec{q}+\frac{R(\theta)\vec{p}}{m}t+\vec{x}~~~~~~~~~~~~~~~~~~~~~~~~~~~~
~~~~~~~~
~~~~~~~~~~~~~~~~~
\end{eqnarray}
\begin{eqnarray}
s^{\prime}=s+\frac{K}{m} t+ \frac{N}{m}\frac{t^2}{2}+(\vec{l}-\frac{\vec{v}t}{2}\pm\frac{\omega^2t^2\vec{x}}{2})\times R(\theta)\vec{p}+
eB\frac{{\vec{x}~^2}}{2}~~~~~~~~~~~~~~~~~~~~
\end{eqnarray}
\begin{eqnarray}
\vec{I}^{~\prime}=R(\theta)\vec{I}+\frac{1}{C}[R(\theta)\vec{q}+\vec{x}~]t\pm \omega^2R(\theta)\vec{p}~\frac{t^2}{2}
+eB\epsilon(\vec{x})-m\vec{v}~~~~~~~~~~~~~~~~~~~~~~~~~~
\end{eqnarray}
where $\frac{K}{m}$ and $\frac{N}{m}$
are respectively an energy and a power and where $K$ and $N$ are given by
\begin{eqnarray}
K=\vec{I}\times \vec{p}-eB\vec{p}.\vec{q}~,~N=\frac{1}{C}\vec{q}\times \vec{p}\mp eB\frac{\vec{p}^{~2}}{2m}
\end{eqnarray}
where we have used relation (\ref{compliance}).\\

 The evolution with respect to the time $t$ is
\begin{eqnarray}\label{configurationevolution2}
\alpha(t)=\alpha~,~\vec{q}~(t)=\vec{q}+\frac{\vec{p}}{m}t~~~~~~~~~~~~~~~~~~~~~~~~~~~~~
\end{eqnarray}
and
\begin{eqnarray}\label{momentaevolution2}
s(t)=s+\frac{K}{m}t+\frac{N}{m}\frac{t^2}{2} ~,~\vec{I}(t)=
\vec{I}+\frac{1}{C}\vec{q}~t+\frac{1}{C}\frac{\vec{p}}{m}~\frac{t^2}{2}
\end{eqnarray}
where we have used relation (\ref{compliance}).\\

It follows that the corresponding Hamiltonian vector field is
\begin{eqnarray}
X_H=\frac{\vec{p}}{m}.\frac{\partial}{\partial \vec{q}}+\frac{\vec{q}}{C}.\frac{\partial}{\partial \vec{I}}+
\frac{K}{m}\frac{\partial}{\partial s}
\end{eqnarray}
and then the Hamiltonian function (energy) is
\begin{eqnarray}
H=\vec{I}.\frac{\vec{p}}{m}-\frac{\vec{q}^{~2}}{2C}-\frac{K}{m}\alpha+e\vec{p}.(\vec{B}\times \vec{q})
\end{eqnarray}
that is a sum of a kinetic term $T(\vec{p}~)=\vec{I}.\frac{\vec{p}~}{m}$, of a potential energy\\ $V_0(\vec{q},\alpha)=
-\frac{\vec{q}^{~2}}{2C}-\frac{K}{m}\alpha$~ depending on the position-angle variables $(\vec{q},\alpha)$ and of an exotic energy
$E_{exotic}=e\vec{p}.( \vec{B}\times \vec{q})$ depending on the magnetic field.\\

The equations of motion are then given by
\begin{eqnarray}\label{positionimpulsepara-gal}
\frac{d}{dt}\left(
\begin{array}{c}
\vec{I}(t)\\ s(t)
\end{array}
\right)=\frac{1}{C}\left(
\begin{array}{c}
\vec{q}\\\frac{K(t)}{m C}
\end{array}
\right)~,~
\frac{d}{dt}\left(
\begin{array}{c}
\vec{q}(t)\\ \alpha(t)
\end{array}
\right)=\frac{1}{m}\left(
\begin{array}{c}
\vec{p}\\0
\end{array}
\right)
\end{eqnarray}
The equations (\ref{positionimpulsepara-gal}) define the linear momentum  $\vec{p}~(t)=m\frac{d\vec{q}~(t)}{dt}$, the position\\
 $\vec{q}=C\frac{d\vec{I}}{dt}$ and
the quantity $K(t)=m\frac{ds}{dt}$ whose dimension is that of a linear momentum squared.\\

Moreover,
\begin{eqnarray}\label{accellerationb}
\frac{d^2}{dt^2}\left(
\begin{array}{c}
\vec{I}\\s
\end{array}
\right)=\frac{1}{m}\left(
\begin{array}{c}
\frac{1}{C}\vec{p}\\N
\end{array}
\right)
,
\frac{d^2}{dt^2}\left(
\begin{array}{c}
\vec{q}\\ \alpha
\end{array}
\right)=\left(
\begin{array}{c}
\vec{0}\\0
\end{array}
\right)
\end{eqnarray}
We interpret the equations (\ref{accellerationb}) as the modified Euler-Newton's second law \cite{wei} associated to the
Para-Galilei groups.\\

So with the noncentrally abelian extended Para-Galilei group, we obtain
 a six-dimensional phase space where one coordinates commute with the others and where the momenta do not
commute.\;\;This noncommutativity
is measured by the presence of the magnetic field $B$ through the Poisson brackets (\ref{possonparagalilei}).
\subsubsection{Comparative analysis between noncentrally Galilei and Para-Galilei cases}
In both cases the coadjoint orbit is a six-dimensional symplectic manifold i.e the product of
a two-dimensional cylinder (parametrized by an angle $\alpha$ and a conjugate action $s$) with a four-dimensional phase space
parametrized by the position-momentum coordinates.\;\;These coadjoint orbits are denoted
 ${\cal{O}}_{(m,f,B^*,U)}$ for the Galilei group and ${\cal{O}}_{(m,p,B,U_{\pm})}$ for the Para-Galilei groups where $B$
and $B^*$ are respectively a magnetic and a dual magnetic field.\;\;The polar coordinates of the points on the basis circle of
the cylinder are $(f,\alpha)$ for the orbit ${\cal{O}}_{(m,f,B^*,U)}$  and $(p,\alpha)$  for ${\cal{O}}_{(m,p,B,U_{\pm})}$
where $f$ is a constant force intensity while $p$ is a constant linear momentum.\;\;The orbit
${\cal{O}}_{(m,f,B^*,U)}$ is equipped with the symplectic form $\sigma_1=ds\wedge d\alpha+dp_i\wedge dq^j+e^*B^*dp_i\wedge dp_j$
and is then a noncommutative symplectic manifold where the
positions $q^i$ do not commute while the orbit ${\cal{O}}_{(m,p,B,U_{\pm})}$ endowed with the symplectic
form $\sigma_1^{\prime}=ds\wedge d\alpha+dI_i\wedge dq^j+eBdq^i\wedge dq^j$ is a noncommutative symplectic
manifold where the momenta $I_i$ do not commute. \\

Kinematically, the positions $q^i$ on ${\cal{O}}_{(m,f,B^*,U)}$
behave as the momenta $I_i$ on ${\cal{O}}_{(m,p,B,U_{\pm})}$ (\ref{configurationevolution1}) and (\ref{momentaevolution2}))
whereas the momenta $p_i$ on ${\cal{O}}_{(m,f,B^*,U)}$ behave as the positions $q^i$
on ${\cal{O}}_{(m,p,B,U_{\pm})}$ (\ref{momentaevolution1}) and (\ref{configurationevolution2})).\;\;The former
are quadratic in time while the last ones are linear in time.\;\;The action $s$ conjugated to
the angle is quadratic in time in the two cases.\;\;Note that the angle $\alpha$ is constant in time in the two cases.\\

The dynamic equations for the system described by the Galilean orbit are
\begin{eqnarray}
\frac{d^2s}{dt^2}=\frac{N^*}{m}~,~\frac{d\vec{p}}{dt}=m\frac{d^2\vec{q}}{dt^2}
\end{eqnarray}
 while they are
  \begin{eqnarray}
\frac{d^2s}{dt^2}=\frac{N}{m}~,~\frac{d\vec{q}}{dt}=C\frac{d^2\vec{I}}{dt^2}
\end{eqnarray}
for the system described by the Para-Galilean orbits, where $C$ is given by relation (\ref{compliance}), the inverse of the
spring constant.\\

 As the equation $\frac{d\vec{p}}{dt}=m\frac{d^2\vec{q}}{dt^2}$ is called the Galilei-Newton law \cite{nzo1} for a massive
particle with mass $m$, the
law  $\frac{d\vec{q}}{dt}=C\frac{d^2\vec{I}}{dt^2}$ can be called the Para-Galilei-Newton law for a spring whose constant
 $C$ is the inverse of the usual Hooke's constant.

The following table summarizes these constructions in both cases.
\\
\begin{table}[htbp]
\begin{tabular}{|c|c|c|}
\hline
\textbf{Group}&\textbf{Galilei~group}&\textbf{Para-Galilei~groups}\\
\hline
 Invariants(non trivial)& $f,~U=E-\frac{\vec{p}^{~2}}{2m}+\vec{f}.\vec{q}+e^*B^*\vec{f}\times\vec{p}$& $p,~U=E+\frac{\vec{q}^{~2}}{2C}$\\
&& $-\vec{p}.\frac{\vec{I}}{m}+eB\vec{p}\times\vec{q}$\\
\hline
magnetic~fields &$e^*B^*=\frac{1}{m\omega}~,~\vec{B}^*=B^*\vec{e}_3$&$eB=m\omega~,~\vec{B}=B\vec{e}_3$\\
&$\vec{A}^*=\frac{1}{2}\vec{B}^*\times \vec{p}$~(\textbf{dual potential})&$\vec{A}=\frac{1}{2}\vec{B}\times
 \vec{q}$~(\textbf{potential})\\
\hline
noncommutative &$(s,p_i,\alpha,q^i)$&$(s,I_i,\alpha,q^i)$\\
phase~coordinates~&\textbf{with}~$\vec{q}=\frac{\vec{k}}{m},~\alpha=arctg(\frac{f_2}{f_1})$~&$\textbf{with}~\vec{q}=\frac{\vec{k}}{m},
~\alpha=arctg(\frac{p_2}{p_1})$~\\
~&~~~~~ $s=j+\vec{p}\times \vec{q}-e^*B^*\frac{\vec{p}^{~2}}{2}$&~~~~~~$s=j+\vec{I}\times \vec{q}-eB\frac{\vec{q}^{~2}}{2}$\\
\hline
symplectic~form&$\sigma_1=\sigma_0+G^{ij}dp_i\wedge dp_j$&$\sigma_1^{\prime}=\sigma_0^{\prime}+F_{ij}dq^i\wedge dq^j$\\
&\textbf{with}~$G^{ij}=e^*B^*\epsilon^{ij}$&\textbf{with}~$F_{ij}=eB\epsilon_{ij}$\\
\hline
Poisson~ brackets&$\{s,p_j\}=p_i\epsilon^i_j$~&$\{s,I_j\}=(I_i-eA_i)\epsilon^i_j~$\\
~&$\{s,q^i\}=\epsilon^i_j(q^j-e^*A^{*j})$&$\{s,q^i\}=\epsilon^i_jq^j$\\
~&$\{p_i,p_j\}=0$&$\{I_i,I_j\}=-eB\epsilon_{ij}$\\
~&$\{p_i,q^j\}=\delta^j_i$&$\{I_i,q^j\}=\delta^j_i$\\
~&$\{q^i,q^j\}=-e^*B^*\epsilon^{ij}$&$\{q^i,q^j\}=0$\\
\hline
Hamiltonian function&$H=\frac{\vec{p}^{~2}}{2m}+V(\vec{q},\alpha)+e^*\vec{f}.(\vec{p}\times \vec{B}^*)$&$H=\vec{I}.\frac{\vec{p}}{m}+
V(\vec{q},
\alpha)+e\vec{p}.(\vec{B}\times \vec{q})$\\
\hline
potential energy&$V(\vec{q},\alpha)=-\vec{f}.\vec{q}-\frac{K^*}{m}\alpha$&$V(\vec{q},\alpha)=\frac{\vec{q}^{~2}}{2C}-\frac{K}{m}\alpha$\\
\hline
equations of motion&$\frac{d\alpha}{dt}=0$~,$\frac{ds}{dt}=\frac{K^*(t)}{m}$&$\frac{d\alpha}{dt}=0$,~$\frac{ds}{dt}=\frac{K(t)}{m}$\\
~&$\frac{d\vec{q}}{dt}=\frac{\vec{p}}{m}$~,$\frac{d\vec{p}}{dt}=\vec{f} (constant)~~~~~$ & $\frac{d\vec{q}}{dt}=\frac{\vec{p}}{m}$
(constant)~,
$\frac{d\vec{I}}{dt}=\frac{\vec{q}}{C}$\\
~&\textbf{with}~&\textbf{with}\\
~&$K^*(t)=K^*+N^*t$&$~K(t)=K+Nt$\\
~&$K^*=m\vec{q}\times \vec{f}+e^*mB^*\vec{p}.\vec{f}$&$~K=\vec{I}\times \vec{p}-eB\vec{q}.\vec{p}$\\
~&~~~~~~$N^*=\vec{f}\times \vec{p}-e^*B^*m\frac{f^2}{2}$&~~~~~~~$N=\frac{\vec{q}}{C}\times \vec{p}\mp eB\frac{\vec{p}^2}{2m}$\\
\hline
"Newton's equations"&$\frac{d^2s(t)}{dt^2}=\frac{N^*}{m}~,~\frac{d^2\vec{q}(t)}{dt^2}=\frac{\vec{f}}{m}$&$\frac{d^2s(t)}{dt^2}=
\frac{N}{m}~,~
\frac{d^2\vec{I}(t)}{dt^2}=\frac{\vec{p}}{mC}$\\
\hline
canonical&$(s,p_i,\alpha,\tilde{q}^i)$&$(s,\tilde{I}_i,\alpha,q^i)$\\
phase~coordinates&\textbf{with}~$\tilde{q}^i=q^i-e^*A^{*i}$&$\textbf{with}~\tilde{I}_i=I_i-eA_i$\\
\hline
\end{tabular}
\caption{\bf Comparative analysis between noncentrally Galilei and Para-Galilei cases}
\end{table}
\subsection{Newton-Hooke noncommutative phase spaces}
 As it has been already said, we have found that noncentral abelian extensions of absolute time Newton-Hooke Lie algebras
lead to nonvanishing commutator of two boosts and two momenta as
their corresponding two-parameter centrally extended Lie algebras as well as their corresponding centrally extended anisotropic Lie
algebras as detailed in \cite{ancilla}.\;\;This means that totally noncommutative phase spaces can be realized with
the Newton-Hooke groups in both cases (i.e anisotropic and absolute time cases). \\

However, noncommutative phase spaces obtained with noncentral abelian extensions of Newton-Hooke groups
are algebraically more general than those obtained with central extensions of the same groups: their Poissonian structures
(and hence their symplectic two-forms on the orbits) contain
additional terms.\;\;This case is not reviewed here.
\subsection{Static noncommutative phase space in the absolute time case}
With the planar noncentrally abelian extended Static group, we realize a completely noncommutative phase space equipped with a
modified symplectic structure.\\

 Indeed, the noncentrally abelian extended Static Lie algebra of the planar Static group satisfies the nontrivial Lie brackets
(\ref{anisotropicnoncentralstatic}).

Let
 $$g=exp(v^iK_i+x^iP_i+tH)exp(\theta J)~~~~~~~~~~~~~~~~~~~~~~~~~~~~~$$
be the general element of the Static group
and $$\hat{g}=exp(\xi M+bS+\varphi M^{\prime}+a\Lambda)exp(\eta^iF_i+{\cal{l}}^i\Pi_i)g~~~~$$
be the general element of the connected Lie group associated to the noncentrally
extended Static Lie algebra ${\cal{G}}$.\;\;By using the Baker-Hausdorff formulas \cite{hall} and by
identifying $\hat{g}$ with $(\beta,\vec{\nu}, g)$
where $\beta=(\xi,\varphi,b,a)$, $\vec{\nu}=(\vec{\eta},\vec{l})$,
we obtain that the multiplication law of the corresponding extended Lie group
is
\begin{eqnarray*}
 (\beta,\vec{\nu},g)(\beta^{~\prime},\vec{\nu}^{~\prime},g^{~\prime})=(\beta+\beta^{~\prime}+c(g,g^{~\prime}),R(\theta)
\vec{\nu}^{~\prime}+
\vec{\nu}+\vec{c}~(g,g^{~\prime}),gg^{~\prime})
\end{eqnarray*}
with
\begin{eqnarray}\label{static2law}
gg^{~\prime}=(\theta,\vec{v},\vec{x},t)(\theta~^{~\prime},\vec{v}~^{\prime},\vec{x}~^{\prime},t~^{\prime})=(\theta+\theta~^{\prime},
R(\theta)\vec{v}~^{\prime}+\vec{v},R(\theta)\vec{x}~^{\prime}+\vec{x},t+t~^{\prime})
\end{eqnarray}
 and
where
\begin{eqnarray*}
c(g,g^{\prime})=(-\frac{1}{2}R(-\theta)\vec{v}.\vec{x}^{~\prime}+\frac{1}{2}R(-\theta)\vec{x}.\vec{v}~^{\prime},
-\frac{1}{2}R(-\theta)\vec{v}.\vec{l}^{~\prime}+\frac{1}{2}R(-\theta)\vec{l}.\vec{v}~^{\prime}\\-\frac{1}{2}\vec{l}.\vec{v}
-\frac{1}{2}R(-\theta)\vec{l}^{~\prime}.R(-\theta)\vec{v}^{~\prime},\\
-\frac{1}{2}R(-\theta)\vec{v}.\vec{\eta}^{~\prime}+\frac{1}{2}R(-\theta)\vec{\eta}.\vec{v}~^{\prime}
-\frac{1}{2}R(-\theta)\vec{x}.\vec{l}^{~\prime}+\frac{1}{2}R(-\theta)\vec{l}.\vec{x}~^{\prime}\\-\frac{1}{2}\vec{\eta}.
\vec{v}-\frac{1}{2}
\vec{l}.\vec{x}-\frac{1}{2}R(-\theta)\vec{l}^{~\prime}.R(-\theta)\vec{x}^{~\prime}-\frac{1}{2}R(-\theta)\vec{\eta}^{~\prime}.R(-\theta)
\vec{v}^{~\prime},\\
-\frac{1}{2}R(-\theta)\vec{x}.\vec{\eta}^{~\prime}+\frac{1}{2}R(-\theta)\vec{\eta}.\vec{x}~^{\prime}-\frac{1}{2}\vec{\eta}.\vec{x}
-\frac{1}{2}R(-\theta)\eta^{~\prime}.R(-\theta)\vec{x}^{~\prime})\\
\end{eqnarray*}
while
\begin{eqnarray*}
\vec{c}~(g,g^{\prime})=(\frac{1}{2}[\vec{x}t^{~\prime}-tR(\theta)\vec{x}~^{\prime}],\frac{1}{2}[\vec{v}t^{~\prime}-
tR(\theta)\vec{v}~^{\prime}])
\end{eqnarray*}
 $\theta$ being an angle of rotations, $\vec{v}$  a boost
vector, $\vec{x}$ a space translation vector and $t$ being a time
translation parameter.\\

Let $jJ^*+mM^*+\beta S^*+\mu M^{\prime*}+k\Lambda^*+f_i F^{*i}+I_i\Pi^{*i}+k_iK^{*i}+p_iP^{*i}+EH^*$, ($ i=1,2$)
be the general element of the dual of the noncentrally abelian extended Static Lie algebra where
the observables are an angular momentum $j$, a static momentum $\vec{k}$, an energy $E$, a force $\vec{f}$, two linear
momenta $\vec{p}$ and $\vec{I}$, two
masses $m$ and $\mu$, a frequency $\omega=\frac{\beta}{\mu}$ and a Hooke's constant $k$.\\

It follows that the adjoint action of the quotient groups $Q=\cal{G}/Z({\cal{G}})$ on the above extended
 Lie algebra
${\cal{G}}$ is such that:
\begin{eqnarray*}
\delta \theta^{~\prime}=\delta{\theta}~,~\delta t^{~\prime}=\delta t~,~\delta\vec{x}^{~\prime}=R(\theta)\delta \vec{x}+
\epsilon(\vec{x})\delta \theta~~~~~~~~~~~~~~~~~
\end{eqnarray*}
\begin{eqnarray*}
\delta \vec{v}^{~\prime}=R(\theta)\delta \vec{v}+\epsilon(\vec{v})\delta \theta~~~~~~~~~~~~~~~~~~~~~~~~~~~~~~~~~~~~~~~~~~~~~~~~~~~~
\end{eqnarray*}
\begin{eqnarray*}
\delta {\vec{\eta}}^{~\prime}=R(\theta)\delta{\vec{\eta}}-t R(\theta)\delta \vec{x}+
\vec{x}\delta t+\epsilon({\vec{\eta}}-\frac{t}{2}(\vec{x}))\delta \theta~~~~~~~~~~~~\\
\delta {\vec{l}}^{~\prime}=R(\theta)\cal\delta{\vec{l}}-t R(\theta)\delta \vec{v}+
\vec{v}\delta t+\epsilon({\vec{l}}-\frac{t}{2}(\vec{v}))\delta \theta~~~~~~~~~~~~~
\end{eqnarray*}
\begin{eqnarray*}
\delta \varphi^{~\prime}=\delta \varphi+R(-\theta)\vec{v}\times \delta \vec{l}
- R(-\theta)\vec{l}\times \delta \vec{v}+\vec{v}\times\vec{l}~\delta \theta~~~~\\
~\delta \xi^{\prime}=\delta \xi-R(-\theta)\vec{x}.\delta \vec{v}+R(-\theta)\vec{v}.\delta \vec{x}+\vec{v}\times \vec{x}~
\delta \theta~~~~~~~~~~\\
\delta b^{~\prime}=\delta b+R(-\theta)\vec{v}. \delta \vec{\eta}
+ R(-\theta)\vec{x}. \delta \vec{l}- R(-\theta)\vec{\eta}. \delta \vec{v}\\-
 R(-\theta)\vec{l}. \delta \vec{x}+\vec{v}\times\vec{\eta}~\delta \theta +\vec{x}\times\vec{l}~\delta \theta~~~~~~~~~~~~~~~~~~~~~~~~~\\
\delta a^{~\prime}=\delta a+R(-\theta)\vec{x}.\delta \vec{\eta}-R(-\theta)\vec{\eta}.\delta \vec{x}+\vec{x}\times \vec{\eta}~
\delta \theta~~~~~~~
\end{eqnarray*}
where $\epsilon(\vec{v}~)$ is given by the relation (\ref{epsilo}).\\

If the duality between ${\cal{G}}$ and its dual ${\cal{G}}^{*}$ gives rise
to the action \\$j\delta \theta+\vec{k}.\delta \vec{v}+\vec{f}.\delta {\vec{\eta}}+
\vec{p}.\delta \vec{x}+E\delta t+\vec{I}.\delta {\vec{l}}+m\delta \xi+\mu\delta \varphi+\beta\delta b+k\delta a$, then
the coadjoint action
\begin{eqnarray}
(m^{\prime},\mu^{\prime},\beta^{\prime},k^{\prime},\vec{f}^{~\prime},j^{\prime},\vec{k}^{~\prime},\vec{p}^{~\prime},E^{~\prime},
\vec{I}^{~\prime})=
Ad^*_{(\vec{\eta},\vec{l},g)}(m,\mu,\beta,k,\vec{f},j,\vec{k},\vec{p},E,\vec{I})
\end{eqnarray}
is such that
\begin{eqnarray}\label{massactionforcemomentum}
m^{\prime}=m~,~\mu^{\prime}=\mu,~ \beta^{\prime}=\beta,~k^{\prime}=k~~~~~~~~~~~~~~~~~~~~~~~~~~~~~~~~~~~~~~~~~~~
\end{eqnarray}
\begin{eqnarray}\label{forcestatic}
\vec{f}^{~\prime}=R(\theta)\vec{f} -k {\vec{x}}-\beta {\vec{v}}~~~~~~~~~~~~~~~~~~~~~~~~~~~~~~~~~~~~~~~~~~~~~~~~~~~~~~~~~~
\end{eqnarray}
\begin{eqnarray}\label{linearmomentum}
\vec{p}^{~\prime}=R(\theta)\vec{p} +tR(\theta)\vec{f}-t k \vec{x}-t\beta \vec{v}-m\vec{v}+k \vec{\eta}+\beta \vec{l}~~~~~~~~~~
\end{eqnarray}
\begin{eqnarray}\label{linearmomentum1}
\vec{I}^{~\prime}=R(\theta)\vec{I} -\mu {\vec{v}}-\beta {\vec{x}}~~~~~~~~~~~~~~~~~~~~~~~~~~~~~~~~~~~~~~~~~~~~~~~~~~~~~~~~~
\end{eqnarray}
\begin{eqnarray}\label{passagestatic}
\vec{k}^{~\prime}=R(\theta)\vec{k}+tR(\theta)\vec{I}+m\vec{x}-t \mu \vec{v}-t \beta \vec{x}+\mu \vec{l}+\beta \vec{\eta}~~~~~~~~~~~~~~
\end{eqnarray}
\begin{eqnarray}\label{energy}
E^{~\prime}=E-\vec{v}.R(\theta)\vec{I}-\vec{x}.R(\theta)\vec{f}+\frac{1}{2}k(\vec{x}^{~2})+\frac{1}{2}\mu (\vec{v}^{~2})+
\beta \vec{x}.\vec{v}~~~
\end{eqnarray}
\begin{eqnarray}\label{angularmomentum}
j^{~\prime}=j+\vec{x}\times R(\theta)\vec{p}+\vec{v}\times R(\theta)\vec{k}+{\vec{l}}\times R(\theta)\vec{I}
+{\vec{\eta}}\times R(\theta)\vec{f}\\\nonumber
+\frac{\vec{v}t}{2}\times R(\theta)\vec{I}+\frac{\vec{x}t}{2}\times R(\theta)\vec{f}+m\vec{v}\times \vec{x}+\beta\vec{l}\times \vec{x}
+\mu\vec{l}\times \vec{v}
\end{eqnarray}
Then  $m,\mu, \beta$ and $ k$ are trivial invariants under coadjoint action of the Static group in two-dimensional space.
The nontrivial invariants $s$ and $U$, solutions of (\ref{kirillovsystem}), are explicitly given by:
\begin{eqnarray}
s=j-(\vec{k}-\frac{\beta}{k}\vec{p}~) \times\vec{u}+(\vec{p}-\frac{\beta}{\mu}\vec{k}~) \times\vec{q}\\
 U=E- \frac{\mu_e\vec{u}^{~2}}{2} -\frac{k_e\vec{q}^{~2}}{2}-\frac{\beta\mu_e}{\mu}\vec{q}.\vec{u}-\nu~h
\end{eqnarray}
where $\nu$ is a frequency and where we have used the relations
\begin{eqnarray}
 \vec{I}=\mu_e\vec{u},~\vec{f}=-k_e\vec{q},~h=j-s
\end{eqnarray}
defining the velocity vector $\vec{u}$, the position vector $\vec{q}$ and the action $h$
where
\begin{eqnarray}\label{effectivehook}
k_e=k-\frac{\beta^2}{\mu},~
\mu_e=\mu-\frac{\beta^2}{k}
\end{eqnarray}
are the effective Hooke's constant and the effective mass respectively.\\

The the restriction of the Kirillov form in the basis \\($J,P_1,P_2,K_1,K_2,F_1,F_2,\Pi_1,\Pi_2,
H,M,M^{\prime},S,\Lambda$) to the orbit is
\begin{eqnarray}\label{kirillovnoncstatic}
\Omega=\left(
\begin{array}{cccccccc}
0&0&-m&0&k&0&\beta&0\\
0&0&0&-m&0&k&0&\beta\\
m&0&0&0&\beta&0&\mu&0\\
0&m&0&0&0&\beta&0&\mu\\
-k&0&-\beta&0&0&0&0&0\\
0&-k&0&-\beta&0&0&0&0\\
-\beta&0&-\mu&0&0&0&0&0\\
0&-\beta&0&-\mu&0&0&0&0\\
\end{array}
\right)
\end{eqnarray}
Its inverse is given by:
\begin{eqnarray}
\Omega^{-1}=\frac{1}{\beta^2-\mu k}\left (\begin{array}{cccccccc}
0&0&0&0&\mu&0&-\beta&0\\
0&0&0&0&0&\mu&0&-\beta\\
0&0&0&0&-\beta&0&k&0\\
0&0&0&0&0&-\beta&0&k\\
-\mu&0&\beta&0&0&0&m&0\\
0&-\mu&0&\beta&0&0&0&m\\
\beta&0&-k&0&-m&0&0&0\\
0&\beta&0&-k&0&-m&0&0\\
\end{array}
\right)
\end{eqnarray}
The symplectic form on the orbit ${\cal{O}}_{(m,\mu,\beta,k,s,U)}$
 is then given by
\begin{eqnarray}\label{symplecticformncstatic2}
\sigma= d{\vec{p}}\wedge d{\vec{q}}+ ~d{\vec{k}}\wedge d{\vec{u}}+\frac{\beta}{k}~ d{\vec{p}}\wedge {d \vec{u}}
-\frac{\beta}{\mu} ~d{\vec{k}}\wedge d{\vec{q}}
\end{eqnarray}
The Poisson brackets of two functions $H$ and $f$ on the orbit corresponding to the symplectic form
 (\ref{symplecticformncstatic2}) is
\begin{eqnarray}\nonumber
\{H,f\}=\frac{\partial H}{\partial p_i}\frac{\partial f}{\partial q^i}-\frac{\partial H}{\partial q^i}\frac{\partial f}{\partial p_i}+
\frac{\partial H}{\partial k_i}\frac{\partial f}{\partial u^i}-\frac{\partial H}{\partial u^i}\frac{\partial f}{\partial k_i}\\-
\frac{\beta}{k}{\epsilon_{i}}^j\frac{\partial H}{\partial p_i}\frac{\partial f}{\partial u^j}+
\frac{\beta}{\mu}{\epsilon^{i}}_j\frac{\partial H}{\partial q^i}\frac{\partial f}{\partial k_j}
\end{eqnarray}
and the nontrivial Poisson brackets within the coordinates are
\begin{eqnarray}\label{possonstaticabs}\nonumber
\{p_j,q^i\}=\delta^i_j~~,~\{k_j,u^i\}=\delta^i_j~,~\{p_j,u^i\}=\frac{\beta}{k}{\epsilon_{i}}^j,\\~~~~~~~~~~~~~
\{q^i,k_j\}=\frac{\beta}{\mu}
{\epsilon^{i}}_j,
~\{p_i,k_j\}=0,~\{q^i,u^j\}=0~~~~~~~
\end{eqnarray}
This means that the linear momentum $\vec{p}$ is canonically conjugate to the position $\vec{q}$ as well as the static
 momentum $\vec{k}$
is canonically conjugate to the the velocity $\vec{u}$.\;\;Moreover the linear momentum does not commute with
the velocity as well as the static momentum does not commute with the position.\;\;Note that in this
case, $(\vec{q},\vec{u})$ is an element of the tangent space (evolution space) while $(\vec{p},\vec{k})$ is its dual.
\subsubsection{Symplectic realizations and equations of motion}
 Let the symplectic realizations of the noncentrally abelian extended Static group on its coadjoint orbit be given by
$(\vec{p}^{~\prime},\vec{k}^{~\prime},\vec{q}^{~\prime},\vec{u}^{~\prime})=D_{(\vec{\eta},\vec{l},\theta,\vec{v},\vec{x},t)}
(\vec{p},\vec{k},\vec{q},\vec{u})$.\;\;By using coadjoint action, we verify that
\begin{eqnarray*}
\vec{q}^{~\prime}=R(\theta)\vec{q}+\vec{v}\tau+\frac{k}{k_e}\vec{x}~~~~~~~~~~~~~~~~~~~~~~~~~~~~~~~~~~~~~~~~~~~~~~~~~~~~~~\\
\vec{u}^{~\prime}=R(\theta)\vec{u}-\nu_e\vec{x}-\frac{\mu}{\mu_e}\vec{v}~~~~~~~~~~~~~~~~~~~~~~~~~~~~~~~~~~~~~~~~~~~~~~~~~~~~\\
\vec{p}^{~\prime}=R(\theta)\vec{p} -tk_e [R(\theta)\vec{q}-\vec{v}\tau -\frac{k}{k_e} \vec{x}~]-m\vec{v}+
\kappa \vec{\eta}+\beta \vec{l}~~~~~~\\
\vec{k}^{~\prime}=R(\theta)\vec{k}+t\mu_e [R(\theta)\vec{u}-\nu_e\vec{x}- \frac{\mu}{\mu_e} \vec{v}~]+m \vec{x}+
\mu \vec{l}+\beta \vec{\eta}~~~
\end{eqnarray*}
$\tau=\frac{\beta}{k_e},~\nu_e=\frac{\beta}{\mu_e}$ being a duration and a frequency respectively.\\

 The evolution with respect to the time  $t$ is
\begin{eqnarray}\label{configurationevolution2st}
\vec{q}~(t)=q~,~\vec{u}~(t)=\vec{u}
\end{eqnarray}
and
\begin{eqnarray}\label{momentaevolution2st}
\vec{p}~(t)=\vec{p}-t k_e\vec{q}\\
\vec{k}~(t)=\vec{k}+t\mu_e\vec{u}
\end{eqnarray}

The equations of motion are then given by
\begin{eqnarray}\label{positionimpulse}
\frac{d}{dt}\left(
\begin{array}{c}
\vec{p}(t)\\ \vec{k}(t)
\end{array}
\right)=\left(
\begin{array}{c}
-k_e\vec{q}\\\mu_e\vec{u}
\end{array}
\right)~,~
\frac{d}{dt}\left(
\begin{array}{c}
\vec{q}(t)\\ \vec{u}(t)
\end{array}
\right)=\left(
\begin{array}{c}
0\\0
\end{array}
\right)
\end{eqnarray}
\\
With the planar noncentrally abelian extended Static group, we have obtained a completely noncommutative phase space equipped with
modified symplectic structure defined by
relation (\ref{symplecticformncstatic2}) and dynamics given by equations (\ref{positionimpulse}).\\
\\
\\
\section{Conclusions and Classification}
We can summarize our findings.  From the group theoretical discussion above, we see that the coadjoint orbit method applied to
the two-parameter central extensions of anisotropic Lie groups and to
the noncentral abelian extensions of
absolute time Lie groups gives rise to three kinds of noncommutative phase spaces:
\begin{itemize}
 \item A phase space whose only positions do not commute, i.e: generators of pure kinematical group transformations are
noncommutative in extended algebras. This arrives with
the Galilei algebra $ G$ by considering both the anisotropic and the absolute time cases, recovering the so called exotic
model \cite{Duval-horvathy}.
\item A phase space whose only momenta are noncommutative due to the noncommutativity of generators of space translations in
extended Lie algebras.
This is the case of the Para-Galilei group $ G^{\prime}$, by considering both the anisotropic and the absolute time cases.
The same result can be obtained also with the Aristotle group ( which is a subgroup of the Galilei group) by considering
one of it noncentral abelian extensions.
\item A phase space which is completely noncommutative, meaning that both generators of pure kinematical group transformations
and generators of space translations do not commute.\;\;This is obtained with the Newton-Hooke groups $ NH_{\pm}$,
the Static group and the Carroll group in both anisotropic and absolute time cases.\\
\end{itemize}
 Note also that the noncommutative phase spaces realized with noncentral abelian extensions are algebraically more
general than those obtained with central extensions.\;\;Indeed, we have remarked
that with noncentral abelian extensions, it is always possible to realize a noncommutative phase space by the coadjoint
orbit method while when uses a central extension, noncommutative phase space shall not exist.\;\;This remark has been well explained
with the Aristotle group.\;\;This will be also clarified in the next Chapter where a comparative analysis between
the centrally and the noncentrally abelian extended results will be given.\;\;Furthermore, in the
noncentrally abelian extended results, we have additional terms in the two-form
symplectic structures.\\

Through these constructions, we have found that each type of noncommutative phase space
corresponds to a minimal coupling of :
\begin{itemize}
 \item  the position with a group theoretically (naturally) introduced dual magnetic field in the position noncommutativity sector,
\item the momentum with a group theoretically (naturally) introduced magnetic field in the momentum noncommutativity sector,
\item the momentum with a group theoretically (naturally) introduced magnetic field and the position with a naturally introduced
dual magnetic field in the noncommutativity of positions as well as momenta cases.
\end{itemize}
\chapter{NONCOMMUTATIVE PHASE SPACES BY LINEAR DEFORMATION OF POISSON BRACKETS}
The mathematical background for the physics on noncommutative spacetimes is the noncommutative geometry \cite{connes10}.\;\;The idea of
noncommutative spacetimes has been used extensively in general classical and quantum field theory
\cite{nekrasov, acatrinei1, falomir}.\;\;In canonical quantization for
instance, one would replace the spacetime coordinates by noncommuting operators and so consider
a noncommutative geometry.\;\;It was shown how the passage from classical
to quantum mechanics or from Newtonian mechanics to special relativity can be understood
as a deformation of algebraic structures \cite{snyder}. \\

A generalization of this concept is to consider deformation of
 Poissonian structure of spaces.\;\;Under some assumptions this generalization yields noncommutative spaces and in the
 particularly case of phase spaces, i.e
 described as symplectic manifolds, we obtain noncommutative symplectic structures.\;\;Furthermore, as it was demonstrated
 in \cite{walczyk}, the spacetime noncommutativity can be distinguished into three kinds
in accordance with the Hopf-algebraic classification
 \cite{zakrzewski}, that is, there exists the canonical, the Lie-algebraic and quadratic noncommutativity respectively.\\

 In this Chapter,
we see in detail the linear deformation of the Poisson bracket as a particular case of the Lie-algebraic noncommutativity and
demonstrate that under some conditions, this gives rise to the noncommutative phase spaces constructed group theoretically
(i.e by the coadjoint orbit method) in the previous Chapter.\\

Our generalization focuses on the Poisson
brackets between different spatial coordinates and momenta.\;\;We point out that the planar
systems with (anisotropic) kinematical group type symmetries are (anisotropic) noncommutative phases
spaces in constant magnetic backgrounds.\\

This Chapter is organized as follows.\\

In the next section, we determine the Poissonian structures associated to the centrally extended planar anisotropic kinematical groups
defined in the first Chapter.\;\;By using the Kirillov-Kostant-Souriau bracket,
the noncommutative spaces obtained are reduced to noncommutative phase spaces on the maximal coadjoint orbits of the corresponding
kinematical groups.\\

In section two, we determine the Poissonian structures associated to the noncentrally abelian extended
absolute time kinematical groups.\;\;Through these constructions, it is also highlighted
that the noncommutative spaces obtained are algebraically general than those obtained in the centrally extensions framework. \\

The comparative analysis between the centrally and the noncentrally abelian extended results is given in the next
section.\;\;Section four is devoted to the linear deformation of the Poisson bracket in a four-dimensional
space.\;\;With some assumptions, we prove that
this gives rise to the same noncommutative phase spaces as those obtained group theoretically, i.e by the coadjoint orbit method. \\
Finally, section five is devoted to the general case, that is the linear deformation of a $2n$-dimensional space and this provides the
fundamental commutation relations and equations of motion of a general noncommutative space.
\section{Poisson manifolds associated to the planar anisotropic kinematical groups}
In Chapter $2$, it has already mentioned that the dual of the Lie algebra of a Lie group has a natural Poisson structure (called
Kirilov-Kostant-Souriau bracket) given by the relation (\ref{kirillov-poissonbracket}).\;\;In the following section, we determine
Poisson manifolds or Poisson-Lie structures associated to the centrally extended planar anisotropic kinematical
groups considered in this thesis.
\subsection{Poisson-Lie structure associated to Carroll group}
Consider the centrally extended Carrollian algebra defined by the brackets (\ref{carrolbrackets}).
The dual (Lie algebra) ${\hat{\cal{G}}}^*$ of the centrally extended Carrollian Lie algebra equipped with the Poisson structure
(\ref{kirillov-poissonbracket}) is a Poisson manifold.\;\;It is
a presymplectic manifold which turns in a symplectic structure on the coadjoint orbit where the Poisson structure corresponding to
the symplectic structure (\ref{sympcarroll}) takes the form (\ref{carrolleanpoissons}).\\

More specifically, let
$X=(\delta v^i, \delta x^i,\delta t, \delta \phi)$ and
 $\mu=(k_i,p_i, E,h)$ be respectively an infinitesimal displacement on Carrollian centrally extended Lie algebra and an element of
its dual Lie algebra
such that
\begin{eqnarray}
 \langle \mu,X \rangle=k_i \delta v^i+p_i \delta x^i+E \delta t+h\delta \phi
\end{eqnarray}
is the associated scalar whose physical dimension is action where $\langle.,.\rangle$ stands for the pairing between the
Lie algebra and its dual.\\

The Poisson bracket (\ref{poissonbrackets}) of two functions $f$ and $g$ in
$C^{\infty}({\hat{\cal{G}}}^*,\Re)$ is in this case
given by :
\begin{eqnarray}
 \{f,g\}=-\frac{h}{m^2c^2}\epsilon_{ij}\frac{\partial f}{\partial q^i}\frac{\partial g}{\partial q^j}+(\frac{\partial f}{\partial p_i}
\frac{\partial g}{\partial q^i}-\frac{\partial f}{\partial q^i}\frac{\partial g}{\partial p_i})
-\kappa^2 h\epsilon_{ij}\frac{\partial f}{\partial p_i}\frac{\partial g}{\partial p_j}
\end{eqnarray}
Then, we obtain a noncommutative space whose coordinates satisfy the following Poisson brackets:
\begin{eqnarray}\label{carrollcpoissons}\nonumber
\{p_i,p_j\}=-\kappa^2 h\epsilon_{ij}~~~~~~~~~~~~~~~~~~~~~~~~~~~~~~~~~~~~~~~~~~~~~~~~~~~~~~~~~~~~~~~~\\
\{p_i,q^j\}=\delta^{j}_{i}~,~ \{q^i,q^j\}=-\frac{h}{m^2 c^2}\epsilon_{ij}~~~~~~~~~~~~~~~~~~~~~~~~~~~~~~~~~~~\\\nonumber
\{p_i,E\}=0~,~\{q^i,E\}=0~~~~~~~~~~~~~~~~~~~~~~~~~~~~~~~~~~~~~~~~~~~~~~~~~~~\\\nonumber
\end{eqnarray}
As it has been already concluded, with the Carroll group, a dynamical system which corresponds to a noncommutative space
can be constructed and its restriction to the coadjoint orbit permits us to recover the noncommutative phase space obtained
in section ($3.3.5$) and equipped with the symplectic form (\ref{sympcarroll}).

\subsection{Poisson-Lie structures associated to centrally extended absolute time anisotropic kinematical groups}

The centrally extended absolute time anisotropic kinematical algebras are defined by the nontrivial Lie brackets
(\ref{anisotropext}) where the parameters are related by the relation
 (\ref{relatio}).

The Kirillov form, in the basis ($K_j,P_j,H,M,S$), is in this case given by:
\begin{eqnarray*}
K_{ij}(a)=\left (\begin{array}{ccccc}
-\frac{h}{m^2 c^2}\mu\epsilon_{ij}&\gamma m\delta_{ij}&\lambda p_j&0&0\\
-\gamma m\delta_{ij}&\kappa^2\alpha h\epsilon _{ij}&\beta k_{i}&0&0\\
-\lambda p_j&-\beta k_j&0&0&0\\
0&0&0&0&0\\
0&0&0&0&0\\
\end{array}
\right)
\end{eqnarray*}
where $k_i, p_i, E, m, h$ span the corresponding dual Lie algebras.\\

The Poisson bracket (\ref{poissonbrackets}) of two functions $f$ and $g$ on dual Lie algebras is then given collectively by:
\begin{eqnarray}\label{globalpoisson}\nonumber
 \{f,g\}=-\frac{h}{m^2c^2}\mu\epsilon_{ij}\frac{\partial f}{\partial q^i}\frac{\partial g}{\partial q^j}+\gamma
(\frac{\partial f}{\partial p_i}
\frac{\partial g}{\partial q^i}-\frac{\partial f}{\partial q^i}\frac{\partial g}{\partial p_i})\\\nonumber
-\kappa^2 \alpha h\epsilon_{ij}\frac{\partial f}{\partial p_i}\frac{\partial g}{\partial p_j}+
\lambda \frac{p_i}{m}(\frac{\partial f}{\partial E}
\frac{\partial g}{\partial q^i}-\frac{\partial f}{\partial q^i}\frac{\partial g}{\partial E})\\
+\beta k_i(
\frac{\partial f}{\partial E}\frac{\partial g}{\partial p_i}-\frac{\partial f}{\partial p_i}\frac{\partial g}{\partial E})~~~~~~
~~~~~~~~~~~~~~~~~
\end{eqnarray}
where we have used the relation (\ref{q^ik^i m}).\\

We then obtain a noncommutative space whose coordinates satisfy the following Poisson brackets:
\begin{eqnarray}\label{carrollpoissons}\nonumber
 \{p_i,p_j\}=-\kappa^2 \alpha h\epsilon_{ij}~~~~~~~~~~~~~~~~~~~~~~~~~~~~\\
\{p_i,q^j\}=\gamma\delta^{j}_{i}~,~ \{q^i,q^j\}=-\frac{h}{m^2 c^2}\mu\epsilon_{ij}\\\nonumber
\{p_i,E\}=-\beta mq^i~,~\{q^i,E\}=-\lambda\frac{p_i}{m}~~~~\nonumber
\end{eqnarray}
Using the normalization $\gamma=1$, we distinguish the following cases:
\begin{itemize}
 \item $ \alpha=-1$, $\beta=\pm \omega^2$ and $\lambda=\mu=1$:\\

In this case, (\ref{globalpoisson}) becomes
\begin{eqnarray}\label{globalpoissonnh}\nonumber
 \{f,g\}=-\frac{h}{m^2c^2}\epsilon_{ij}\frac{\partial f}{\partial q^i}\frac{\partial g}{\partial q^j}+ (\frac{\partial f}{\partial p_i}
\frac{\partial g}{\partial q^i}-\frac{\partial f}{\partial q^i}\frac{\partial g}{\partial p_i})\\\nonumber
+\kappa^2 h\epsilon_{ij}\frac{\partial f}{\partial p_i}\frac{\partial g}{\partial p_j}+
 \frac{p_i}{m}(\frac{\partial f}{\partial E}
\frac{\partial g}{\partial q^i}-\frac{\partial f}{\partial q^i}\frac{\partial g}{\partial E})\\
\pm m\omega^2 q^i(
\frac{\partial f}{\partial E}\frac{\partial g}{\partial p_i}-\frac{\partial f}{\partial p_i}\frac{\partial g}{\partial E})~~~~~~~~~~
~~~~~~~~
\end{eqnarray}
and then
\begin{eqnarray}\label{globalpoissoncoordnh1}\nonumber
 \{p_i,p_j\}=-\kappa^2 h\epsilon_{ij}~~~~~~~~~~~~~~~~~~~~~~~~~~~~\\
\{p_i,q^j\}=\delta^{j}_{i}~,~ \{q^i,q^j\}=-\frac{h}{m^2 c^2}\epsilon_{ij}\\\nonumber
\{p_i,E\}=\mp m\omega^2 q^i~,~\{q^i,E\}=-\frac{p_i}{m}\nonumber
\end{eqnarray}
The system above provides planar noncommutative spaces on Newton-Hooke groups and their restrictions to the coadjoint orbits give rise
to the noncommutative phase spaces obtained in section ($3.3.3$) i.e equipped with the modified symplectic structures given
 by (\ref{symplnh}).

\item  $\alpha =\beta=0$ and $\lambda=\mu=1$:\\

In this case, (\ref{globalpoisson}) becomes
\begin{eqnarray}\label{globalpoissongal}\nonumber
 \{f,g\}=-\frac{h}{m^2c^2}\epsilon_{ij}\frac{\partial f}{\partial q^i}\frac{\partial g}{\partial q^j}+
 (\frac{\partial f}{\partial p_i}
\frac{\partial g}{\partial q^i}-\frac{\partial f}{\partial q^i}\frac{\partial g}{\partial p_i})+
\frac{p_i}{m}(\frac{\partial f}{\partial E}\frac{\partial g}{\partial q^i}-\frac{\partial f}
{\partial q^i}\frac{\partial g}{\partial E})
\end{eqnarray}
and then
\begin{eqnarray}\label{globalpoissongalc}\nonumber
 \{p_i,p_j\}=0~~~~~~~~~~~~~~~~~~~~~~~~~~~~~~~~~~~~~~~~~\\
\{p_i,q^j\}=\delta^{j}_{i}~,~ \{q^i,q^j\}=-\frac{h}{m^2 c^2}\epsilon_{ij}\\\nonumber
\{p_i,E\}=0~,~\{q^i,E\}=-\frac{p_i}{m}~~~~~~~~~\nonumber
\end{eqnarray}
The system above provides planar noncommutative space on Galilei group and its restriction to the orbit gives rise to the noncommutative
 phase space obtained in section ($3.3.1$) i.e equipped with the modified symplectic structure given by (\ref{symplgal}).

 \item $\alpha=1$, $\beta=\pm\omega^2$ and $\lambda=\mu=0$: \\

In this case, (\ref{globalpoisson}) becomes
\begin{eqnarray}\label{globalpoissonparagal}\nonumber
 \{f,g\}= (\frac{\partial f}{\partial p_i}
\frac{\partial g}{\partial q^i}-\frac{\partial f}{\partial q^i}\frac{\partial g}{\partial p_i})
-\kappa^2 h\epsilon_{ij}\frac{\partial f}{\partial p_i}\frac{\partial g}{\partial p_j}\\
\pm m\omega^2 q^i(
\frac{\partial f}{\partial E}\frac{\partial g}{\partial p_i}-\frac{\partial f}{\partial p_i}\frac{\partial g}{\partial E})
\end{eqnarray}
and then
\begin{eqnarray}\label{globalpoissonparagalc}\nonumber
  \{p_i,p_j\}=-\kappa^2 h\epsilon_{ij}~~~~~~~~~~~~~~~~~~~~~~~~~~\\
\{p_i,q^j\}=\delta^{j}_{i}~,~ \{q^i,q^j\}=0~~~~~~~~~~~~~\\\nonumber
\{p_i,E\}=\mp m\omega^2 q^i~,~\{q^i,E\}=0~~~\nonumber
\end{eqnarray}
The system above provides planar noncommutative spaces on Para-Galilei groups and their restrictions to the orbit give rise to
the noncommutative phase spaces obtained in section ($3.3.2$) i.e equipped with the modified symplectic structures
given by (\ref{symplecticpara-1}).

 \item $\mu=\alpha=1$ and $\lambda=\beta=0$:\\

In this case, (\ref{globalpoisson}) becomes
\begin{eqnarray}\label{globalpoissostatic}
 \{f,g\}=-\frac{h}{m^2c^2}\epsilon_{ij}\frac{\partial f}{\partial q^i}\frac{\partial g}{\partial q^j}+ (\frac{\partial f}{\partial p_i}
\frac{\partial g}{\partial q^i}-\frac{\partial f}{\partial q^i}\frac{\partial g}{\partial p_i})
-\kappa^2 h\epsilon_{ij}\frac{\partial f}{\partial p_i}\frac{\partial g}{\partial p_j}
\end{eqnarray}
and then
\begin{eqnarray}\label{globalpoissoncoordstatic}\nonumber
 \{p_i,p_j\}=-\kappa^2 h\epsilon_{ij}~~~~~~~~~~~~~~~~~~~~~~~~~~~~~~~~~~~~\\
\{p_i,q^j\}=\delta^{j}_{i}~,~ \{q^i,q^j\}=-\frac{h}{m^2c^2}\epsilon^{ij}~~~~~~~~~\\\nonumber
\{p_i,E\}=0~,~\{q^i,E\}=0~~~~~~~~~~~~~~~~~~~~~~~~\nonumber
\end{eqnarray}
\end{itemize}
The system above provides planar noncommutative space on Static group and its restriction to the orbit gives rise to the noncommutative
 phase space obtained in section ($3.3.4$) i.e equipped with the modified symplectic structures given by (\ref{symplstatic}).

\section{Poisson-Lie structures associated to noncentrally abelian extended absolute time kinematical groups}
Noncentrally extensions of absolute time kinematical groups have been defined in the last section of the first Chapter.

In this section, we determine their associated Poisson structures and hence the corresponding noncommutative spaces.\;\;In
general, the obtained structures are those obtained in the previous sections i.e with central extensions of anisotropic
kinematical groups with additional terms due to the presence of rotational generator (isotropy case).\\

For example, for the noncentrally abelian extended Newton-Hooke absolute time Lie algebras defined by the Lie brackets
(\ref{anisotropicextncnh}), the
associated Poissonian structures are given by
\begin{eqnarray*}\nonumber
 \{f,g\}=(\ref{globalpoissonnh})+
p_j\epsilon^{ij}(\frac{\partial f}{\partial j}\frac{\partial g}{\partial p_i}-
\frac{\partial f}{\partial p_i}\frac{\partial g}{\partial j})
+q^j\epsilon_{ij}(\frac{\partial f}{\partial j}\frac{\partial g}{\partial q^i}-
\frac{\partial f}{\partial q^i}\frac{\partial g}{\partial j})
\end{eqnarray*}
where the Kirillov form in the basis ($J, K_i, P_i,H,M,S$) is given by
\begin{eqnarray*}\label{kirillovnhnc}
(K_{ij})=\left(
\begin{array}{cccccc}
0&k_i\epsilon^{i}_j&p_i\epsilon^{i}_j&0&0&0\\
-k_j\epsilon^{j}_i&\frac{h}{c^2}\epsilon^{ij}&m\delta_{ij}&p_i&0&0\\
-p_j\epsilon^{j}_i&-m\delta_{ij}&\kappa^2 h\epsilon^{ij}&\pm \omega^2 k_i&0&0\\
0&-p_j&\mp\omega^2 k_j&0&0&0\\
0&0&0&0&0&0\\
0&0&0&0&0&0\\
\end{array}
\right)
\end{eqnarray*}
 and where we have used the relation (\ref{q^ik^i m}).\\
It holds that:
\begin{eqnarray}\label{globalpoissoncoordnh}\nonumber
 \{j,p_k\}=p_i\epsilon^i_k~~,\{p_i,p_j\}=-\kappa^2h\epsilon_{ij}~~~~~~~~~~~~~~~~~~~~~~~~~~~~~~~~\\
 \{j,q^k\}=q^i\epsilon^k_i~~,\{p_i,q^j\}=\delta^j_i~~,~~\{q^i,q^j\}=-\frac{h}{m^2c^2}\epsilon^{ij}~~~~\\\nonumber
 \{j,E\}=0~~~~,\{p_i,E\}=\mp m\omega^2q^i,~~\{q^i,E\}=-\frac{p_i}{m}~~~~~~\\\nonumber
\end{eqnarray}
For the noncentrally abelian extended Galilei absolute time Lie algebra defined by the Lie brackets
(\ref{galileinoncentral}), the associated Poissonian structure is
\begin{eqnarray*}\nonumber
  \{f,g\}=(\ref{globalpoissongal})+
p_i\epsilon^{i}_j(\frac{\partial f}{\partial j}\frac{\partial g}{\partial p_j}-
\frac{\partial f}{\partial p_j}\frac{\partial g}{\partial j})
+q^i\epsilon^{i}_j(\frac{\partial f}{\partial j}\frac{\partial g}{\partial q^j}-
\frac{\partial f}{\partial q^j}\frac{\partial g}{\partial j})+
f_i\epsilon^{i}_j(\frac{\partial f}{\partial j}\frac{\partial g}{\partial f_j}-
\frac{\partial f}{\partial f_j}\frac{\partial g}{\partial j})
\end{eqnarray*}
where the Kirillov form in the basis ($J, K_i, P_i,F_i,H,M,S$) is given by
\begin{eqnarray*}\label{kirillovnhnc}
(K_{ij})=\left(
\begin{array}{ccccccc}
0&k_i\epsilon^{i}_j&p_i\epsilon^{i}_j&f_i\epsilon^{i}_j&0&0&0\\
-k_j\epsilon^{j}_i&\frac{h}{c^2}\epsilon^{ij}&m\delta_{ij}&0&p_i&0&0\\
-p_j\epsilon^{j}_i&-m\delta_{ij}&0&0&f_i&0&0\\
-f_j&0&0&0&0&0&0\\
0&-p_j&-f_j&0&0&0&0\\
0&0&0&0&0&0&0\\
0&0&0&0&0&0&0\\
\end{array}
\right)
\end{eqnarray*}
 and where we have used the relation (\ref{q^ik^i m}).\\

It holds that:
\begin{eqnarray}\label{globalpoissoncoordgal}\nonumber
 \{j,p_i\}=p_k\epsilon ^{k}_i,~~\{p_i,p_j\}=0~~~~~~~~~~~~~~~~~~~~~~~~~~~~~~~~~~~~~~~~~~~~~~~~~~~~~~~~~~~~~~\\
\{j,q^i\}=q^k\epsilon ^{i}_k,~~\{p_i,q^j\}=\delta^{j}_{i},~~\{q^i,q^j\}=-\frac{h}{m^2 c^2}\epsilon^{ij}~~~~~~~~~~~~~~~~~~~~~~~\\\nonumber
\{j,f_i\}=f_k\epsilon ^{k}_i,~~\{p_i,f_j\}=0,~~\{q^i,f_j\}=0,~~\{f_i,f_j\}=0~~~~~~~~~~~~~~\\\nonumber
\{j,E\}=0,~~\{p_i,E\}=-f_i,~~\{q^i,E\}=-\frac{p_i}{m},~\{f_i,E\}=0~~~~~~~~~~~\\\nonumber
\end{eqnarray}
For the noncentrally abelian extended Para-Galilei absolute time Lie algebras defined by the Lie brackets
(\ref{anisotropicnoncentralparag}), the
associated Poissonian structures are given by
\begin{eqnarray*}\nonumber
  \{f,g\}=(\ref{globalpoissonparagal})+p_i\epsilon^{i}_j(\frac{\partial f}{\partial j}\frac{\partial g}{\partial p_j}-
\frac{\partial f}{\partial p_j}\frac{\partial g}{\partial j})
+q^i\epsilon^{i}_j(\frac{\partial f}{\partial j}\frac{\partial g}{\partial q^j}-
\frac{\partial f}{\partial q^j}\frac{\partial g}{\partial j})+
\pi_i\epsilon^{i}_j(\frac{\partial f}{\partial j}\frac{\partial g}{\partial \pi_j}-
\frac{\partial f}{\partial \pi_j}\frac{\partial g}{\partial j})
\end{eqnarray*}
where the Kirillov form in the basis ($J, K_i, P_i,\Pi_i,H,M,S$) is given by
\begin{eqnarray*}\label{kirillovnhnc}
(K_{ij})=\left(
\begin{array}{ccccccc}
0&k_i\epsilon^{i}_j&p_i\epsilon^{i}_j&\pi_i\epsilon^{i}_j&0&0&0\\
-k_j\epsilon^{j}_i&0&m\delta_{ij}&0&\pi_i&0&0\\
-p_j\epsilon^{j}_i&-m\delta_{ij}&\kappa^2 h \epsilon_{ij}&\pm \omega^2 k_i&\pm\omega^2k_i&0&0\\
-\pi_j\epsilon^{j}_i&0&0&0&0&0&0\\
0&-\pi_j&\mp \omega^2 k_j&0&0&0&0\\
0&0&0&0&0&0&0\\
0&0&0&0&0&0&0\\
\end{array}
\right)
\end{eqnarray*}
 and where we have used the relation (\ref{q^ik^i m}).\\

It holds that:
\begin{eqnarray}\label{globalpoissoncoordparagal}\nonumber
 \{j,p_i\}=p_k\epsilon ^{k}_i,~~\{p_i,p_j\}=-\kappa^2h\epsilon_{ij}~~~~~~~~~~~~~~~~~~~~~~~~~~~~~~~~~~~~~~~~~~~~~~~~~~
 ~~~~~\\
\{j,q^i\}=q^k\epsilon ^{i}_k,~~\{p_i,q^j\}=\delta^{j}_{i},~~\{q^i,q^j\}=0~~~~~~~~~~~~~~~~~~~~~~~~~~~~~~~~~~~~~~
~~~\\\nonumber
\{j,\pi_i\}=\pi_k\epsilon ^{k}_i,~~\{p_i,\pi_j\}=0,~~\{q^i,\pi_j\}=0,~~\{\pi_i,\pi_j\}=0~~~~~~~~~~~~~~~~\\\nonumber
\{j,E\}=0,~~\{p_i,E\}=\mp m\omega^2q^i,~~\{q^i,E\}=-\frac{\pi_i}{m},~~~~\{\pi_i,E\}=0~~~~\\\nonumber
\end{eqnarray}
For the noncentrally abelian extended Static absolute time Lie algebra defined by the Lie brackets
(\ref{anisotropicnoncentralstatic}), the
associated Poissonian structures are given by
\begin{eqnarray*}\nonumber
  \{f,g\}=p_i\epsilon^{i}_j(\frac{\partial f}{\partial j}\frac{\partial g}{\partial p_j}-
\frac{\partial f}{\partial p_j}\frac{\partial g}{\partial j})
+q^i\epsilon^{i}_j(\frac{\partial f}{\partial j}\frac{\partial g}{\partial q^j}-
\frac{\partial f}{\partial q^j}\frac{\partial g}{\partial j})\\+f_i\epsilon^{i}_j(\frac{\partial f}{\partial j}\frac{\partial g}
{\partial f_j}-\frac{\partial f}{\partial f_j}\frac{\partial g}{\partial j})
+\pi_i\epsilon^{i}_j(\frac{\partial f}{\partial j}\frac{\partial g}{\partial \pi_j}-
\frac{\partial f}{\partial \pi_j}\frac{\partial g}{\partial j})+(\frac{\partial f}{\partial p_i}
\frac{\partial g}{\partial q^i}-\frac{\partial f}{\partial q^i}\frac{\partial g}{\partial p_i})\\
+\kappa(\frac{\partial f}{\partial f_i}
\frac{\partial g}{\partial p_i}-\frac{\partial f}{\partial p_i}\frac{\partial g}{\partial f_i})+
B(\frac{\partial f}{\partial \pi_i}
\frac{\partial g}{\partial p_i}-\frac{\partial f}{\partial p_i}\frac{\partial g}{\partial \pi_i})+
\frac{m^{\prime}}{m}(\frac{\partial f}{\partial \pi_i}
\frac{\partial g}{\partial q^i}-\frac{\partial f}{\partial q^i}\frac{\partial g}{\partial \pi_i})\\
+\frac{\pi}{m}(\frac{\partial f}{\partial E}
\frac{\partial g}{\partial q^i}-\frac{\partial f}{\partial q^i}\frac{\partial g}{\partial E})+
f_i(\frac{\partial f}{\partial E}
\frac{\partial g}{\partial p_i}-\frac{\partial f}{\partial p_i}\frac{\partial g}{\partial E})
\end{eqnarray*}
where the Kirillov form in the basis ($J, K_i, P_i,F_i,\Pi_i,H,M,S,\Lambda,M^{\prime}$) is given by
\begin{eqnarray*}\label{kirillovnhnc}
(K_{ij})=\left(
\begin{array}{cccccccccc}
0&k_i\epsilon^{i}_j&p_i\epsilon^{i}_j&f_i\epsilon^{i}_j&\pi_i\epsilon^{i}_j&0&0&0&0&0\\
-k_j\epsilon^{j}_i&0&m\delta_{ij}&B\delta_{ij}&m^{\prime}\delta_{ij}&\pi_i&0&0&0&0\\
-p_j\epsilon^{j}_i&-m\delta_{ij}&0&\kappa\delta_{ij}&B\delta_{ij}&f_i&0&0&0&0\\
-f_j\epsilon^{j}_i&-B\delta_{ij}&-\kappa\delta_{ij}&0&0&0&0&0&0&0\\
-\pi_j\epsilon^{j}_i&-m^{\prime}\delta_{ij}&-B\delta_{ij}0&0&0&0&0&0&0\\
0&-\pi_j&-f_j&0&0&0&0&0&0&0\\
0&0&0&0&0&0&0&0&0&0\\
0&0&0&0&0&0&0&0&0&0\\
0&0&0&0&0&0&0&0&0&0\\
0&0&0&0&0&0&0&0&0&0\\
\end{array}
\right)
\end{eqnarray*}
 and where we have used the relation (\ref{q^ik^i m})\\

It holds that:
\begin{eqnarray}\label{globalpoissoncoordst}\nonumber
\{j,p_i\}=p_k\epsilon ^{k}_i,~~\{p_i,p_j\}=-\kappa^2h\epsilon_{ij}~~~~~~~~~~~~~~~~~~~~~~~~~~~~~~~~~~~~~~~~~~~~~~~~~~~~~~~
~~~~~~~~~~~~~~~~~~~~~~~~~~~~~~\\
\{j,q^i\}=q^k\epsilon ^{i}_k,~~\{p_i,q^j\}=\delta^{j}_{i},~~\{q^i,q^j\}=0~~~~~~~~~~~~~~~~~~~~~~~~~~~~~~~~~~~~~~~~~~~~~~
~~~~~~~~~~~~~~~~~~~~~~~~~\\\nonumber
\{j,f_i\}=f_k\epsilon^k_i,~\{p_i,f_j\}=-\kappa \delta_{ij},~\{q^i,f_j\}=-\frac{B}{m}\delta^i_j,~~~\{f_i,f_j\}=0~~~~~~~~~~~~~~~~~~
~~~~~~~~~~~~~~~\\\nonumber
\{j,\pi_i\}=\pi_k\epsilon ^{k}_i,~~\{p_i,\pi_j\}=-B\delta_{ij},~~\{q^i,\pi_j\}=-\frac{m^{\prime}}{m}\delta^i_j,~~\{f_i,\pi_j\}=0~,
~~\{\pi_i,\pi_j\}=0~~\\\nonumber
\{j,E\}=0~~~~,~~\{p_i,E\}=-f_i,~~\{q^i,E\}=-\frac{\pi_i}{m},~~\{f_i,E\}=0~,~\{\pi_i,E\}=0~~~~~~~~~~~~~~\\\nonumber
\end{eqnarray}
Let us compare the results of the two previous sections in order to highlight the fact that noncommutative spaces
constructed on kinematical groups by considering their noncentral abelian extensions are algebraically more general than those obtained with
central extensions.
\section{Comparative analysis of the centrally and the noncentrally abelian extended results}
Note that the motion equations are given by the last line in each system of the Poisson brackets
(\ref{globalpoissoncoordnh}),(\ref{globalpoissoncoordgal}),(\ref{globalpoissoncoordparagal}) and
(\ref{globalpoissoncoordst}). \\

First of all in the centrally extended cases, the momentum $p_i$ changes
due to a position dependent force while the position $q^i$ changes due to a momentum dependent velocity in
the Newton-Hooke groups case,
it is only the position $q^i$ which changes due to a momentum dependent velocity in the Galilei group case, it is only
the momentum $p_i$ which changes due to a position dependent force in the Para-Galilei groups case and the position $q^i$
as well the momentum $p_i$ are constant in the Static and Carroll groups cases.\\

In the noncentrally abelian extended groups cases, the momentum $p_i$ which was constant in the Galilei and Static cases is now
changing due to a constant force $f_i$, the position $q^i$ which was constant in the Para-Galilei and Static cases
is now changing due to a constant momentum $\pi_i$.\;\;Moreover the first column of (\ref{globalpoissoncoordnh}),
(\ref{globalpoissoncoordgal}), (\ref{globalpoissoncoordparagal}) and
(\ref{globalpoissoncoordst}) shows that the momenta $p_i$ and $\pi_i$, the positions $q^i$ and the forces $f_i$ behave
as components of vectors while the energy $E$ behaves as a scalar under rotation through the Poisson brackets with an angular
momentum $j$.\\

Furthermore, comparing relations: (\ref{globalpoissoncoordnh1} and (\ref{globalpoissoncoordnh}) (for Newton-Hooke groups case),
(\ref{globalpoissongalc}) and (\ref{globalpoissoncoordgal}) (for Galilei group case),
(\ref{globalpoissonparagalc}) and (\ref{globalpoissoncoordparagal}) (for Para-Galilei groups case),
(\ref{globalpoissoncoordstatic}) and (\ref{globalpoissoncoordst}) (for Static group case),
we can conclude that noncentrally abelian extended results are algebraically more general than those obtained
with centrally extended structures.
\section{Possible four-dimensional noncommutative phase spaces by linear deformation}
Let us consider the phase space with coordinates ($ p_1,p_2,q^1,q^2)$ such that the Poisson brackets be given by:
\begin{eqnarray}\label{generalpoisson}
 \{p_i,p_j\}= (a_{k}q^k)\epsilon_{ij}+n\epsilon_{ij},~
\{q^i,q^j\}=(p_k b^{k})\epsilon^{ij}+d\epsilon^{ij},~\{p_j,q^i\}=\delta^i_j\\\nonumber
\end{eqnarray}
and such that the equations of motion are
\begin{eqnarray}\label{motioneq}
\{E,p_i\}=w_{ik}q^k ,~\{E,q^i\}=p_k r^{ki}
\end{eqnarray}
It is a linear deformation (of the form Lie-algebraic deformation given in \cite{walczyk}) of the
canonical Poisson brackets of phase space coordinates.\\

The dimensions of the structure constants also called
the noncommutative parameters are fixed: $[a_k]=ML^{-1}T^{-1}$, $[b^k]=M^{-2}LT^2$, $[w_{ij}]=MT^{-2}$, $[r^{ij}]=M^{-1}$,$ [n]=
 MT^{-1}$, $[d]=M^{-1}T$ where $M$, $L$ and $T$ stand for mass, length and time respectively.\\

The Jacobi identities with two $p$'s and one $q$ give rise to the constraints $a_k=0$.\;\;The Jacobi identities with
two $q$'s and one $p$ give rise to the constraints $b^k=0$.\;\;The Jacobi identity with two
$p$'s and $H$ adds that the matrix $(W=(w_{ij}))$ is symmetric while that with two $q$'s and $E$
adds that the matrix $(R=(r^{ij}))$ is also symmetric.\\
Lastly the Jacobi identities with one $p$, one $q$ and $E$ imply that the two matrices $(w_{ij})$ and $(r^{ij})$ are related by
\begin{eqnarray}
 d w_{kl}=n\epsilon_{ki}\epsilon_{jl}r^{ij}
\end{eqnarray}
or equivalently:
\begin{eqnarray*}
d\left(\begin{array}{cc}
w_{11}&w_{12}\\ w_{12}&w_{22}
\end{array}
\right)=n\left(\begin{array}{cc}
-r^{22}&r^{12}\\r^{12}&-r^{11}
\end{array}
\right)
\end{eqnarray*}
As the matrices  $W$ and $R$ are diagonalizable, let us suppose that $W=diag(w,w)$ and
$R=diag(\frac{1}{m},\frac{1}{m})$ respectively.\\

Then the Poisson brackets (\ref{generalpoisson}) become
\begin{eqnarray}
  \{p_i,p_j\}= n \epsilon_{ij},~
\{q^i,q^j\}=d\epsilon^{ij},~\{p_j,q^i\}=\delta^i_j\\\nonumber
\end{eqnarray}
while the equations of motion (\ref{motioneq}) are written as:
\begin{eqnarray}
\{E,p_i\}=-\frac{n}{md}\delta_{ij}q^j ,~\{E,q^i\}=\frac{p_j}{m}\delta^{ij}
\end{eqnarray}
We distinguish five interesting cases according to the possible values of the nontrivial noncommutative parameters, i.e:
case I stands for $n\neq0, d\neq0, r\neq0$, case II stands for $n\neq0, d\neq0, r=0$,
case III stands for $n\neq0, d=0, r\neq0$, case IV stands for $n=0, d\neq0, r\neq0$ and case V stands for
 $n=0, d=0, r=0$, as summarized in the following table:
\begin{table}[htbp]
\begin{tabular}{|c|c|c|c|c|c|}
\hline
$ ~ $& case~I&case~II&case~III&case IV&case V\\
\hline
$\{p_i,p_j\}$&$n\epsilon_{ij}$&$n\epsilon_{ij}$&$n\epsilon_{ij}$&$0$&$0$\\
\hline
$\{q^i,q^j\}$&$d\epsilon_{ij}$&$d\epsilon_{ij}$&$0$&$d\epsilon_{ij}$&$0$\\
\hline
$\{p_j,q^i\}$&$\delta^{i}_j$&$\delta^{i}_j$&$\delta^{i}_j$&$\delta^{i}_j$&$\delta^{i}_j$\\
\hline
$\{E,q^i\}$&$\frac{p_j}{m}\delta^{ij}$&$0$&$0$&$\frac{p_j}{m}\delta^{ij}$&$\frac{p_j}{m}\delta^{ij}$\\
\hline
 $\{E,p_i\}$&$-\frac{n}{m d}\delta_{ij}q^j$&$0$&$wq^i$&$0$&$w q^i$\\
 \hline
\end{tabular}
{\bf \caption{\it Lie-algebraic noncommutative phase spaces in a two-dimensional space }}
\end{table}
\\
Setting $n=eB$, $d=e^*B^*$  and $w=\pm m\omega^2$ where $B$ and $B^*$ are the magnetic and dual magnetic fields, $e$ is the
electric charge while $e^*$ is its dual charge, then
the case I corresponds to the two anisotropic Newton-Hooke noncommutative phase spaces
(i.e by considering the central extensions of the two Newton-Hooke groups), the case II corresponds to the both anisotropic Static and
Carrollian noncommutative phase spaces with the assumption that $E=mc^2$ (i.e Einstein's formula) in the Carroll group case.\;\;The
case III corresponds to the anisotropic Para-Galilean
noncommutative phase spaces while case IV corresponds to the anisotropic Galilean
noncommutative phase space.\;\;Finally, the case V corresponds to the canonical one (commutative case).\\

Furthermore, using the wave-particle duality (\ref{wave-partiduality}) and the equality $c=\frac{\omega}{\kappa}$ in relations
 (\ref{carrollcpoissons}), (\ref{globalpoissoncoordnh1}), (\ref{globalpoissongalc}), (\ref{globalpoissonparagalc}),
(\ref{globalpoissoncoordstatic}), the above results become equivalent to the Poissonian structures constructed in the previous section
and their restrictions to the corresponding coadjoint orbits give rise to the same symplectic structures as those constructed group
theoretically.\\
\\
\\
\\

We have thus proved the following theorem:
\begin{thm}
 The planar systems with anisotropic kinematical group type symmetries are noncommutative anisotropic phase spaces in uniform
 magnetic backgrounds.\;\;Full planar kinematical group symmetries (with rotation) give rise to noncommutative phase spaces
 in the isotropic case.
\end{thm}

As we have already seen, we can realize these noncommutative phase spaces by the coadjoint orbit method or, in the anisotropic case,
by linear deformation of the Poisson bracket.\;\;Every type corresponds to a minimal coupling.\\

Among them, the most general ones are the anisotropic noncommutative Newton-Hooke phase spaces (case $I$ in the above table)
which model effectively an anisotropic oscillator.\;\;This result recovers the result obtained in \cite{zhang} in the oscillating
Newton-Hooke group case.

\section{$2n$-dimensional possible noncommutative phase spaces by linear deformation}

We start with Poisson brackets of coordinates on a $2n$-dimensional noncommutative phase space ($n\geq 3$) defined by
\begin{eqnarray}\label{genpoisson}
 \{p_i,p_j\}=a_{ijk}q^k+\alpha_{ij}~,~\{p_j,q^i\}=\delta_j^i~,~\{q^i,q^j\}=p_k b^{kij}+\beta^{ij}
\end{eqnarray}
where the structure constants $a_{ijk}$, $b^{kij}$, $\alpha_{ij}$, $\beta^{ij}$ also called the noncommutative parameters are
characterized by
\begin{eqnarray}
a_{ijk}=-a_{jik}~,~b^{kij}=-b^{kji}~,~\alpha_{ij}=-\alpha_{ji}~,~\beta^{ij}=-\beta^{ji}
\end{eqnarray}
We also suppose that the motion equations are
\begin{eqnarray}\label{motionequations}
\{E,p_i\}=\lambda_{ik}q^k~,~\{E,q^i\}=p_k r^{ki}
\end{eqnarray}
  $\lambda_{ij}$ and $r^{ij}$ being also constants.\;\;Note that $E$ stands for the Hamiltonian function or the total energy.\\
The Jacobi identity with three $p$ implies the constraint equations
\begin{eqnarray}\label{threep}
a_{ijk}+a_{jki}+a_{kij}=0
\end{eqnarray}
while that with three q implies
\begin{eqnarray}\label{threeq}
b^{ijk}+b^{jki}+b^{kij}=0
 \end{eqnarray}
The Jacobi identity with two $p$ and one $q$ implies that
\begin{eqnarray}\label{twoponeq}
a_{ijl}\beta^{lk}=0~,~a_{ijl}b^{mkl}=0
\end{eqnarray}
for each fixed value of $m$.\;\;Similarly the Jacobi identity with two $q$ and one $p$ implies that
\begin{eqnarray}\label{twoqonep}
\alpha_{il}b^{ljk}=0~,~a_{ilm}b^{ljk}=0
\end{eqnarray}
for each fixed value of $m$.\\The Jacobi identity with two $p$ and $E$ implies that
\begin{eqnarray}\label{twoponeH}
\lambda_{ij}=\lambda_{ji},~a_{ijk}r^{kl}=0
\end{eqnarray}
for each fixed value of $l$.\;\;Similarly, the Jacobi identity with two $q$ and $E$ implies that
\begin{eqnarray}\label{twoqoneH}
r^{ij}=r^{ji},~\lambda_{kl}b^{kij}=0
\end{eqnarray}
for each fixed value of $l$.\\Finally the Jacobi identity with one $p$, one $q$ and $E$ implies that
\begin{eqnarray}
\alpha_{ik}r^{kj}=\lambda_{ik}\beta^{kj}~,~a_{ikl}r^{kj}=0~,~\lambda_{ik}b^{lkj}=0
\end{eqnarray}
for each fixed value of $l$.\;\;From the symmetry of $r^{ij}$ and $\lambda_{ij}$, we
can set that\\ $r^{ij}=\frac{1}{m}\delta^{ij}$ , $\lambda_{ij}=w\delta_{ij}$.\\

Then, it follows that
$\alpha_{ij}=mw\beta^{ij}$ and that $a_{ijk}=b^{ijk}=0$.\\The Poisson brackets (\ref{genpoisson}) become in this case
\begin{eqnarray}\label{genpoisson1}
 \{p_i,p_j\}=mw\beta_{ij}~,~\{p_j,q^i\}=\delta_j^i~,~\{q^i,q^j\}=\beta^{ij}
\end{eqnarray}
while the motion equations (\ref{motionequations}) become
\begin{eqnarray}\label{motionequations1}
\{E,p_i\}=w q^i~,~\{E,q^i\}=\frac{p_i}{m}
\end{eqnarray}
The Poisson brackets (\ref{genpoisson1}) realize the case where the momenta as well the positions do not
commute, the momenta change due to a position dependent force while the positions change due
to a momenta dependent velocity.\;\;This provides the general commutation relations for the $2n$-dimensional
noncommutative space where $mw\beta_{ij}$ and $\beta^{ij}$ are the noncommutative parameters.
\chapter*{Conclusion and outlook}
To conclude, let us revisit the main results we can across in this thesis.\\
Recall that the main focus was to provide a detailed investigation on noncommutative phase spaces
by group theory in a universe with two space and one time dimensions.\;\;We have first provided
a detailed account of the algebraic structures of the centrally and noncentrally abelian extended planar kinematical
groups (which act transitively on manifolds).\;\;We then went on to
describe the construction of noncommutative phase spaces by introducing minimal couplings and turning to the
extended structures above, we applied the coadjoint orbit method
to construct and classify noncommutative phase spaces of planar anisotropic and absolute time kinematical Lie groups.\;\;We
have distinguished between the following three cases: noncommutative phase spaces whose only positions do not commute,
noncommutative phase spaces whose only momenta do not commute and noncommutative phase spaces whose both positions and momenta do not
commute.\;\;Each case was realized group theoretically
and corresponded to a specific minimal coupling.\;\;We have proved also
that the group theoretical discussion, which allows the study of dynamical systems and in the same time
eliminates the non minimal
couplings in them, can be compared with the linear deformation of the Poisson bracket in a $2n$-dimensional phase space.\\

We have therefore described the methodology and the details relative to noncommutative phase spaces in $(2+1)-$dimensional spaces.\\

While we hope this analysis is complete, it would be unfair to label it fully conclusive.
There are some open questions that the author feels are left unanswered and therefore deserve further researching.\;\;In particular, we
would like to mention three of them:
\begin{itemize}
 \item Since the kinematical groups discussed here are contractions from the de Sitter or the anti-de Sitter group and therefore
 still related themselves through the contraction process \cite{inonu}, it would be interesting to verify if the
 extended structures and hence the obtained noncommutative phase spaces are also related via contractions;
\item It would be also a great step to find a star product corresponding to the noncommutative momenta sector.\;\;Indeed, it
has been argued that when the momenta of the phase space do not commute, one would have to envisage the corresponding star-product
which is still unknown \cite{yan}.
\item Noncommutative phase spaces constructed on some kinematical groups in this thesis provide an interpretation of the modified
 Newton's equations (e.g: as a damping force (\cite{wei}).\;\;It would also be important
to see a more in depth analysis of the associated dynamics and possible physical applications, as
it has been done with the Hill's problem for the
Newton-Hooke groups (\cite{zhang-horvathy1}).
\end{itemize}

\bibliographystyle{plain}

\end{large}
 \end{document}